\documentclass[12pt,aip,jcp,showpacs,floatfix,nofootinbib,preprintnumbers,superscriptaddress,amsmath,amssymb]{revtex4-2}
\usepackage{graphicx}% Include figure files
\usepackage{dcolumn}% Align table columns on decimal point
\usepackage{bm}% bold math
\usepackage{xcolor}
\usepackage[utf8]{inputenc}
\usepackage[T1]{fontenc}
\usepackage{mathptmx}
\usepackage{etoolbox}
\usepackage{appendix}
\usepackage{enumerate}
\usepackage{orcidlink}

\definecolor{myred}{rgb}{0.5, 0, 0.13}
\definecolor{myblue}{rgb}{0, 0.2, 1.0}
\definecolor{mycor}{rgb}{1, 0.6, 0}
\definecolor{mycom}{rgb}{1, 0.4, 0}

\newcommand{\ABNew}[1]{{\textcolor{myblue}{#1}}}

\usepackage{amsthm} %%% to make the \begin{proof} and \end{proof}. Only used once.
\newtheorem{Theorem}{Theorem}[section] %%% to make the \begin{Theorem} and \end{Theorem}. Only used once.
\newcommand{\ud}{\mathrm{d}} %%% For use in integrals, e.g., $\int f(x) \ud x$

\begin{document}

%\preprint{AIP/123-QED}

\title{Exact lattice summations for Lennard-Jones potentials coupled to a three-body Axilrod-Teller-Muto term applied to cuboidal phase transitions}

%\title[Cuboidal Lattices]{The Cuboidal Lattices and their Lattice Sums}
% Force line breaks with \\
\author{Andres Robles-Navarro\,\orcidlink{0000-0001-9586-1003}}\email{andres.robles.n@gmail.com}
\affiliation{Centre for Theoretical Chemistry and Physics, The New Zealand Institute for Advanced Study (NZIAS), Massey University Albany, Private Bag 102904, Auckland 0745, New Zealand}

\author{Shaun Cooper\,\orcidlink{0000-0001-5103-0400}}\email{s.cooper@massey.ac.nz}
\affiliation{School of Natural and Computational Sciences, Massey University Albany, Private Bag 102904,
Auckland 0745, New Zealand.}

\author{Andreas A. Buchheit\,\orcidlink{0000-0003-4004-713X}}\email{buchheit@num.uni-sb.de}
\affiliation{%
Department of Mathematics, Saarland University, PO 15 11 50, D-66041, Saarbr\"ucken, Germany}

\author{Jonathan Busse\,\orcidlink{0009-0001-3323-3455}}\email{jonathan@jbusse.de}
\affiliation{%
Department of Mathematics, Saarland University, PO 15 11 50, D-66041, Saarbr\"ucken, Germany}
\affiliation{%
German Aerospace Center (DLR), 51147 Cologne, Germany}

\author{Antony Burrows}\email{antony@yottabyte27.com}
\affiliation{%
Centre for Theoretical Chemistry and Physics, The New Zealand Institute for Advanced Study (NZIAS), Massey University Albany, Private Bag 102904, Auckland 0745, New Zealand.}

\author{Odile Smits\,\orcidlink{0000-0003-1259-147X}}\email{smits.odile.rosette@gmail.com}
\affiliation{%
Centre for Theoretical Chemistry and Physics, The New Zealand Institute for Advanced Study (NZIAS), Massey University Albany, Private Bag 102904, Auckland 0745, New Zealand.}

\author{Peter Schwerdtfeger\,\orcidlink{0000-0003-4845-686X}}\email{peter.schwerdtfeger@gmail.com}
\affiliation{%
Centre for Theoretical Chemistry and Physics, The New Zealand Institute for Advanced Study (NZIAS), Massey University Albany, Private Bag 102904, Auckland 0745, New Zealand.}

\date{\today}

\begin{figure}[b]
TOC figure
\centering
    \includegraphics[scale=0.8]{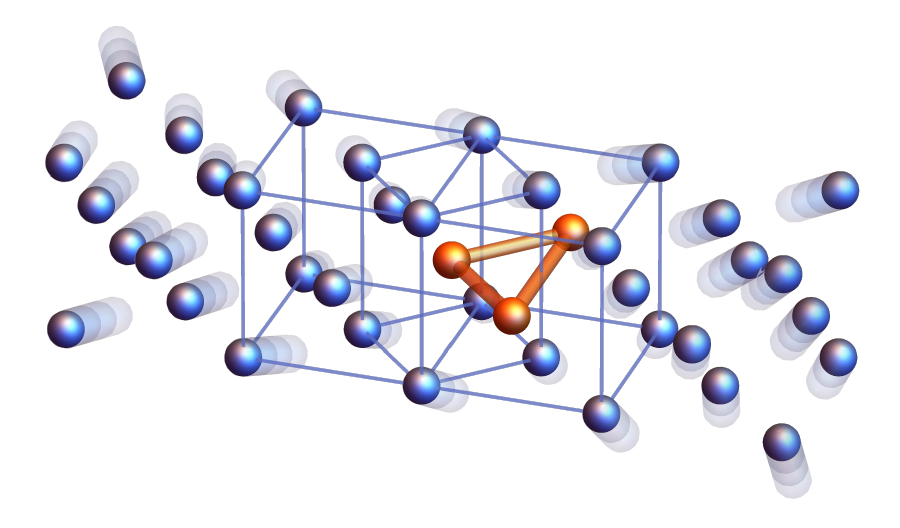}
\end{figure}

\begin{abstract}
Three-body interactions have long been conjectured to play a crucial role in the stability of matter. However, rigorous studies have been scarce due to the computational challenge of evaluating small energy differences in high-dimensional lattice sums. This work provides a rigorous analysis of Bain-type cuboidal lattice transformations, which connect the face-centered cubic (fcc), mean-centered cubic (mcc), body-centered cubic (bcc) and axially centered cubic (acc) lattices. Our study incorporates a general $(n,m)$ Lennard-Jones two-body potential and a long-range repulsive Axilrod-Teller-Muto (ATM) three-body potential. The two-body lattice sums and their meromorphic continuations are evaluated to full precision using super-exponentially convergent series expansions. Furthermore, we introduce a novel approach to computing three-body lattice sums by converting the multi-dimensional sum into an integral involving products of Epstein zeta functions. This enables us to evaluate three-body lattice sums and their meromorphic continuations to machine precision within minutes on a standard laptop.  Using our computational framework, we analyze the stability of cuboidal lattice phases relative to the close-packed fcc structure along a Bain transformation path for varying ATM coupling strengths. We analytically demonstrate that the ATM cohesive energy exhibits an extremum at the bcc phase and show numerically that it corresponds to a minimum for repulsive three-body forces along the Bain path. Our results indicate that strong repulsive three-body interactions can destabilize the fcc phase and render bcc energetically favorable for soft LJ potentials. However, even in this scenario, the bcc phase remains susceptible to further cuboidal distortions. These results suggest that the stability of the bcc phase is, besides vibrational, temperature, and pressure effects, strongly influenced by higher than two-body forces. Because of the wrong short-range behavior of the triple-dipole ATM model the LJ potential is limited to exponents $n>9$ for the repulsive wall, otherwise one observes distortion into a set of linear chains collapsing to the origin.
\end{abstract}

\maketitle

\clearpage
\section{Introduction}
\begin{figure*}[htbp]
\begin{center}
\includegraphics[width=9cm]{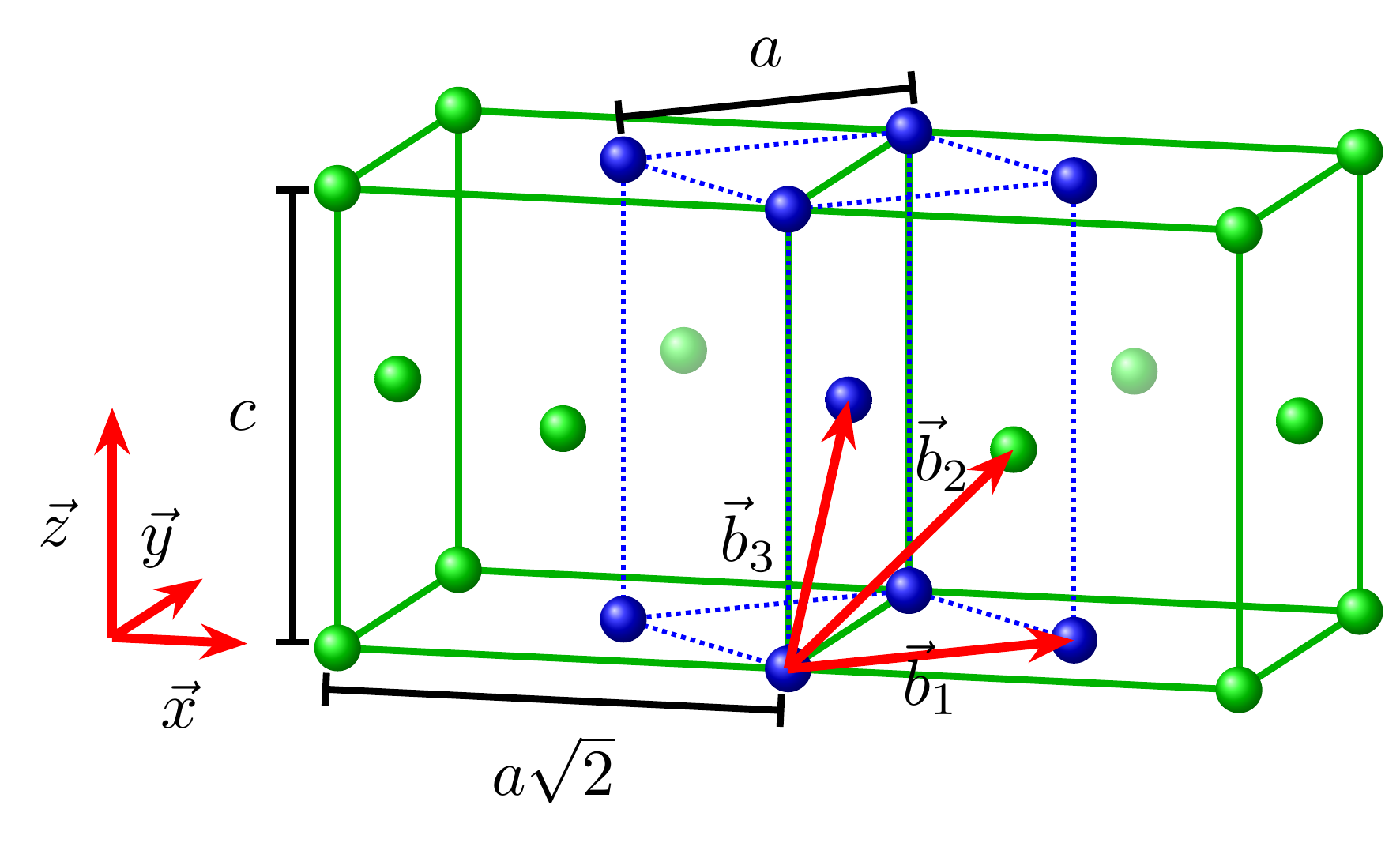}
\caption{
Body-centered tetragonal lattice shown in blue with lattice constants $a$ and $c$. For $a=c$ we have the bcc lattice. The usual fcc unit cell with additional green atoms is shown as well with lattice constants $a'=a\sqrt{2}=c$. }
\label{fig:bct}
\end{center}
\end{figure*}

Crystalline solid-to-solid phase transitions are induced by temperature or pressure change and often involve symmetry breaking away from the original space group of the starting phase along the minimum energy transition path towards the final crystal phase.\cite{stanley1971phase} The special class of martensitic phase transformations is described by diffusionless transitions induced by lattice strain and a collective movement of the atoms in the lattice.\cite{falk1982landau,Izyumov1994,otsuka1999shape,Ackland2008,Grimvall2012} Such martensitic transitions are found in many important materials such as steel or oxide ceramics,\cite{Torrents2017} but also for some elements in the Periodic Table such as lithium.\cite{young1991phase,Ackland2017} The body-centered cubic (bcc) to face-centered cubic (fcc) phase transition belongs to the class of martensitic transformations.\cite{Caspersen2005} Both the bcc lattice ($c/a=1$) and the face-centered cubic (fcc) lattice ($c/a=\sqrt{2}$) have the body-centered tetragonal (bct) lattice (crystallographic group $\#139$ or $I4/mmm$) in common defined by the lattice constants $a_1=a_2=a$ and $a_3=c$ and right angles $\alpha_1=\alpha_2=\alpha_3=90^o$) as shown in Figure \ref{fig:bct}. 

Concerning the interactions between the atoms or molecules in a lattice, the associated infinite lattice sums describing such interactions have a long history in solid-state physics and discrete mathematics.\cite{borwein-2013} They connect lattices  to observables such as the equation of state for a bulk system with inverse power potentials $V(r)=r^{-k}$ acting between lattice points.\cite{Gruneisen1912,furth1944,Stillinger2001,Smits2021} Most notable cases 
for such interaction potentials are the Lennard-Jones potential\cite{Jones-1925} (see Ref.~\onlinecite{Wales2024} for a historical account), which in its most general case is given by
\begin{equation} \label{eq:VLJ}
E_{\rm LJ}(r)=\epsilon\frac{ nm}{n-m} \;  \left[ \frac{1}{n}\left( \frac{r_e}{r} \right)^{n} - \frac{1}{m} \left( \frac{r_e}{r} \right)^{m} \right] \,,
\end{equation}
and the Coulomb potential leading to the famous Madelung constant derived as early as 1918 by Madelung.\cite{madelung1918} In Eq.~\eqref{eq:VLJ}, $r_e$ is the equilibrium distance for a diatomic molecule, $\epsilon$ the corresponding dissociation energy, and we have the condition $n>m>d$ with $d$ the dimension of the lattice.
In the following, we consider $d$-dimensional Bravais lattices $\Lambda=B^\top \mathbb Z^d=\{B^\top \vec i~|~\vec i\in \mathbb Z^d\}$, with $d=1,2,3$, where the generator matrix $B^\top=(\vec{b_1},\cdots,\vec{b_d})$ contains the basic lattice vectors.\cite{burrows-2020} When evaluating energies or forces in such long-range interacting lattices, we encounter lattice sums of the  form
\begin{equation} \label{eq:latticesum}
L=\sum_{\vec x\in \Lambda} f(\vec x)=\sum_{\vec{i}\in\mathbb{Z}^d} f\left(B^\top\vec{i} \right),
\end{equation}
where $f$ is a scalar or vector-valued function that decreases sufficiently fast such that the sum is absolutely convergent. An important special case is given by an inverse power-law potential $f(\vec x)=\vert \vec x \vert^{-\nu}$, where the resulting lattice sum is a special case of the Epstein zeta function, a generalization of the Riemann zeta function to multidimensional lattice sums.\cite{Epstein-1903}
These lattice sums are often slowly convergent, and their efficient and precise computation poses significant challenges. Moreover, meaning can be given to 
conditionally convergent or even divergent series through techniques such as meromorphic continuation.\cite{borwein1998convergence} The theory of converting lattice sums, including their meromorphic continuations, into fast converging series has become a research field on its own.\cite{borwein-2013}

It is widely recognized that the expansion of the total interaction energy in terms of many-body interaction contributions in a cluster or bulk system is often slowly convergent.\cite{N-body.AH2007} The dominant long-range three-body interaction contribution comes from the triple-dipole interaction and is described approximately by the Axilrod-Teller-Muto (ATM) potential.\cite{AxilrodTeller1943,Muto1943} For a trimer of atoms at positions $\vec r_1,\vec r_2,\vec r_3$, the ATM potential reads\cite{attard1992simulation}
\begin{equation}\label{eq:MAT}
E^{(3)}_{\rm ATM}=  \lambda \frac{ r_{12}^2r_{13}^2 r_{23}^2+3(\vec r_{12}\cdot \vec r_{13})(\vec r_{21} \cdot \vec r_{23})(\vec r_{31}\cdot \vec r_{32})}{\left(r_{12}r_{13}r_{23}\right)^5}.
\end{equation}
Here, $\lambda > 0$ represents the ATM coupling constant, while $\vec{r}_{ij} = \vec{r}_i - \vec{r}_j$ denotes the relative position vector between distinct atoms $i$ and $j$ with norm $r_{ij} = \vert \vec{r}_{ij} \vert$. The coupling strength \(\lambda\) depends on the polarizabilities of the interacting atoms, where three-body interactions can become highly relevant, among others, for the heavier and more polarizable noble gases. Notably, for solid argon at \(0\) K, three-body forces have been shown to contribute approximately \(8.9\%\) of the total cohesive energy.\cite{Schwerdtfeger-2016}

The precise simulation of solid-solid phase transitions can be highly challenging due to the movement of many atoms within the simulation cell.\cite{Torrents2017,Bingxi2017} Martensitic transformations are less complex in nature but are nevertheless difficult to predict theoretically and to measure experimentally.\cite{Caspersen2004,Carter2008a,Grimvall2012} For a general $(n, m)$ LJ two-body potential, we recently showed by exact lattice summations that the bcc phase is at an extremum along the cuboidal distortion path and becomes either energetically unstable or metastable. This bcc instability persists into the high-pressure regime for a LJ solid.\cite{Burrows2022a} This result can most likely be generalized to all physically relevant two-body interactions. Thus, the existence of the bcc phase, known for a number of elements in the periodic table, likely results from vibrational and temperature effects, and/or from dominant higher than two-body forces. We note that Landau theory predicts that the bcc phase becomes dominant near the melting line.\cite{Alexander1978}

In this work, we analyze a smooth connection between the cuboidal body-centered tetragonal (bct) lattices through a martensitic Bain transformation including both general two-body LJ interactions and a three-body ATM potential.\cite{bain1924,Zener1948,rifkin1984} We first write the lattice $\Lambda$ along the transition path as a function of a single parameter $A$,\cite{Jerabek2022a,burrows2021b} and collect the basic properties of the resulting lattice. We then present efficient methods to evaluate both the arising two-body and three-body lattice sums to full precision. For the two-body potential, we re-express the algebraically decaying sum in terms of a series of super-exponentially decaying sums, which can be efficiently evaluated. 
We then present a novel efficient method for computing general three-body lattice sums, based on integrals involving zeta functions on multidimensional lattices. Using these advanced numerical techniques, we offer a rigorous study of the stability of the bcc phase relative to the fcc phase as a function of the ATM coupling constant $\lambda$.

This work is structured as follows: In Sec.~\ref{subsec:lattices}, we provide basic definitions for general Bravais lattices. Subsequently, we discuss cuboidal lattices in Sec.~\ref{subsec:cuboidal}. We then introduce the Bain transformation in Sec.~\ref{subsec:bain_transformation} and discuss the resulting lattice sums for the cohesive energy, including both two- and three-body contributions in Sec.~\ref{subsec:lj_and_atm}. We present our novel method for precisely evaluating three-body lattice sums in Sec.~\ref{sec:Epstein zeta}. After discussing the optimization of the nearest neighbor distance in Sec.~\ref{sec:optimization}, we apply our methods first to a one-dimensional chain in Sec.~\ref{sec:Linear Chain}, and subsequently to two-dimensional square and hexagonal lattices in \ref{sec:2D-Lattices}. Finally, we study three-dimensional lattices along the Bain path in \ref{sec:Bain} and discuss qualitatively new physical behavior caused by the inclusion of the three-body ATM potential. We draw our conclusions and provide an outlook in Sec.~\ref{sec:conclusion}.

%%%-----------------------------------------------------------------------

%Theory
\section{Theory}

%The cuboidal lattices
\subsection{General lattice properties}
\label{subsec:lattices}
We begin our treatment by defining lattices and important associated quantities. We call a point set $\Lambda\subseteq \mathbb R^d$ a (Bravais) lattice, if $\Lambda=B^\top \mathbb{Z}^d=\{ \vec{i}^\top B = B^\top\vec{i}~ |~ \vec{i}\in\mathbb{Z}^d \}$ for some nonsingular matrix $B\in \mathbb R^{d\times d}$. The matrix $B = (\vec{b}_1,\cdots,\vec{b}_d)^\top$, called the generator matrix, contains the set of linearly independent basis vectors $\vec{b}_i^\top$ as its rows. Lattices exhibit discrete translational invariance, meaning that $\Lambda+\vec x =\Lambda$ for any $\vec x\in \Lambda$.
The Gram matrix $G$ is defined in terms of the generator matrix as $G=B B^\top$ and appears in the computation of lattice vector norms.  
We further define the elementary lattice cell $B^\top(-\frac{1}{2},\frac{1}{2})^d$. 
The lattice volume is defined by $V_\Lambda=|\det B|=\sqrt{ \det G}$. An important lattice quantity is the minimum or nearest neighbor distance $R_\Lambda$ with 
\begin{equation}
\label{eq:mindist}
R_\Lambda=\textrm{min}\{|\vec{x}-\vec{y}|~ |~ \vec{x}, \vec{y} \in \Lambda,~\vec{x}\ne \vec{y}\}= \min_{\vec x\in \Lambda\setminus\{0\}} \vert \vec x\vert,
\end{equation}
due to translational invariance of the lattice, where $|\vec x| $ denotes the Euclidean distance.
In terms of the Gram matrix this is equivalent to 
\begin{equation}\label{eq:nearestneighbor}
R_\Lambda=\min_{\vec i\in \mathbb Z^d\setminus\{0\}} \sqrt{\vec{i}^{\top}G\vec{i}}.
\end{equation}

The packing density $\Delta_\Lambda$ describes the ratio between the volume of particles with radius $\rho$ and the volume of the elementary lattice cell,
\begin{equation}
\label{eq:sphere_packing}
\Delta_\Lambda=\frac{\pi^{d/2}}{\Gamma(d/2+1)}\frac{\rho^d}{V_\Lambda},
\end{equation}
with  the gamma function $\Gamma$. For dense hard sphere packings we have $\rho=R_\Lambda /2$. 
%Thus, we find in three dimensions that $\Delta_\Lambda=\pi R_\Lambda ^2/3$. 
Finally, the kissing number for dense hard sphere packings is defined as the number of nearest neighbors of an arbitrary lattice point,
\begin{equation}
\label{eq:kiss}
\textrm{kiss}(\Lambda)=\#\{\vec{v}\in \Lambda ~|~ |\vec{v}|=R_\Lambda \}.
\end{equation}

\subsection{Properties of cuboidal lattices}
\label{subsec:cuboidal}

In case of the three-dimensional cuboidal lattices, we start from the work of Conway and Sloane~[\onlinecite[Sec.~3]{Conway1994}] and consider the lattice generated by the vectors $(\pm u, \pm v, 0)^\top\quad\text{and}\quad (0,\pm v, \pm v)^\top$, where $u$ and $v$ are non-zero real numbers. We now use the basis vectors $\vec{b}_1^\top = (u,v,0),~ \vec{b}_2^\top = (u,0,v),~ \vec{b}_3^\top = (0,v,v)$, where $u$ and $v$ are non-zero real numbers. Let  $A=u^2/v^2$. The generator matrix $B^\top$ and the Gram matrix $G$ are 
\begin{equation}
\label{eq:lattice_and_gram}
B^\top = \begin{pmatrix}
\vec{b}_1 &
\vec{b}_2 &
\vec{b}_3 
\end{pmatrix}
= v \begin{pmatrix}
\sqrt{A} & \sqrt{A} & 0 \\
1 & 0 & 1 \\
0 & 1 & 1 
\end{pmatrix},\qquad 
G = B\,B^\top = v^2\begin{pmatrix}
A+1 & A & 1 \\
A & A+1 & 1 \\
1 & 1 & 2
\end{pmatrix}.
\end{equation}
The determinant of the generator matrix reads
% $-2uv^2=-2A/u$, % JB: maybe this A is from a different paper?
$\det B =-2 v^3 \sqrt{A}$ and thus $V_\Lambda  = 2 \vert v^3\vert \sqrt{A}$.

Different lattice phases are obtained depending on the choice of the argument $A$. These are, in decreasing numerical order,
\begin{enumerate}[(i)]
\item $A=1$: the face-centered cubic (fcc) lattice,
\item $A=1/\sqrt2$: the mean centred-cuboidal (mcc) lattice,
\item $A=1/2$: the body-centred cubic (bcc) lattice,
\item $A=1/3$: the axial centred-cuboidal (acc) lattice.
\end{enumerate}
The resulting Gram matrices for the fcc and bcc lattices
are identical to the ones shown in our previous work on lattice sums,\cite{burrows-2020} whereas the mcc and acc lattices
occur in Refs.~[\onlinecite{Conway1994}] and~[\onlinecite{conway2007optimal}]. The mcc lattice is the densest isodual lattice in three-dimensional space, but beside being of theoretical interest, has not been observed in nature so far. However, this lattice is expected to play a role in the dynamics of the cuboidal fcc to bcc transition, as we investigate in detail in this work.

\begin{figure*}[hbp!]
\begin{center}
\includegraphics[width=0.45\textwidth]{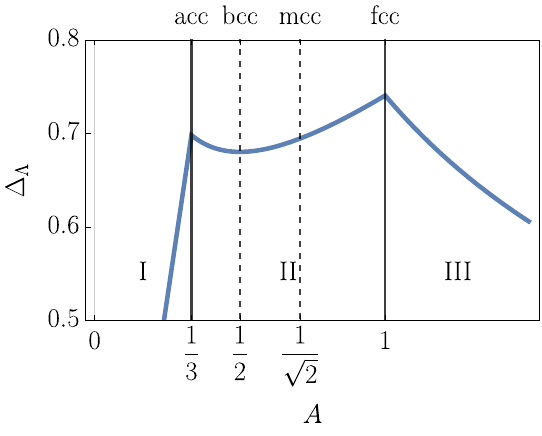}
\caption{
Graph of the packing density $\Delta_\Lambda$ versus~$A$. The regions I, II and III, divided by the solid black lines, correspond to the different kissing numbers. Explicit formulas are given in Table~\ref{tab:lattice}. The location of the fcc, mcc, bcc and acc lattices are indicated the solid and dashed black lines.}
\label{fig:ALattice}
\end{center}
\end{figure*}

Inserting either the generator matrix or the Gram matrix in Eq.~\eqref{eq:nearestneighbor}  yields the nearest neighbor distance as a function of $A$,
\begin{equation}
\label{pd1}
R_\Lambda
= \begin{cases}
2v\sqrt{A}, & 0<A<1/3, \\
v\sqrt{A+1}, & 1/3\leq A \leq 1, \\
v\sqrt2, & A>1.
\end{cases} 
\end{equation}
From Eq.~\eqref{eq:sphere_packing} then follows the packing density $\Delta_\Lambda$ for dense sphere packings,
\begin{equation}
\Delta_\Lambda
= \begin{cases}
(2\pi/3) A, & 0<A<1/3, \\ 
(\pi/12)\sqrt{(A+1)^3/A}, & 1/3\leq A \leq 1, \\
(\pi/6)\sqrt{2/A}, & A>1
\end{cases} 
\end{equation}
which is displayed in Fig.~\ref{fig:ALattice}.
On the interval $1/3\leq A \leq 1$, which includes the acc, bcc, mcc, and fcc phases, the packing density has a maximum of $\pi\sqrt{2}/6 \approx 0.74$ at $A=1$ corresponding to fcc, and a minimum of \mbox{$\pi\sqrt{3}/8 \approx 0.68$} at $A=1/2$ corresponding to bcc. The acc lattice has a rather large packing density of $\Delta_\text{acc}=\frac{2\pi}{9}\approx 0.698$, but is the least dense packing with kissing number 10\cite{Patterson1941,Fields1980}. However, it is most likely strictly jammed according to the definition by Torquato and Stillinger.\cite{Torquato2007} It consists of linear chains of touching spheres surrounded by four neighboring linear chain arranged within a bct cell. It is the starting point of separated linear chain formation within region I ($A\le\frac{1}{3}$). 

Finally Eq.~\eqref{eq:kiss} yields the kissing number for dense sphere packings,
\begin{equation}\label{eq:kiss2}
\mathrm{kiss}\big(\Lambda\big)=\#\{\vec{v}\in \Lambda ~|~ |\vec{v}|=R_\Lambda\}= \begin{cases}
2, & A<1/3, \\
10, & A=1/3~\text{(acc)}, \\
8, & 1/3<A<1\\
12 & A = 1~\text{(fcc)},\\
4, & A >  1.
\end{cases}
\end{equation}
The limiting case $A\rightarrow\infty$ corresponds to infinitely separated two-dimensional square lattice layers with kissing number~$4$, while in the other extreme case, the limit $A\rightarrow 0$, we obtain infinitely dense 1D chains with kissing number~$2$ repeated on a two-dimensional grid.

Figure \ref{fig:ALattice} shows a graph of the packing density as a function of the parameter $A$.
Further information is recorded in Table~\ref{tab:lattice}.
\begin{table}[htp]
\setlength{\tabcolsep}{0.1cm}
\caption{\label{tab:lattice} Kissing number $\textrm{kiss}(\Lambda)$ and packing density $\Delta_\Lambda$ for the lattice defined in Eq.~\eqref{eq:lattice_and_gram}. The values in the table depend only on $A$
and are independent of $v$.}
\begin{center}
\begin{tabular}{l|l|l|l}
\hline
Region & $A$ & $\textrm{kiss}(\Lambda)$ & $\Delta_\Lambda$ \\
\hline
I 		& $(0,\frac{1}{3})$	& 2 		& $\frac{2\pi A}{3}$	 \\
acc 		& $\frac{1}{3}$		& 10 & $\frac{2\pi}{9}$	\\
II 		& $(\frac{1}{3},1)$	& 8 	& $\frac{\pi}{12}\sqrt{\frac{(A+1)^3}{A}}$	\\
fcc 		& $1$			& 12 & $\frac{\pi\,\sqrt{2}}{6}$	 \\
III 		& $(1,\infty)$		& 4 	& $\frac{\pi}{6}\sqrt{\frac{2}{A}}$	\\
\hline
\end{tabular}
\end{center}%
%\scriptsize{$^a$ The integer combinations $\vec{i}$, given in the quadratic form, which determine $R $ in~\eqref{pd1} for the different regions are as follows: $\vec{i}_1^\textrm{I}=(-1,-1,1)$, $\vec{i}_2^\textrm{I}=(1,1,-1)$, $\vec{i}_3^\textrm{II}=(-1,0,0)$, $\vec{i}_4^\textrm{II}=(-1,0,1)$, $\vec{i}_5^\textrm{II}=(0,-1,0)$, $\vec{i}_6^\textrm{II}=(0,-1,1)$, $\vec{i}_7^\textrm{II}=(0,1,-1)$, $\vec{i}_8^\textrm{II}=(0,1,0)$, $\vec{i}_9^\textrm{II}=(1,0,-1)$, $\vec{i}_{10}^\textrm{II}=(1,0,0)$, $\vec{i}_{11}^\textrm{III}=(-1,1,0)$, $\vec{i}_{12}^\textrm{III}=(0,0,-1)$, $\vec{i}_{13}^\textrm{III}=(0,0,1)$, $\vec{i}_{14}^\textrm{III}=(1,-1,0)$. }
\end{table}

The cuboidal lattices belong to the body-centered tetragonal lattices (bct) usually defined by the two lattice constants $a$ and $c$, see Figure \ref{fig:bct}. We can easily transform our two parameter space $(u,v)$ used by Conway in terms of $(R_\Lambda,A)$ used here and $(a,c)$ used for bct lattices in the interval $1/3\leq A \leq 1$ by
\begin{equation}\label{eq:relation}
(u,v)= \left( R_\Lambda \sqrt{\frac{A}{A+1}} , \frac{R_\Lambda }{\sqrt{A+1}} \right) \quad \text{and} \quad (R_\Lambda,A)=\left( \frac{a}{2}\sqrt{2+\gamma^2}\;,\; \frac{1}{2}\gamma^2\right)
\end{equation}
for the range $1/3\leq A \leq 1$ (region II) and $\gamma=c/a$. For example, if we use for the bct lattice in Figure \ref{fig:bct} the lattice constants $a$ and $\gamma=\sqrt{2}$, we get $A=1$ (fcc) and $R_\Lambda=a$, which is the distance from the origin of the lattice to the nearest face-centered point, whilst for $\gamma=1$ we get $A=\frac{1}{2}$ (bcc) and $R_\Lambda=a\frac{\sqrt{3}}{2}$, which is the distance from the origin to the nearest body-centered (bc) point, i.e. $R_\Lambda=R_\text{bc}$. From Eq. \eqref{eq:relation} we see that for $A<\frac{1}{2}$ we have $\gamma<1$ which implies $c<a$. For $A\ge 1$ (region III) we have $R_\Lambda=a$, the distance between nearest neighbors in the base layer.  For $A<\frac{1}{3}$ we enter region I for which we get $\gamma<\sqrt{\frac{2}{3}}$ and therefore $c<R_\text{bc}$. Using the two lattice parameters $(R_\Lambda,A)$ has the advantage that $R_\Lambda\in a[\frac{\sqrt{3}}{2},1]$ in region II varies only slowly, and the Bain transformation introduced in the next subsection is mostly described by one single dimensionless lattice parameter $A$.

%The Bain transformation
\subsection{The Bain transformation}
\label{subsec:bain_transformation}
The Bain transformation is a diffusionless smooth transformation from bcc to fcc and vice versa. If we start conveniently from the fcc generator matrix we find a smooth transformation in terms of a diagonal matrix
\begin{equation}
\label{Bain1}
\tilde{B}^\top(A)=\lambda(A) T_\text{Bain}(A)  \tilde{B}^\top_\text{fcc}=\lambda(A)
\frac{1}{\sqrt{2}}\begin{pmatrix}
\sqrt{A} & \sqrt{A} & 0 \\
1 & 0 & 1 \\
0 & 1 & 1 
\end{pmatrix},
\end{equation}
with $\tilde{B}(A)=B(A)/R(A)$ and $\tilde B_\mathrm{fcc}=\tilde B(1)$. The diagonal Bain matrix reads
\begin{equation}
T_\mathrm{Bain}(A) = \begin{pmatrix}
\sqrt{A} & 0 & 0 \\
0 & 1 & 0 \\
0 & 0 & 1 
\end{pmatrix}
\end{equation}
%or 
%\begin{equation}
%\tilde{B}_\text{cub}=\lambda(A)  \tilde{B}_\text{fcc}  T_\text{Bain}(A)
%\end{equation}
and the prefactor $\lambda(A)$ is given by
\begin{equation}
%\label{pd1}
\lambda(A)
= \begin{cases}
1/\sqrt{2A}, & 0<A<1/3, \\
\sqrt{2/(A+1)}, & 1/3\leq A \leq 1, \\
1, & A>1.
\end{cases} 
\end{equation}
In the particularly relevant range $1/3\le A\le 1$, the rescaled generator matrix takes the form
\begin{equation}
\label{eq:bain_transform_final}
\tilde B^\top(A) =  \frac{1}{\sqrt{A+1}} \begin{pmatrix}
\sqrt{A} & \sqrt{A} & 0 \\
1 & 0 & 1 \\
0 & 1 & 1 
\end{pmatrix}.
\end{equation}
In later sections, we will extend the above definition of $\tilde B^\top(A)$ in Eq.~\eqref{eq:bain_transform_final} to the whole range $0<A\le 1$, where the nearest neighbor distance for $A\ge 1/3$ is given by the lattice constant $a$ (see Figure \ref{fig:bct}) and to the distance between the origin and the body-centered atom for $0<A<\frac{1}{3}$. As always, the Bain transformation matrix depends on the particular choice of lattice basis vectors.

\subsection{Cohesive energies from a Lennard-Jones potential coupled to a three-body Axilrod-Teller-Muto term}
\label{subsec:lj_and_atm}

Using translational invariance of the lattice, the static cohesive energy for a lattice can be expressed in terms of a many-body perturbative expansion of the interaction energy from a chosen atom at the origin, 
\begin{align}
E_\text{coh} &= \sum\limits_{k = 2}^{\infty} E_\text{coh}^{(k)} 
=  \frac{1}{2}\sum\limits_{i\in\mathbb{N}} {E^{(2)} (\vec{r}_{0i})}
+ \frac{1}{3}\sum\limits_{\substack{i,j\in\mathbb{N} \\ j>i}} {E^{(3)} (\vec r_{0i},\vec r_{0j},\vec r_{ij})} + \mathrm{h.o.t},
 \label{eq:nbody}
 \end{align}
with  $\vec r_{ij}=\vec r_i-\vec r_j$, $r_{ij}=|\vec r_{ij}|$ and where $i=0$ denotes the index of the atom at the chosen origin in the solid. The perturbative expansion is formally exact, but is often slowly converging, especially for metallic systems.\cite{N-body.AH2007}
In this work we focus our studies on the two- and three-body interactions, neglecting vibrational and temperature effects, as well as higher order terms ($\mathrm{h.o.t.}$) such as four-body interactions. In the following, we adopt dimensionless units, writing length scales in units of the equilibrium distance $r_e$ of the LJ potential and energies in units of the LJ dissociation energy $\epsilon$.

The dimensionless two-body potential in \eqref{eq:nbody} then takes the form,
\begin{equation} \label{eq:VLJ2}
E^{(2)}_{\rm LJ}(\vec r)=\frac{ nm}{n-m} \;  \left( \frac{1}{n} {\vert\vec r\vert}^{-n} - \frac{1}{m} {\vert\vec r\vert}^{-m} \right). 
\end{equation}
with $n>m>3$. The resulting cohesive energy can be written in terms of the Epstein zeta function, a generalization of the Riemann zeta function to higher-dimensional lattices. For a lattice $\Lambda$, an interaction exponent $\nu>d$, and a wavevector $\vec k$, it reads\cite{Epstein-1903} 
\begin{equation}
Z_{\Lambda,\nu}(\vec k) = \,\sideset{}{'}\sum_{\vec x\in \Lambda}\frac{e^{-2\pi i \vec x \cdot \vec k}}{\vert \vec x\vert^\nu},
\end{equation}
where the lattice sum can be meromorphically continued to $\nu\in \mathbb C$. For the LJ lattice sum, the Epstein zeta function is evaluated at $\vec k=0$ only, where we omit the argument $Z_{\Lambda,\nu}=Z_{\Lambda,\nu}(0)$ to simplify the notation. General wavevectors will, however, become crucial in the evaluation of three-body lattice sums.
The two-body term in the cohesive energy for a LJ potential can then be rewritten as
\begin{equation}\label{eq:EcohLJ}
    E_{\text{coh}}^{(2)} =\frac{nm}{2(n-m)} \,\sideset{}{'}\sum_{\vec x\in \tilde \Lambda} \left(\frac{\vert \vec x \vert^{-n}}{n R ^{n}}  - \frac{\vert \vec x\vert^{-m}}{m R ^{m}} \right)=\frac{nm}{2(n-m)}  \left(\frac{Z_{\tilde \Lambda,n}}{n R ^{n}}  - \frac{Z_{\tilde \Lambda,m}}{m R ^{m}} \right)
\end{equation}
where we use the normalized lattice $\tilde \Lambda=\Lambda/R$. This normalization is useful, as the distance $R$ (e.g. the nearest neighbor distance) will become a tuning parameter depending along the Bain path on the exponents $n$ and $m$ and on the parameter $A$ as specified in the next sections. It also shows more clearly the link to the LJ potential \eqref{eq:VLJ2} for a diatomic.

Different computationally efficient methods for evaluating the Epstein zeta function exist. In Appendix \ref{sec:latticesumsA}, we evaluate the arising sums for particular lattices $\tilde \Lambda(A)=\tilde B^\top(A) \mathbb Z^3$ using Bessel function expansions in Equation \eqref{L3formula1} or Equation \eqref{L3formula2}, with the more common notation
\begin{equation}
    L(A,n/2)=Z_{\tilde \Lambda(A),n}.
\end{equation}
As an alternative, for general $d$-dimensional lattice sums including oscillatory factors and lattice shifts, the recently created high-performance library EpsteinLib 
 (\url{github.com/epsteinlib}) can be used.\cite{buchheit2024epstein} Both approaches allow to compute the two-body term to machine precision.

In a similar way we express the three-body Axilrod-Teller-Muto (ATM) potential in Eq.~\eqref{eq:MAT} in dimensionless units. As the ATM potential only depends on relative distance vectors, we can set $\vec x = \vec r_{0i}$, $\vec y = \vec r_{0j}$ and $\vec z=\vec r_{ij}=\vec y-\vec x$, yielding the potential as a function  of two vectors only,
% and re-express the cosine terms \cite{Chell_1968},
\begin{align}
\label{eq:MAT1}
E^{(3)}_{\rm ATM}&(\vec x,\vec y)=\lambda \;\; \bigg( \frac{1}{\vert \vec{x} \vert^3 \vert  \vec{y}\vert^3 \vert \vec{z} \vert^3}-3 \frac{(\vec{x}\cdot \vec{y})(\vec{y}\cdot \vec{z})(\vec{z}\cdot \vec{x})}{\vert \vec{x} \vert^5 \vert  \vec{y}\vert^5 \vert \vec{z} \vert^5}\bigg)\bigg \vert_{\vec{z}=\vec{y}-\vec{x}},
\end{align}
where the minus sign on the right-hand side arises due to $\vec r_{ji} =-\vec r_{ij}$.
The cohesive energy contribution due to the three-body interactions is  given by the lattice sum
\begin{equation}
    E_{\mathrm{coh}}^{(3)} = \frac{1}{6} \,\sideset{}{'} \sum_{\vec x,\vec y\in \Lambda} E^{(3)}_{\rm ATM}(\vec x,\vec y),
\end{equation}
where the prefactor $1/6$ avoids double counting and where the primed sum excludes the undefined cases $\vec x=0$, $\vec y=0$, and $\vec x=\vec y$. We now normalize the lattice, setting $\tilde \Lambda=\Lambda/R$ and subsequently split the three-body lattice sum into a radially isotropic and an anisotropic part,
\begin{equation}
E_{\mathrm{coh}}^{(3)}=\lambda f_{\mathrm{coh}}^{(3)}R^{-9}=\lambda(f_r^{(3)}+f_a^{(3)})R^{-9}
\end{equation}
with the normalized lattice sums
\begin{equation}
\label{eq:f_coh}
f_{\mathrm{coh}}^{(3)}=f_r^{(3)}+f_a^{(3)},\quad f_r^{(3)}= \frac{1}{6} \,\sideset{}{'}\sum_{\vec{x},\vec{y} \in \tilde \Lambda}\frac{1}{\vert \vec{x} \vert^3 \vert  \vec{y}\vert^3 \vert \vec{z} \vert^3},\quad f_a^{(3)}= -\frac{1}{2} \,\sideset{}{'}\sum_{\vec{x},\vec{y} \in \tilde \Lambda} \frac{(\vec{x}\cdot \vec{y})(\vec{y}\cdot \vec{z})(\vec{z}\cdot \vec{x})}{\vert \vec{x} \vert^5 \vert  \vec{y}\vert^5 \vert \vec{z} \vert^5}
\end{equation}
where we adopt the convention $\vec z=\vec y-\vec x$ from now on. For simplicity, we leave away the tilde in the following, assuming that the lattices have been appropriately normalized. The above form for the three-body lattice form makes it immediately clear that the ATM potential becomes attractive in one dimension, as then $f_a= -3 f_r $ and hence $E_{\mathrm{coh}}^{(3)} = -2 \lambda f_r R^{-9} <0$.

%with 
%\begin{align}
%\label{eq:coeff1}
%\nonumber
%C_1&= {R_{0j}}^2 + {R_{ij}}^2 - {R_{0i}}^2 \\
%C_2&= {R_{0i}}^2 + {R_{ij}}^2 - {R_{0j}}^2 \\
%C_3&= {R_{0i}}^2 + {R_{0j}}^2 - {R_{ij}}^2
%\nonumber
%\end{align}
%and the coupling strength $\lambda$ is in units of $\epsilon$. Here we define $R_{ij}=|\vec{R}_{0i}-\vec{R}_{0j}|$. Because the product $C_1C_2C_3$ contains a product of cosine functions we have $(C_1C_2C_3)\in [-1,+1]$, that is the term containing the product can become negative.

%Efficient computation of the ATM potential through three-body zeta functions
\subsection{Efficient computation of the ATM cohesive energy}
\label{sec:Epstein zeta}
The efficient computation of three-body lattice sums has been an important open problem, which we solve in this work. In the past, elaborate direct summation methods have been used, \cite{Schwerdtfeger-2016,Smits2018,Jerabek2019,Smits2020,schwerdtfeger2020,edison2022} where, however, a single evaluation in three dimensions can demand up to 4 weeks of single core CPU time. In this section, we show how general three-body interactions, including the ATM potential, can be computed from singular integrals that involve products of Epstein zeta functions. For a lattice $\Lambda = B^\top\mathbb Z^d$ with $B\in \mathbb R^{d\times d}$ nonsingular, we consider general lattice sums of the form
\begin{equation}
\zeta_\Lambda^{(3)}(\vec{\nu}) = \,\sideset{}{'}\sum_{\vec{x},\vec{y} \in \Lambda} \vert \vec{x} \vert^{-\nu_1}\vert \vec{y} \vert^{-\nu_2} \vert \vec{y}-\vec{x} \vert^{-\nu_3},
%\quad\text{for all}~i,j\in\{1,2,3\}  ~\text{with}~i\neq j,
\end{equation} with $\vec{\nu}=(\nu_1,\nu_2,\nu_3)^T$, and its meromorphic continuations to $\nu_i\in \mathbb C$ (see Appendix \ref{ATMLSdata} for details), which we call three-body zeta functions. One can show that the above double sum converges absolutely and independently of the summation order if and only if the conditions $\nu_i+\nu_j>d$ for $i\neq j$, and $\nu_1+\nu_2+\nu_3>2d$ hold. We note in passing that this lattice sum can be extended to the more general $n$-body zeta function, which will be addressed in our future work.

We first show that the normalized ATM cohesive energy in Eq.~\eqref{eq:f_coh} can be written as a finite recombination of the above zeta functions. The radially symmetric term $f_r^{(3)}$ is already in the desired form with
\begin{equation}
f_r^{(3)} = \frac{1}{6} \zeta_{\Lambda}^{(3)}(3,3,3). 
\end{equation}
For the anisotropic part $f_a^{(3)}$, we note that the vector products can be rewritten as
\begin{align}
2\vec{x}\cdot \vec{y} = |\vec{x}|^2+|\vec{y}|^2-|\vec{z}|^2,\quad
2\vec{y}\cdot \vec{z} = |\vec{y}|^2+|\vec{z}|^2-|\vec{x}|^2,\quad
2 \vec{z}\cdot \vec{x} = -(|\vec{z}|^2 + |\vec{x}|^2-|\vec{y}|^2).
\end{align}
As the above lattice sums remain unchanged under permutation of $\vec{x}$, $\vec{y}$, and $\vec{z}$, we find,
\[
f_a^{(3)}=-\frac{1}{2}\,\sideset{}{'}\sum_{\vec{x},\vec{y} \in \Lambda} \frac{(\vec{x}\cdot \vec{y})(\vec{y}\cdot \vec{z})(\vec{z}\cdot \vec{x})}{\vert \vec{x} \vert^5 \vert  \vec{y}\vert^5 \vert \vec{z} \vert^5}
= -\frac{1}{16}\,\sideset{}{'}\sum_{\vec{x},\vec{y} \in \Lambda} \Big(3 \frac{\vert \vec{x}\vert}{\vert \vec{y}\vert^5\vert \vec{z}\vert^5}-6\frac{1}{\vert \vec{x}\vert\vert \vec{y}\vert^3\vert \vec{z}\vert^5}+2\frac{1}{\vert \vec{x}\vert^3\vert \vec{y}\vert^3\vert \vec{z}\vert^3}\Big). 
\]
Rewriting the above right-hand side in terms of three-body zeta functions yields
\begin{equation}
    f_a^{(3)}=-\frac{1}{16} \Big(3 \zeta_\Lambda^{(3)}(-1,5,5)-6 \zeta_\Lambda^{(3)}(1,3,5)+2 \zeta_\Lambda^{(3)}(3,3,3)\Big),
\end{equation}
Recombining $f_r^{(3)}$ and $f_a^{(3)}$ finally yields the ATM cohesive energy in terms of three-body zeta functions,
\begin{align}
f_\mathrm{coh}^{(3)} = \frac{1}{24}\zeta_\Lambda^{(3)}(3,3,3) - \frac{3}{16}
\zeta_\Lambda^{(3)}(-1,5,5)+\frac{3}{8} \zeta_\Lambda^{(3)}(1,3,5).
\end{align}

The three-body zeta function can now be recast as an integral over products of Epstein zeta functions. Recall that for a wavevector $\vec{k}\in \mathbb R^d$, the Epstein zeta function reads
\[
Z_{\Lambda,\nu}(\vec{k}) =\,\sideset{}{'}\sum_{\vec{x} \in \Lambda} \frac{e^{-2\pi i \vec{k} \cdot \vec{x}}}{\vert \vec{x}\vert^\nu},\quad \nu>d,
\]
which can be meromorphically continued to $\nu \in \mathbb C$. Note that the properties and efficient computation of the Epstein zeta function have been recently discussed in Ref.~\onlinecite{buchheit2024epstein} with the high-performance library EpsteinLib available on \url{github.com/epsteinlib}\cite{buchheit2024}.  
Using the properties of the Epstein zeta function, one can now show that for any $\nu_i>0, ~i=1,\dots, 3$, 
\begin{equation}
\zeta_\Lambda^{(3)}(\vec{\nu}) = V_\Lambda \int \limits_{\mathrm{BZ}} Z_{\Lambda,\nu_1}(\vec{k})Z_{\Lambda,\nu_2}(\vec{k})Z_{\Lambda,\nu_3}(\vec{k})  \,\mathrm d \vec{k},
\end{equation}
with the Brillouin zone $\mathrm{BZ}=B^{-1}(-1/2,1/2)^d$ and the volume of the elementary lattice cell $V_\Lambda = \vert \det B\vert$. The proof of this formula is based on exchanging summation and integration for sufficiently large $\nu_i$ and then applying the relation 
\[
V_\Lambda \int_{\mathrm BZ}  e^{-2\pi i \vec{k} \cdot \vec{x}}\,\mathrm d \vec{k} = \delta_{\vec{x},\vec{0}}
\]
with $\delta$ the Kronecker delta. A mathematically rigorous proof as well as details on the numerical computation of the integral will be provided elsewhere. 

Special care needs to be taken in evaluating the resulting integral, as the Epstein zeta function exhibits a singularity at $\vec{k} = 0$. 
We can separate the Epstein zeta function into an analytic function and a singularity as follows
\begin{equation}
\label{eq:epstein_decomposition}
Z_{\Lambda,\nu}(\vec{k}) = Z_{\Lambda,\nu}^{\mathrm{reg}}(\vec{k}) +\frac{1}{V_\Lambda} \hat s_{\nu}(\vec{k}), 
\end{equation}
where the regularized Epstein zeta function $Z_{\Lambda,\nu}^{\mathrm{reg}}(\vec k)$ is analytic in the Brillouin zone. The function $\hat s_\nu(\vec{k})$ can be understood as the Fourier transform of $\vert \vec{z} \vert^{-\nu}$ (in the distributional sense). It is defined as
\begin{align*}
    \hat s_\nu(\vec{k}) =  \frac{\pi^{\nu-d/2}}{\Gamma(\nu/2)}\Gamma\big((d-\nu)/2\big)  \vert \vec{k}\vert ^{\nu - d},\quad \nu \not\in (d+2\mathbb N).
\end{align*}
In case that $\nu =d+2n$, $n\in \mathbb N$, the Fourier transform is only uniquely defined up to a polynomial of degree $2n$. We adopt the choice
\[
\hat s_{d+2n}(\vec{k}) =  \frac{\pi^{n+d/2}}{\Gamma(n+d/2)}\frac{(-1)^{n+1}}{n!} ( \pi \vec{k}^2 )^{n} \log (\pi  \vec{k}^{2}).
\]
Hence, the Epstein zeta function equals the sum of an analytic function  and a power-law or logarithmic singularity. Therefore, the integral can be efficiently computed using either a specialized Gauss-Legendre quadrature or a Duffy transformation.\cite{duffy1982quadrature} Our results are benchmarked against a direct summation approach, presented in Appendix~\ref{appendix:direct_sum}, where we reach full precision in one and two dimensions, where the direct sum can still be evaluated to machine precision.

\subsection{Minimizing the cohesive energy}
\label{sec:optimization}

In order to analyze the impact of a long-range three-body ATM potential on the stability of lattices with two-body LJ interactions, we need to determine the optimal nearest neighbor distance $R>0$ that minimizes the cohesive energy

\begin{equation}
\label{eq:ecohATMJL}
E_{\text{coh}} = c_{n,m} \Big( \frac{Z_{\Lambda,n}}{n R^n}- \frac{Z_{\Lambda,m}}{m R^{m}}\Big)+\lambda f_\mathrm{coh}^{(3)} R^{-9},
\end{equation}
with $n>m$ and
\[
c_{n,m}= \frac{n m}{2(n-m)},
\]
for a given lattice $\Lambda$ with distance $R$. The resulting global minimization problem can be easily solved numerically using standard tools. It is, however, instructive to discuss particular special cases some of which allow for an analytic solution. Here we distinguish the cases where the repulsive part of the LJ potential dominates the three-body potential for small nearest neighbor distances or not. 
\begin{enumerate}
    \item $n>9$. After setting $\partial E_\mathrm{coh}/\partial R=0$, this case reduces to solving the following root finding problem,
    \begin{equation}
    \label{eq:root_finding_problem}
        c_{n,m}(Z_{\Lambda,n}-Z_{\Lambda,m}R^{n-m})+9\lambda f_\mathrm{coh}^{(3)} R^{n-9}=0.
    \end{equation}
    For the special case $n=9+k$ and $m=9-k$
 %   \JBCor{for $k\in\mathbb N$, %?
    the energy minimum can be determined analytically as
    \begin{equation}
    \label{eq:special_solution_9plusk}
    R_\mathrm{min}(n,m,\lambda) =\Bigg(\frac{9 \lambda f_\mathrm{coh}^{(3)}}{2c_{n,m}Z_{\Lambda,m}}+\sqrt{\left(\frac{9 \lambda f_\mathrm{coh}^{(3)}}{2c_{n,m}Z_{\Lambda,m}}\right)^2+\frac{Z_{\Lambda,n}}{Z_{\Lambda,m}}}\Bigg)^{1/(n-9)}.
    \end{equation}
    The often used $(12,6)$ LJ potential with $k=3$ belongs to this class.
    \item $n=9$: In this special case, we find from Eq.~\eqref{eq:root_finding_problem} that
    \begin{equation}
        R_{\mathrm{min}}(n,m,\lambda) = \bigg(\frac{Z_{\Lambda,n}}{Z_{\Lambda,m}}+\frac{9\lambda f_\mathrm{coh}^{(3)}}{c_{n,m}Z_{\Lambda,m}}\bigg)^{1/(9-m)}.
        \label{eq:solution_n9}
    \end{equation}
    In case of attractive three-body interactions ($f_\mathrm{coh}^{(3)}<0$), the minimum only exists for sufficiently small ATM coupling strength $\lambda$ with
    \begin{equation}
    \label{eq:lambdacritn9}
    \lambda \le \frac{c_{n,m}Z_{\Lambda,n}}{9 |f_\mathrm{coh}^{(3)}|}.
    \end{equation}
    \item $n<9$: This case requires special care. The ATM potential dominates the LJ term for small $R$. For attractive three-body interactions, this means that the global minimum of the energy is obtained for $R\to 0$, leading to a collapse of the lattice into the origin. Local energy minima can however exist for $R>0$. For the special case $n=9-k$, $m=9-2k$, we find extrema at
    \begin{equation}
        R = \Bigg(\frac{Z_{\Lambda,n}}{2Z_{\Lambda,m}} \pm \sqrt{\left(\frac{Z_{\Lambda,n}}{2Z_{\Lambda,m}}\right)^2+\frac{9\lambda f_\mathrm{coh}^{(3)}}{c_{n,m}Z_{\Lambda,m}}}\,\Bigg)^{1/(9-n)}.
    \end{equation}
    For attractive three-body interactions, a local minimum exists under the condition
    \begin{equation}
    \label{eq:crit9minusk}
        \lambda\le c_{n,m}\frac{Z_{\Lambda,n}^2}{36Z_{\Lambda,m}|f_\mathrm{coh}^{(3)}|}.
    \end{equation}
\end{enumerate}

In the following we introduce instructive toy models in one- and two dimensions where $R$ is chosen to be the nearest neighbor distance. We then analyze the influence of three-body interactions on the stability of cuboidal phases along a Bain path in three dimensions, where we choose $R$ as the distance from the atom at the origin to the body-centered atom, which is the nearest neighbor distance in region I ($\frac{1}{3}\le A \le 1$, see Figure \ref{fig:ALattice}).

\section{Results and Discussion
}
\label{sec:results}

\subsection{LJ+ATM Potential for a equidistant infinite linear chain}
\label{sec:Linear Chain}

We begin our investigation with the effect of a three-body ATM potential coupled to a two-body LJ-potential in one dimension for an equidistant linear chain.  The cohesive energy for the chain with nearest neighbor distance $R$ and normalized lattice $\Lambda=\mathbb Z$ becomes
\begin{equation}\label{eq:LCLJ}
E_\text{coh}(R,n,m) = E_\text{coh}^{(2)}(R,n,m)+E_\text{coh}^{(3)}(R,n,m)= c_{n,m} \Big( \frac{2 \zeta(n)}{n R^n}- \frac{2\zeta(m)}{m R^{m}}\Big)+\lambda f_\mathrm{coh}^{(3)} R^{-9},
\end{equation}
where we have used that the Epstein zeta function in 1D reduces to twice the Riemann zeta function,
$Z_{\mathbb Z,n}= 2\zeta(n)$.
In this 1D case the simple pole is situated at $n=1$ and $Z_{\mathbb Z,n}\rightarrow 2$ for $n\rightarrow\infty$ as each atom has two nearest neighbors.
%  \begin{figure*}[htbp!]
%  \begin{center}
%\includegraphics[scale=0.7]{LinearChain.pdf}
%\caption{\ABComment{(This figure is probably not needed anymore.)} The three-body ATM potential applied to an equidistant linear chain showing the two different classes of interactions between lattices points (0,1,2) and (0,1,3) relating to the two different sums in Eq.~\eqref{eq:ecohATMJL}.}
%  \label{fig:chain}
%  \end{center}
%  \end{figure*}
The three-body ATM potential in 1D is purely attractive, which follows directly from Eq.~\eqref{eq:f_coh},
\begin{equation}
f_{\mathrm{coh}}^{(3)}=f_r^{(3)}+f_a^{(3)}= \frac{1}{6} \,\sideset{}{'}\sum_{{x},{y} \in \mathbb Z}\frac{1}{\vert {x} \vert^3 \vert  {y}\vert^3 \vert {z} \vert^3} -\frac{1}{2} \,\sideset{}{'}\sum_{{x},{y} \in \mathbb Z} \frac{({x}{y})({y} {z})({z} x)}{\vert {x} \vert^5 \vert  {y}\vert^5 \vert {z} \vert^5}=-\frac{1}{3} f_r^{(3)}<0.
\end{equation}
The attractive behavior of the ATM potential for three atoms in a line has been discussed already by Axilrod and Teller \cite{AxilrodTeller1943} and will have important consequences in the following.

%The double sum in  expressions in \eqref{eq:linearATM2} converges to the values $f^{(3)}_\text{LC-1}=0.136150924753844$ and $f^{(3)}_\text{LC-2}=5.44993782529202\times 10^{-3}$ by direct summation over 5000 terms. 

In one dimension, the three-body cohesive energy can still be evaluated to machine precision using exact summation. We obtain 
$f^{(3)}_\mathrm{coh}=-0.2723018495076886$, which is in excellent agreement with the result from the Epstein zeta function treatment 
($f^{(3)}_\text{coh}=-0.27230184950768865$) as outlined in section \ref{sec:Epstein zeta}. This serves as a benchmark for higher dimensional lattices, where exact summation becomes exceedingly numerically expensive. 
% !!! OLD NOTATION: $f^{(3)}_\text{LC}=0.20422638713077$, !!!!
% $f^{(3)}_\mathrm{coh}=-0.15316979034807748$,
% which is in excellent agreement with the result from the Epstein zeta function treatment 
% % !!! OLD NOTATION: ($f^{(3)}_\text{LC}=0.20422638713076$) !!!
% ($f^{(3)}_\text{coh}=-0.15316979034807$)
%\JBCom{Previously $f^{(3)}_\mathrm{coh}=-0.15316979034807748$ and ($f^{(3)}_\text{coh}=-0.15316979034807$). Could not reproduce the values, maybe the normalization changed? Recomputed with direct sum (cutoff 1500 and 100 digits of precision) and Epstein Zeta.}

We now discuss the optimal nearest neighbor distance $R_\mathrm{min}(n,m,\lambda)$ as obtained in the previous Section~\ref{sec:optimization} for different repulsive LJ exponents $n$.
\begin{enumerate}
\item $n>9$: In this regime, the repulsive part of the LJ potential dominates the attractive ATM term. The solution to the root finding problem in Eq.~\eqref{eq:root_finding_problem} can be obtained numerically, with analytical solutions available for special cases such as the $(12,6)$-LJ potential in Eq.~\eqref{eq:special_solution_9plusk}.
We obtain $R_\text{min}(12,6,0.0)
=0.997179263885806$,  $R_\text{min}(12,6,1.0)
=0.964148870884975$, $R_\text{min}(12,6,3.0)
=0.902526982458744$ and $R_\text{min}(12,6,5.0)
=0.847847116323818$.
% $R_\text{min}(12,6,0.0)=0.99717926388581$,  $R_\text{min}(12,6,1.0)=0.964148870884977$, $R_\text{min}(12,6,3.0)=0.902526982458749$ and $R_\text{min}(12,6,5.0)=0.847847116323825$
%\JBCom{Changed values from 15 to 16+ digits of precision by recomputing with high precision and analytic formulas in mathematica}

As expected, the nearest neighbor distance decreases with increasing coupling strength $\lambda$ due to the increasing ATM attraction.

\item $n=9$: Here, the repulsive part of the LJ potential and the attractive ATM potential share the same scaling and their prefactors determine the dominant term. The value of $R$ that minimizes the energy is given by Eq.~\eqref{eq:solution_n9} as long as $\lambda$ obeys the bound
\begin{equation}
\lambda \le \frac{m\zeta(9)}{|(m-9)f_\mathrm{coh}^{(3)}|}.
\end{equation}
If this critical value of $\lambda$ is exceeded, then the energy diverges to $-\infty$ for $R\to 0$ and the chain collapses into the origin.

For example, for the $(9,6)$-LJ potential we get $\lambda\le 7.359541586938727$. 
%\JBCom{previously $\lambda\le 7.35954158683508$ which agrees in 11 digits}
At larger coupling strengths, the minimum vanishes and the interaction becomes purely attractive and collapse occurs, i.e. $R\rightarrow 0$. This has consequences for 2D or 3D lattices as under this model, the crystal may distort into a set of linear chains, as we shall see later on.

\item $n<9$. Here, the attractive ATM potential dominates at small distances $R$ and the global minimum is obtained for $R\to 0$.  Local minima can, however, exist for $R>0$, as described in the previous section, with analytic solutions available for $n=9-k$ and $m=9-2k$. A minimum then exists for sufficiently small $\lambda$ as described by 
% Eq.~\eqref{eq:lambdacritn9}.
Eq.~\eqref{eq:crit9minusk}.
For example, for $m=4 ~(k=\frac{5}{2})$ and $n=\frac{13}{2}$, we obtain 
% $\lambda<1.00389745875095$,
$\lambda<1.003897458750910$
which is a rather small value.
We find that by lowering the exponent for the repulsive force in the LJ potential, the existence of a minimum for the cohesive energy is achieved at lower critical values of the coupling strength $\lambda$.
\end{enumerate}

\begin{figure}[ht!]
    \centering
    a)\includegraphics{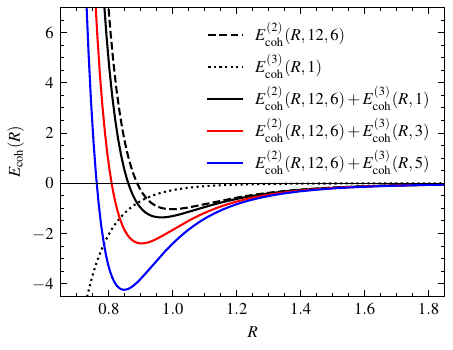}
    b)\includegraphics[]{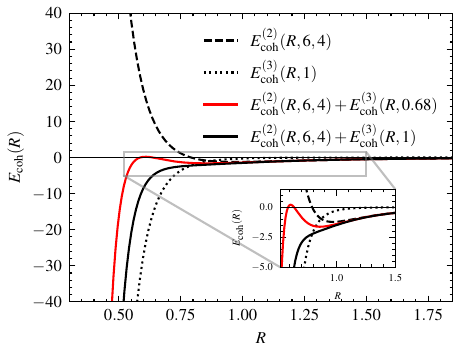}
    \caption{Cohesive energy of a linear chain with atoms interacting through: a) $(12,6)$-LJ coupled to an ATM potential, and b) $(6,4)$-LJ coupled to an ATM potential. Separate two- and three-body contributions are also shown in dashed and dotted lines, respectively.}
    \label{fig:LinearChain_Ecoh}
\end{figure}

Figure \ref{fig:LinearChain_Ecoh} summarizes our results for cases 1 and 3 for two different LJ potentials. When the two-body potential is of $(12,6)$-LJ type, the repulsive LJ term dominates at short distances over the ATM term in the cohesive energy. On the other hand, for a softer two-body potential with exponent $n<9$, as for the $(6,4)$-LJ potential, the ATM potential completely dominates over the repulsive part of the LJ term for $\lambda > 0.9$ making the total cohesive energy behave like $-R^{-9}$ with a singularity at $R=0$. When $\lambda \leq 0.689$, there is a competition between the attractive and repulsive parts of the cohesive energy, leading to a maximum in the short-range region that makes the cohesive energy slightly positive, followed by a divergence towards $-\infty$ due to the dominance of the attractive 3-body term. It is well known that the simple ATM term is valid only in the long-range,\cite{Sherrill2023} and one has to correct the unphysical behavior of the 3-body term in the short range in order to avoid the collapse of all atoms towards the origin. On the other hand one should make sure that the repulsive wall in the two-body potential is described realistically.

\subsection{LJ+ATM Potential for a square and hexagonal lattice}
\label{sec:2D-Lattices}

\begin{figure}[ht!]
    \centering
    \includegraphics[scale=0.55]{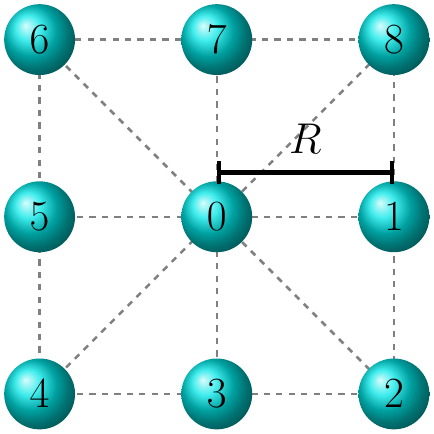}
    \hspace{25pt}
    \includegraphics[scale=0.55]{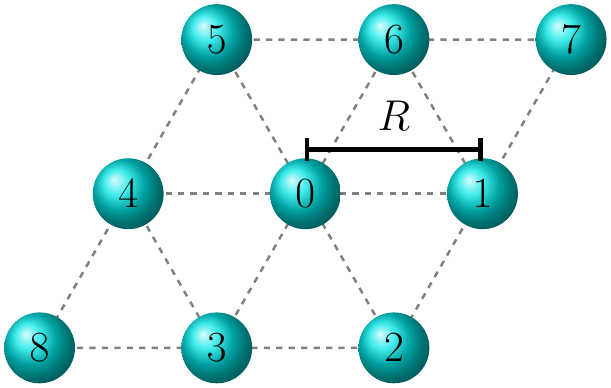}
    \caption{Square and hexagonal lattices of atoms interacting through LJ+ATM potentials with nearest neighbor distance $R$.}
    \label{fig:squarelattice}
\end{figure}

After analyzing the one-dimensional chain, we extend our focus to two-dimensional lattices. Among the five possible Bravais lattices, we restrict ourselves to the case of a square (SL) and hexagonal lattice (HL) with a nearest-neighbor distance $R$, as shown in Figure \ref{fig:squarelattice}, and to the rectangular lattices as a mode to distort the square lattice into a set of linear chains. The hexagonal lattice is the densest packing of circles in a two-dimensional plane.

The two- and three-body terms of the cohesive energy for a square lattice $\Lambda=\mathbb Z^2$ are given by Eq.~\eqref{eq:ecohATMJL}. Note that the Epstein zeta function for the square lattice appearing in the LJ term can be rewritten in terms of elementary functions,\cite{Zucker-1974}
\begin{equation}
    Z_{\mathbb Z^2,n}= 4\zeta(\tfrac{n}{2})\beta(\tfrac{n}{2}),
\end{equation} where $\beta$ denotes the Dirichlet beta function.

The three-body term of the cohesive energy is given by Eq. \eqref{eq:f_coh}. After an evaluation of the lattice sums through direct summation restricting the sums over integers to $N_\text{max}=1600$, the ATM term in the cohesive energy of the square lattice can be written as
\begin{equation}
    E_\text{coh}^{(3)}=  \lambda f^{(3)}_\text{coh} R^{-9}
\end{equation}
with $f^{(3)}_\text{coh}=f_r^{(3)}+f_a^{(3)}=0.77009365051710454$ where 
$f_r^{(3)}=2.2754822858923625$
and 
%$f_A^{(3)}=-24.086218166004134$%
$f_a^{(3)}=-1.5053886353752584$, see Eq.~\eqref{eq:f_coh}. This compares well with the more accurate result from the Epstein zeta treatment as outlined in section \ref{sec:Epstein zeta}, i.e. we get $f^{(3)}_\text{coh}=0.7700936505167162$ with $f_r^{(3)}=2.27548228589309$ and $f_a^{(3)}=-1.5053886353763737$. 
%\ABComment{(Adapted notation, factors $1/6$ and $1/16$ are now included. Rescaled values accordingly.)} 
The three-body contribution from an ATM potential to the cohesive energy is now repulsive for any $R$ in the square lattice in contrast to the 1D case.

One can now perform a similar analysis compared to the 1D case. We only mention two examples here. For the (12,6)-LJ potential, the optimal nearest neighbor distance is given by Eq.~\eqref{eq:special_solution_9plusk}, where the ATM potential is now positive. This results in
$R_\text{min}(12,6,0.0)=0.977489041852768$, \break $R_\text{min}(12,6,1.0)=1.021577293064089$, 
$R_\text{min}(12,6,3.0)=1.112586607942759$ and 
$R_\text{min}(12,6,5.0)=1.202957096531386$.
We see that the distance is increasing rapidly with increasing coupling strength $\lambda$.

The other case we consider here is when the potential becomes completely repulsive over the whole range of $R$ values. This can happen if the attractive $R^{-m}$ term is always dominated by the repulsive ATM term, which can only occur for $m\ge 9$. Consider the case $m=9$ which leads to the minimum distance,
\begin{equation}
R_\text{min}(n,9,\lambda)=\left(\frac{Z_{\Lambda,n}}{Z_{\Lambda,9}-9\lambda f_\mathrm{coh}^{(3)}/c_{n,9}}\right)^{1/(n-9)}
\end{equation}
and to the condition that
\begin{equation}
\lambda<\frac{c_{n,9} Z_{\Lambda,9}}{9 f_\mathrm{coh}^{(3)}}.
\end{equation}
For example, for $n=12$ we get $\lambda<10.88508744343488$.
With increasing exponent $n$ the critical $\lambda$ value decreases as one would expect.

\begin{figure}[ht!]
    \centering
    a)\includegraphics{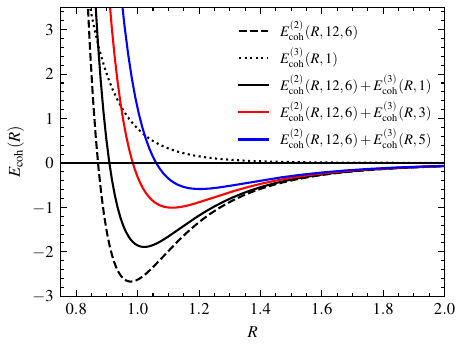}
    % b)\includegraphics{SquareLattice_3body_64LJ}
    b)\includegraphics{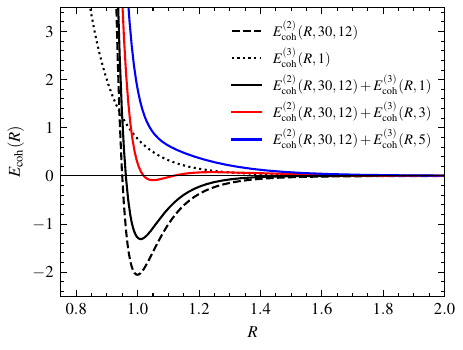}
    \caption{Cohesive energy of the square lattice with atoms interacting through: a) $(12,6)$-LJ potential coupled to an ATM potential, and b) $(30,12)$-LJ coupled to an ATM potential. Separate two- and three-body contributions are also shown in dashed and dotted lines, respectively.}
    \label{fig:Ecoh-squarelattice}
\end{figure}

The cohesive energy for the square lattice as a function of $R$ for two different $(n,m)$-LJ potentials coupled to the ATM potential is shown in Figure \ref{fig:Ecoh-squarelattice}. We depict the two-body term $E_\mathrm{coh}^{(2)}(R,n,m)$, the three-body contribution $E_\mathrm{coh}^{(2)}(R,\lambda)$, and the full cohesive energy. An example of the long-range region becoming dominated by the repulsive ATM term is for the hard $(30,12)$-LJ potential as shown in Figure \ref{fig:Ecoh-squarelattice}. For $\lambda=1$, there is a region in which the square crystal is bounded around the equilibrium distance, however for $\lambda > 3.1834$ the cohesive energy becomes positive at any distance, meaning that we have a purely repulsive potential energy.

There is one small caveat here to consider as the square lattice might distort to a set of weakly interacting linear chains which may collapse to the origin for $n\le 9$ as discussed in the previous section. In order to show this we consider the following generator matrix $B^\top$ and Gram matrix $G$,
\begin{equation}\label{eq:Gramrect}
B^\top=\begin{pmatrix}
1 &0 \\
0 &\gamma
\end{pmatrix},\quad
G=\begin{pmatrix}
1 &0 \\
0 &\gamma^2
\end{pmatrix},
\end{equation}
which allow the square lattice to distort into a rectangular lattice where $\gamma$ is the distortion parameter.
Here, analytical solutions for the corresponding LJ lattice sums exist only for special cases of $\gamma$ values.\cite{Zucker-1975a,Zucker2017,burrows-2020} We therefore use either the Van der Hoff Benson expansion\cite{Hoff-Benson-1953} (see Appendix \ref{sec:2bodylatticesums}) or EpsteinLib for the two-body term, and direct summation for the three-body term, where the latter approximation is sufficiently accurate to demonstrate the effect of such an unphysical distortion.

\begin{figure}
    \centering
    \includegraphics[width=0.6\textwidth]{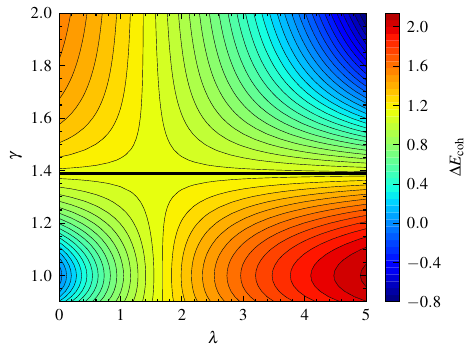}
    \caption{($\gamma,\lambda$)-hypersurface for the difference in the cohesive energy of a rectangular lattice at optimized $R$ with respect to the ideal square lattice at $(\gamma,\lambda)=(1,0)$. The horizontal black line indicates the critical value of $\lambda$ in which the ATM potential is neither repulsive nor attractive.}
    \label{fig:RectangularHypersurface}
\end{figure}

As it turns out, there exists a critical value of $\gamma\approx 1.388$ that makes the three-body term of the cohesive energy neither repulsive nor attractive, i.e. $E_\text{coh}^{(3)}(R,\gamma,\lambda)=0, \lambda>0$. This value can be seen as the limit between the three-body attractive interaction in the linear chain and the three-body repulsion characteristic of the square lattice as shown before. The $(\gamma,\lambda)$-cohesive energy hypersurface for the difference in cohesive energy with respect to the square lattice with $(\gamma,\lambda)=(1,0)$, i.e.
\begin{equation}
    \Delta E_\mathrm{coh}(R,\gamma,\lambda)=E_\mathrm{coh}(R,\gamma,\lambda) - E_\mathrm{coh}(R,1,0),
\end{equation}
is shown in Figure \ref{fig:RectangularHypersurface}. The square lattice without three-body interactions is located at a local minimum of the hypersurface at $(\gamma,\lambda)=(1,0)$, whereas the global minimum in the selected range is found at the upper right corner of the plot. 

The point $(\gamma,\lambda)=(2,5)$ corresponds to a rectangular lattice with one of the sides of its unit cell being twice as large as the other. The reason for the high stability of this structure with respect to the square lattice is due to the fact that it is located at the region where the three-body potential becomes attractive, similar to the case of the linear chain. In fact, the square lattice structure is highly destabilized by the repulsive three-body forces, as shown in the lower right corner of Figure \ref{fig:RectangularHypersurface}, whereas the rectangular lattice in the upper left corner is destabilized by due to two-body forces.

In a similar way, the generator matrix and the Gram matrix for the hexagonal lattice $\Lambda_\mathrm{hex}=B^\top \mathbb Z^2$, depicted in Figure \ref{fig:squarelattice}, are given by
\begin{equation}
B_\mathrm{hex}^\top =\begin{pmatrix}
1 & \frac{1}{2} \\
0 & \frac{\sqrt{3}}{2}
\end{pmatrix},\quad  G_\mathrm{hex}=\begin{pmatrix}
1 &\frac{1}{2} \\
\frac{1}{2} &1
\end{pmatrix}.
\end{equation}

\begin{figure}[ht!]
    \centering
    a)\includegraphics{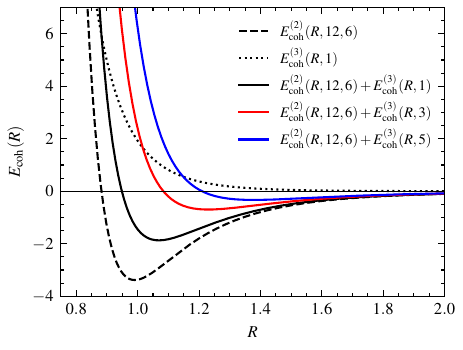}
    b)\includegraphics{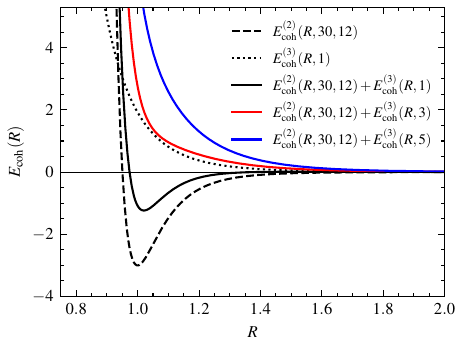}
    \caption{Cohesive energy of the hexagonal lattice with atoms interacting through: a) $(12,6)$-LJ potential coupled to an ATM potential, and b) $(30,12)$-LJ coupled to an ATM potential. Separate two- and three-body contributions are also shown in dashed and dotted lines, respectively.}
    \label{fig:Ecoh-hexlattice}
\end{figure}
The lattice sum of the hexagonal lattice also has an analytical formula given by Zucker and Robertson,\cite{Zucker-1974}
\begin{equation}
    Z_{\Lambda_{\mathrm{hex}},n}= 6\zeta(\tfrac{n}{2})\left[3^{-n/2}\left(\zeta(\tfrac{n}{2},\tfrac{1}{3})-\zeta(\tfrac{n}{2},\tfrac{2}{3})\right)\right]
\end{equation}
where $\zeta(n,x)$ is the Hurwitz zeta function, see Appendix \ref{sec:specialfunctions}. 
%For numerical evaluations, however, the evaluation of the two-body potential using Bessel function expansions or EpsteinLib is preferable due to possible cancellation error. 
The full cohesive energy is then given by Eq.~\eqref{eq:ecohATMJL}.

For the three-body term in Eq.~\eqref{eq:f_coh} we evaluate the 2D lattice sums through direct summation with $N_\text{max}=1600$ to get
\begin{equation}
    E_\text{coh}^{(3)}(R, \lambda) = \lambda f^{(3)}_\text{coh} R^{-9}
\end{equation}
with $f^{(3)}_\text{coh}=1.9183333648489187$ ($f_r^{(3)}=4.263827935989311, f_a^{(3)}=-2.3454945711403923$)
% !!!OLD NOTATION!!! ($f_R^{(3)}=25.582967615935868, f_A^{(3)}=-37.527913138246277$).
From Epstein zeta treatment, we get $f^{(3)}_\text{coh}=1.9183333648478795$ ($f_r^{(3)}=4.263827935991082$ and $f_a^{(3)}=-2.3454945711432025$).
% !!!OLD NOTATION!!! ($f_R^{(3)}=25.582967615946494$ and $f_A^{(3)}=-37.52791313829124$)
As in the square lattice, the three-body term also results in a repulsive contribution to the total cohesive energy, as shown in Figure \ref{fig:Ecoh-hexlattice}. Furthermore, the lattice becomes unstable after a critical value of $\lambda$ is reached, where the total cohesive energy is positive for any value of $R$. For example, this limit is obtained for the $(30,12)$-LJ potential coupled with the ATM when $\lambda>1.8854$. These results show that the hexagonal lattice is more strongly destabilized by adding three-body interactions compared to the square lattice because $f^{(3)}_\text{coh,hex}>f^{(3)}_\text{coh,sq}$. This is due to the hexagonal lattice being a close-packed structure in 2D with the highest packing density and kissing number.

Again we can do the same analysis as for the square lattice case, but mention only here the minimum distances for the (12,6)-LJ potential for which we get $R_\text{min}(12,6,0.0)=0.990193636287356$, $R_\text{min}(12,6,1.0)=1.06923072624940$, $R_\text{min}(12,6,3.0)=1.22992964510945$ and $R_\text{min}(12,6,5.0)=1.37801743468056$. Similar to the square lattice, a distortion into a set of linear chains can occur if $n\le 9$ for large ATM coupling strengths $\lambda$.

\subsection{LJ+ATM Potential for the cuboidal lattices} 
\label{sec:Bain}
We are interested in the Bain minimum energy path $E_\text{coh}(R_\text{min},A)$ along the $A$-dependent cuboidal lattices at an optimized distance $R=R_\text{min}$. The corresponding cohesive energy is obtained from Eq.~\eqref{eq:ecohATMJL} as
\begin{equation}\label{eq:cohLJATM3D}
    E_{\text{coh}}(n,m,A,\lambda,R_\mathrm{bc}) = \frac{nm}{2(n-m)} \bigg( \frac{Z_{\Lambda(A),n}}{n R_\mathrm{bc}^n}- \frac{Z_{\Lambda(A),m}}{m R_\mathrm{bc}^{m}}\bigg)+\lambda f_\mathrm{coh}^{(3)}(A) R_\mathrm{bc}^{-9},
\end{equation}
for the lattice $\Lambda(A)$ along the Bain transformation path
\[
\Lambda(A)=B^\top(A) \mathbb Z^3,\quad B^\top(A)=\frac{1}{\sqrt{A+1}} \begin{pmatrix}
\sqrt{A} & \sqrt{A} & 0 \\
1 & 0 & 1 \\
0 & 1 & 1 
\end{pmatrix},\quad 0<A\le 1.
\]
It is important to notice that the above lattice only exhibits unit nearest neighbor distance for $1/3\le A\le 1$. We here define our measure of distance $R_\mathrm{bc}$ for all  values of $A$ as the distance from the atom in the origin to the body centered atom, otherwise one has to change the lattice sum in region I. In region I in Fig.~\ref{fig:ALattice}, the resulting nearest neighbor distance can easily be obtained from $R_\mathrm{bc}$. This choice is made to assure a smooth behavior of the resulting minimized distance $R_{\text{min}}$ across the whole range of $A$ values and facilitates the exploration of region I, where we investigate the distortion of the cubic lattice to a set of weakly interacting linear chains along the $c$-axis. For ease of notation, we set $R=R_\mathrm{bc}$ in the following.

 %Notice that we will also explore region I (see Figure \ref{fig:ALattice}) to understand the distortion of the cubic lattice to a set of weakly interacting linear chains ($A<\frac{1}{3}$) along the $c$-axis. For this case we keep the Gram matrix as defined for region II ($\frac{1}{3}\le R\le 1$) of Figure \ref{fig:ALattice} and Eq.\eqref{eq:g} to assure a smooth behavior of the minimized distance $R_{\text{min}}$ at the boundary between the two regions, describing $R=R_\text{bc}$ as the distance from the atom in the base layer (origin) to the body-centred atom. We keep in mind that in region I we have $R_{\text{min}}\ne R $, as the nearest neighbor distances in the linear chains become shortest (change in kissing number from 8 to 2 at the boundary). However, in this region $R $ can easily be obtained from $R_\text{bc}$. We will address region I in more detail further below.

The two-body contribution to the cohesive energy depends on the Epstein zeta function $Z_{\Lambda(A),n}=L(\frac{n}{2},A)$, which is either obtained from the Bessel expansion in in Appendix \ref{sec:latticesumsA} or using EpsteinLib. In the following we analyse the Bain phase transition for a range of $(n,m)$-LJ potentials, i.e. $(6,4)$-LJ, $(8,6)$-LJ, $(12,6)$-LJ and $(30,12)$-LJ. Note that the Epstein zeta function   $Z_{\Lambda(A),n}$ becomes minimal for the bcc structure ($A=\frac{1}{2}$), as discussed in detail in Appendix \ref{sec:MinimumProperty}.

The three-body lattice sum $f_\mathrm{coh}^{(3)}(A)$ is depicted in Figure~\ref{fig:f3(A)} as a function of the parameter $A$. The corresponding data points are provided in the supplementary material. For region II the highest repulsive three-body energy occurs for the densely packed fcc lattice, while the energy minimum is reached for the bcc lattice within the studied parameter range, similar to the lattice sums for the Lennard-Jones (LJ) potential. This suggests that the fcc lattice may become unstable relative to the bcc lattice if the coupling parameter $\lambda$ becomes sufficiently large. Moreover, at small $A$-values in region I we see that the three-body has a maximum and starts to go steeply down in energy becoming eventually attractive as discussed for the one- and two-dimensional cases.
 \begin{figure}[htbp!]
  \begin{center}
  \includegraphics[scale=0.6]{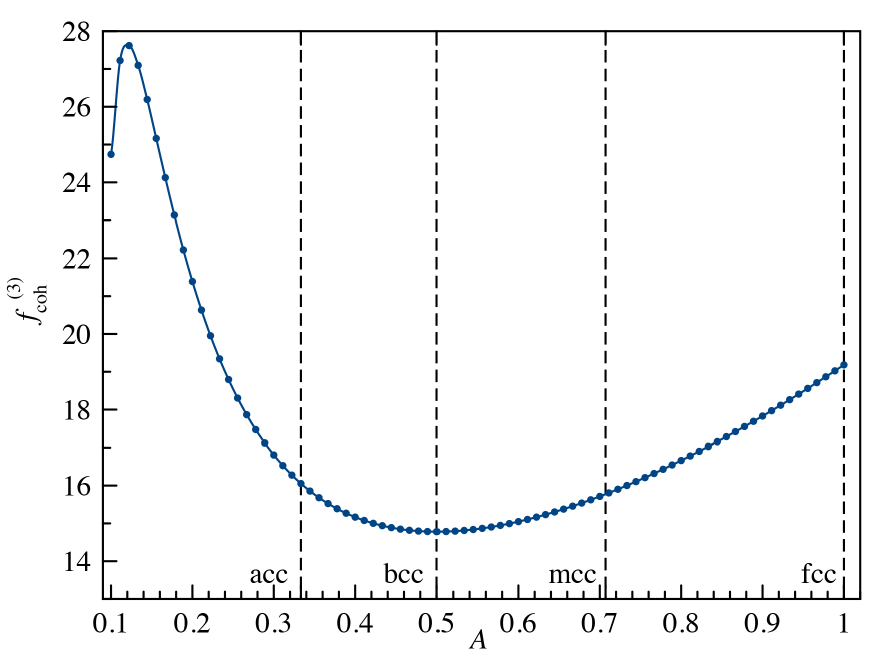}\\
  \caption{
  The normalized ATM cohesive energy $f^{(3)}_{\mathrm{coh}}(A)=f_r^{(3)}(A)+f_a^{(3)}(A)$ is displayed along the Bain path as a function of the lattice parameter $A$.}
  \label{fig:f3(A)}
  \end{center}
  \end{figure}

In order to assess the stability of the bcc with respect to the fcc phase we need to minimize the cohesive energy \eqref{eq:cohLJATM3D} with respect to the distance $R$. The resulting optimization problem is discussed in detail in Sec.~\ref{sec:optimization} and can be easily solved numerically with standard tools. However, as already explained in the previous sections, for the (12,6)-LJ potential coupled to an ATM potential we can derive analytical solutions for the minimum of Eq.\eqref{eq:cohLJATM3D}, which we like to analyse in more detail. 

The optimal distance $R_\mathrm{min}$ to the body-centered atom is obtain from Eq.~\eqref{eq:special_solution_9plusk} as
\begin{equation}
    R_\text{min}(12,6,\lambda,A)=\left( \sqrt{\left( \frac{3\lambda f^{(3)}(A)}{4 Z_{\Lambda(A),6}} \right)^2+\frac{Z_{\Lambda(A),12}}{Z_{\Lambda(A),6}}} + \frac{3\lambda f^{(3)}(A)}{4 Z_{\Lambda(A),6}}\right)^\frac{1}{3}.
\end{equation}
For $\lambda=0$ we obtain the well known result,\cite{Schwerdtfeger-2006}
\begin{align}
R_\text{min}(12,6,\lambda=0,A) = \left( \frac{Z_{\Lambda(A),12}}{Z_{\Lambda(A),6}} \right)^\frac{1}{6}
\end{align}
and for $\lambda\rightarrow\infty$ we see that $R_\text{min}(A)\rightarrow\infty$ for a repulsive three-body term. 
%A few examples illustrate the behavior of the minimum distance $R_\text{min}(\lambda)$ with increasing coupling strength: for $A=1.0$ (fcc) we have $R_\text{min}(0.0)=0.97123369095965$, $R_\text{min}(1.0)=1.32916247386527$, $R_\text{min}(3.0)=1.82812159023424$, $R_\text{min}(5.0)=2.15711563016713$, and for $A=0.5$ (bcc) we have $R_\text{min}(0.0)=0.95186481866244$, $R_\text{min}(1.0)=1.29157038625249$, $R_\text{min}(3.0)=1.77174890457456$, $R_\text{min}(5.0)=2.08990393283754$. 
A few examples illustrate the behavior of the minimum distance $R_\text{min}(\lambda)$ with increasing coupling strength: for $A=1.0$ (fcc) we have $R_\text{min}(0.0)=0.9712336909596462$, $R_\text{min}(1.0)=1.3291651590715157$, $R_\text{min}(3.0)=1.8281263889243278$, $R_\text{min}(5.0)=2.1571214550726303$, and for $A=0.5$ (bcc) we have $R_\text{min}(0.0)=0.9518648186624387$, $R_\text{min}(1.0)=1.2915727206984038$, $R_\text{min}(3.0)=1.7717531211610782$, $R_\text{min}(5.0)=2.0899090593393477$.

\begin{figure}[htb!]
  \begin{center}
\includegraphics[scale=0.55]{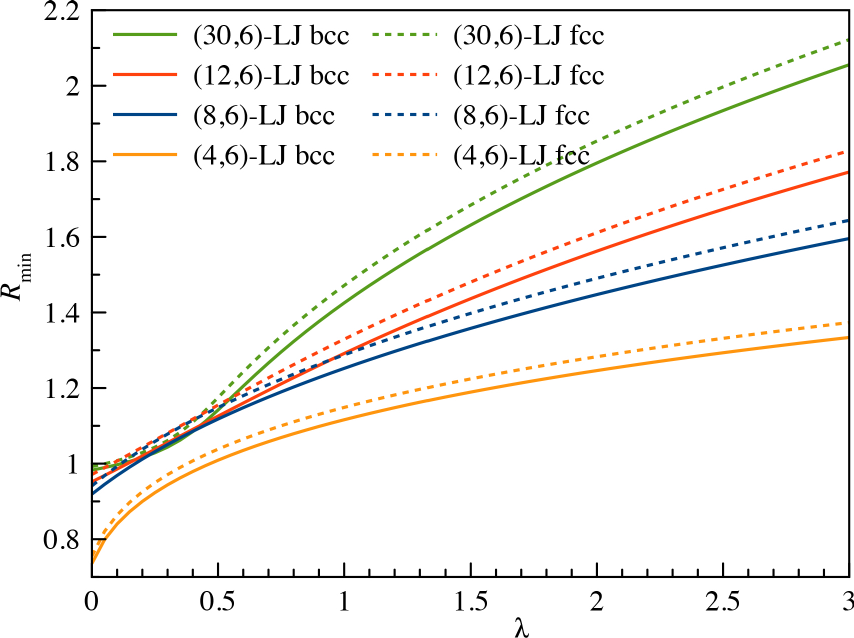}
  \caption{
  Minimum distances $R_\text{min}(A,\lambda)$ for the bcc ($A=\frac{1}{2}$) and fcc ($A=1$) structures for different $(n,m)$-LJ potentials as a function of the ATM coupling parameter $\lambda$.
  }
  \label{fig:distance}
  \end{center}
  \end{figure}

Because of  $f_\mathrm{coh}^{(3)}$(fcc)$>f_\mathrm{coh}^{(3)}$(bcc) the minimum distance is more rapidly increasing for fcc than for bcc with increasing coupling strength $\lambda$. This is shown for different $(n,m)$-LJ potentials in Figure \ref{fig:distance}. The minimum properties are also shown in Table \ref{tab:LJ}. We note that for the (30,6)-LJ potential we have $R_\text{min}=$ 0.9828 (bcc) and 0.9923 (fcc) at $\lambda=0$, which is close to the unit distance for hard spheres. This is expected for a hard-wall potential that approaches the sticky hard-sphere limit.\cite{Baxter1968} Further, soft potentials (low $n$ and $m$ values) lead to larger contractions in $R_\text{min}$ when moving along the cuboidal distortion path from fcc to bcc. 
  
\begin{table}[htb!]
\setlength{\tabcolsep}{0.2cm}
\caption{\label{tab:LJ} Minimum distances and cohesive energies at $\lambda=0$ derived analytically from the lattice sums (see also Ref.~\onlinecite{burrows2021b}). For the bcc structure we have the general condition that \break $\partial E_\text{coh}^{(2)}(A\!=\!\frac{1}{2})/\partial A=0$. For the (8,6), (12,6) and (30,6) LJ potentials the bcc structure is a maximum along the Bain path.
}
\begin{center}
\begin{tabular}{|l|r|r|r|r|}
\hline
                                                                & (6,4)    & (8,6)   & (12,6)  & (30,6) \\
\hline 
$R_\text{min}(A\!=\!\frac{1}{2})$ 			                    & 0.7357107511 & 0.9192764815 & 0.9518648187 & 0.9827992166\\
$R_\text{min}(A\!=\!1)$ 					                    & 0.7552731838	& 0.9411200107 & 0.9712336910 & 0.9922781478\\
$E_\text{coh}^{(2)}(A\!=\!\frac{1}{2})$ 	                    & -38.636118884 & -10.152177739 &-8.237291910	& -6.799035350\\
$E_\text{coh}^{(2)}(A\!=\!1)$ 			                        & -38.934203192 &-10.401252415  &-8.610200157 	& -7.571032638\\
$\partial^2 E_\text{coh}^{(2)}(A\!=\!\frac{1}{2})/\partial R^2$ & 3426.26165602 & 1153.28899200 & 1309.17106453 & 2534.07927284\\
$\partial^2 E_\text{coh}^{(2)}(A\!=\!1)/\partial R^2$           &3276.15467791 & 1127.37098543 & 1314.40215104 & 2768.15729574\\
$\partial^2 E_\text{coh}^{(2)}(A\!=\!\frac{1}{2})/\partial A^2$ & 1.3106119526 & -4.0072840086 & -8.6586541684 & -19.701773034\\
$\partial^2 E_\text{coh}^{(2)}(A\!=\!1)/\partial A^2$           & 7.2332403470 &  4.8703884251 & 6.8704958970  & 17.713990956\\
\hline
\end{tabular}
\end{center}
\end{table}

To estimate the range of typical coupling strengths $\lambda$, we consider the formula derived from the Drude model describing the triple dipole interactions between three identical atoms at equilibrium distance $r_e$,\cite{Bade-1957,Polymeropoulos1985}
\begin{align}
\lambda=\frac{9}{16} \frac{I}{\epsilon}\frac{\alpha^3}{r_e^9}
\end{align}
where $I$ is the first ionization potential of the atom and $\alpha$ the static dipole polarizability. For example, taking known experimental or theoretical values \cite{moore1949,Huber-Herzberg1979,Nagle2019} we get for argon $\lambda_\text{Ar}=0.025$, for xenon $\lambda_\text{Xe}=0.034$, for the heaviest noble gas atom $\lambda_\text{Og}=0.101$, and for lithium (due to its large polarizability and small equilibrium distance) $\lambda_\text{Li}=6.0$. However, for bulk lithium the many-body expansion is not converging smoothly, as this is generally the case for metallic solids.\cite{N-body.AH2007,burrows2021b} This implies that the ATM term is applicable only for small coupling parameters $\lambda$, as larger values suggest that higher-order terms in the expansion \eqref{eq:nbody} become important as well. Based on these $\lambda$ values we chose the following grid in our computations: $A\in[\frac{1}{10},1]$ with step size $\Delta A=\frac{1}{60}$, and  $\lambda\in[0,6.0]$ with step size $\Delta \lambda=0.05$.

\begin{figure}[htb!]
  \begin{center}
a) \includegraphics[scale=0.48]{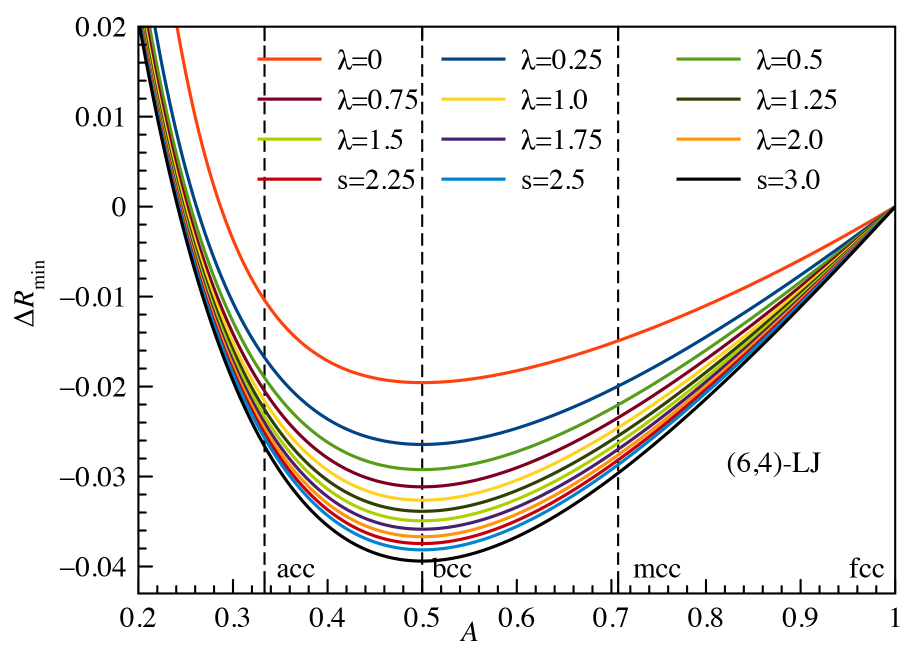}
b) \includegraphics[scale=0.48]{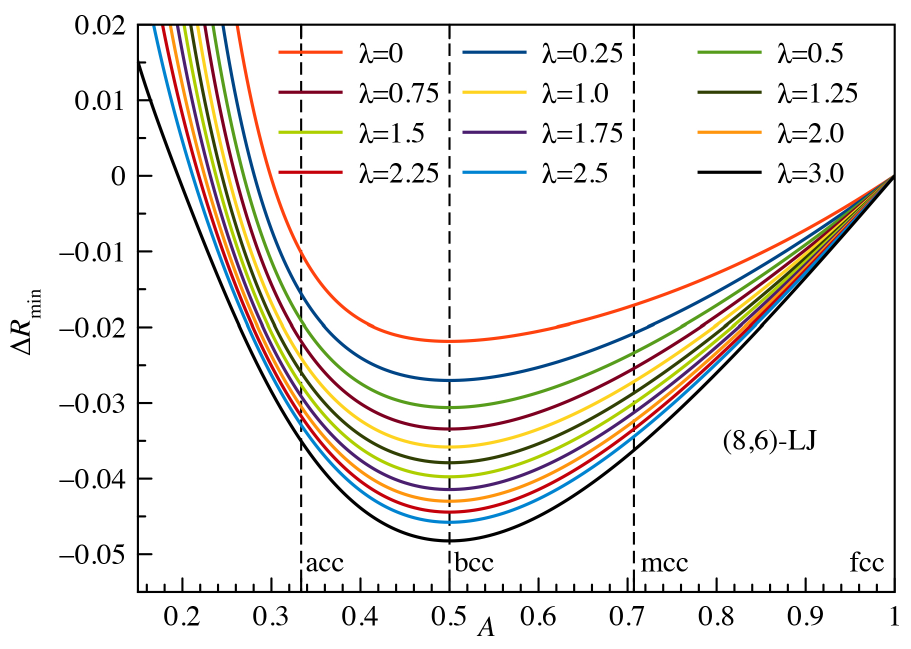}\\
c) \includegraphics[scale=0.48]{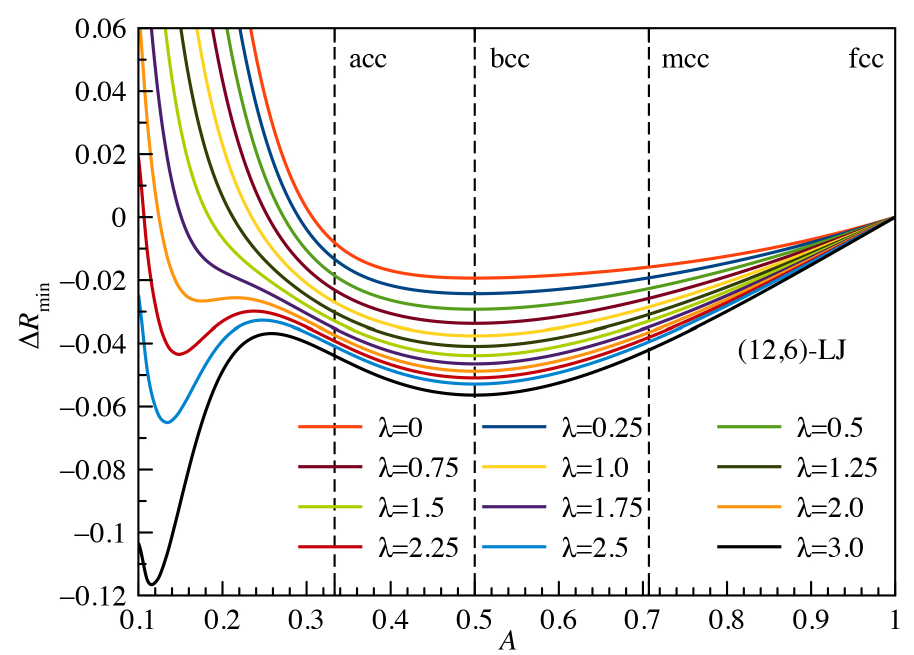}
d) \includegraphics[scale=0.48]{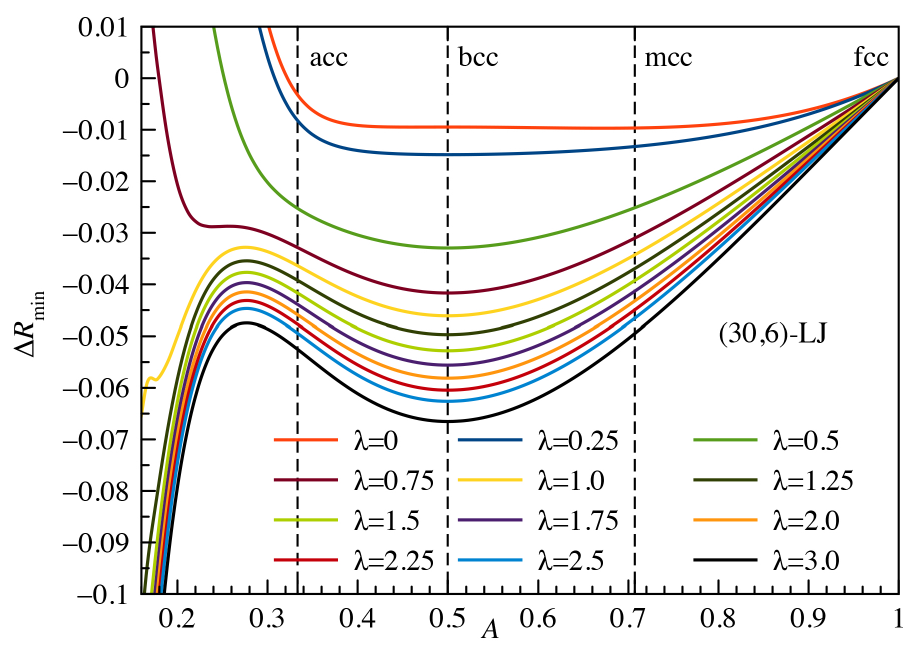}
  \caption{
  Difference $\Delta R_\text{min}(A,\lambda) = R_\text{min}(A,\lambda)-R_\text{min}(A=1,\lambda)$ for the $(6,4)$ (a) and $(12,6)$-LJ (b) potentials for different coupling parameters $\lambda$. The four distinct lattices acc ($A=\frac{1}{3}$ on the left, bcc ($A=1$) and mcc ($A=\frac{1}{\sqrt{2}}$) at dashed lines, and fcc ($A=1$) on the right are indicated.
  }
  \label{fig:distance2}
  \end{center}
  \end{figure}

As the differences between the bcc and fcc minimum distances are relatively small compared to $R_\text{min}$ at constant $\lambda$, we consider the difference in the minimum distances between the cuboidal and the fcc structures, i.e. $\Delta R_\text{min}(A,\lambda)= R_\text{min}(A,\lambda)-R_\text{min}(A=1,\lambda)$. Figure \ref{fig:distance2} shows $\Delta R_\text{min}(A,\lambda)$ values for the four different LJ potentials. They all show a qualitatively similar behavior in the region $\frac{1}{3}\le A \le 1$. The smallest distance is always found at the bcc structure ($A=\frac{1}{2}$). However, we see some significant changes to lower $\Delta R_\text{min}$ values with minima occurring in the region $A<\frac{1}{3}$ for larger coupling strengths $\lambda$ and larger repulsive walls (exponents $n=12$ and 30), which is due to linear chain formation as will be discussed in the following.
  
\begin{figure}[htb!]
  \begin{center}
\includegraphics[width=\linewidth]{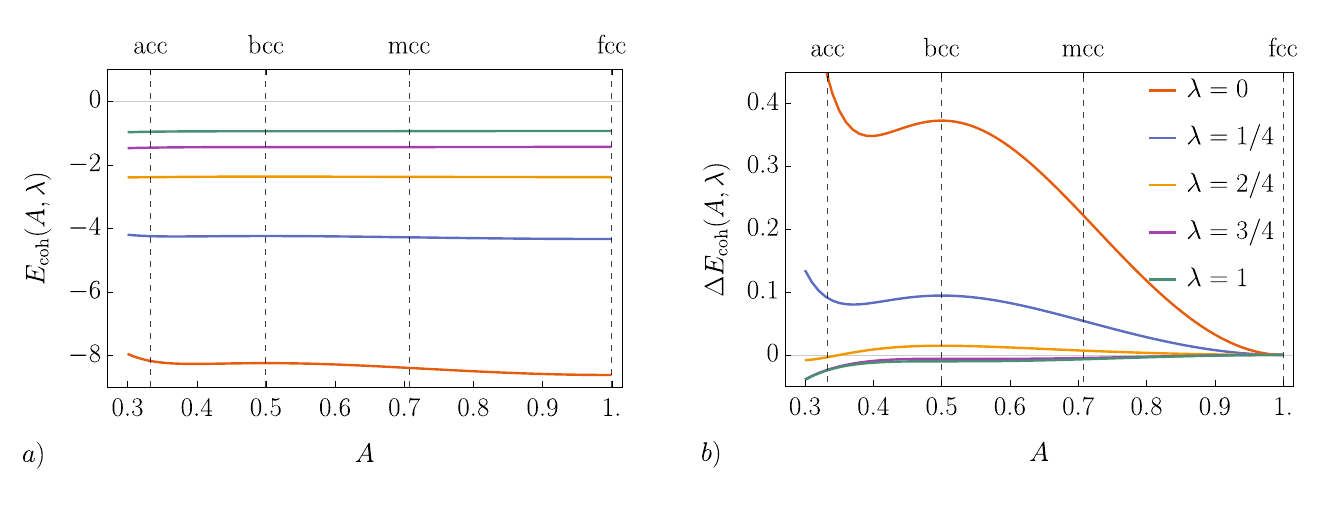}
  \caption{
  Cohesive energies, $E_\text{coh}(A,\lambda)$ (a) at the corresponding $R_\text{min}(A,\lambda)$ values dependent on the parameters $A$ and $\lambda$ for the $(12,6)$-LJ potential. The four distinct lattices acc ($A=\frac{1}{3}$ on the left, bcc ($A=1$) and mcc ($A=\frac{1}{\sqrt{2}}$) at dashed lines, and fcc ($A=1$) on the right are indicated. The difference $\Delta E_{\rm coh}(A,\lambda)=E_{\rm coh}(A,\lambda)-E_{\rm coh}(A=1,\lambda)$ is shown in (b).} 
  \label{fig:Ecohlambda}
  \end{center}
  \end{figure}

Figure \ref{fig:Ecohlambda} shows the cohesive energies for a few selected $\lambda$ values for the (12,6)-LJ potential. The energy curves are shifted towards higher energies with increasing $\lambda$ value as we expect, and become very flat at high energies. At the optimized distance $R_\text{min}(A,\lambda)$ it consistently holds that $\Delta E_\text{coh}(A,\lambda)<0$, which is below the atomization limit as expected. With increasing $\lambda$, $R_\text{min}$ becomes larger to the point that at long range the dispersive $R^{-m}$ term ($m=4,6$) in the LJ potential dominates over the repulsive ATM force.

\begin{figure}[htb!]
  \begin{center}
a)  \includegraphics[scale=0.5]{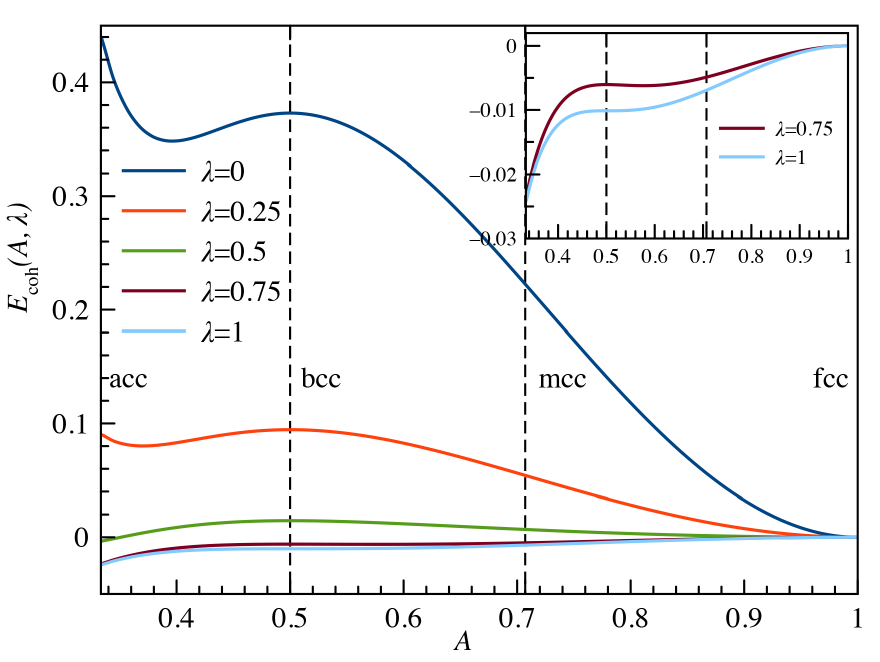} 
b) \includegraphics[scale=0.5]{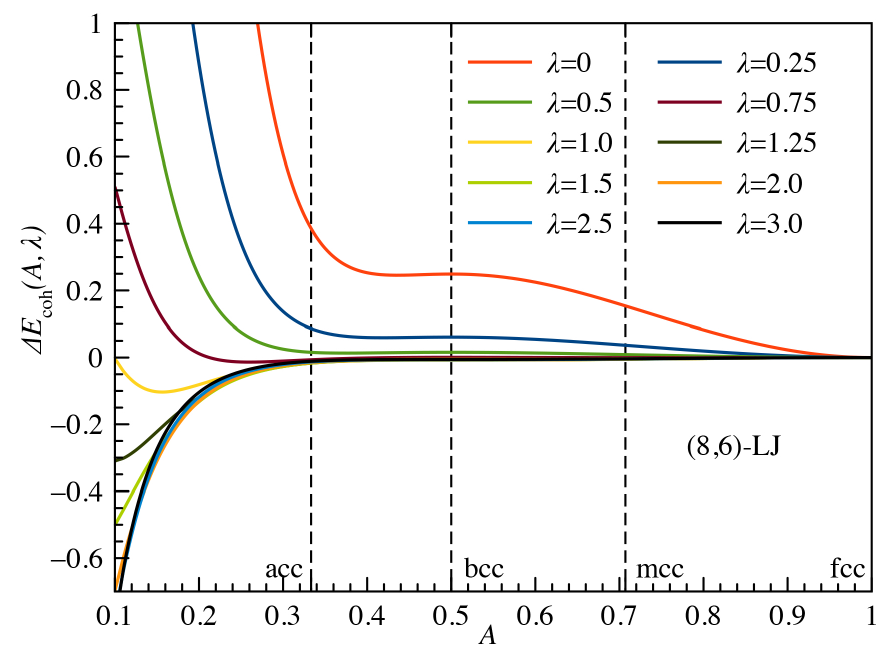}\\
c)    \includegraphics[scale=0.5]{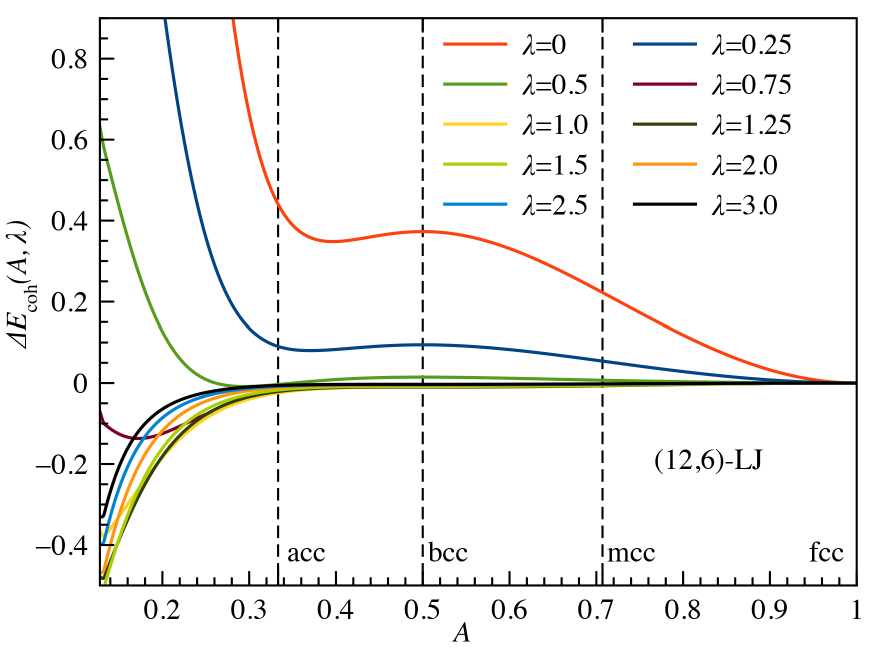}
d)    \includegraphics[scale=0.5]{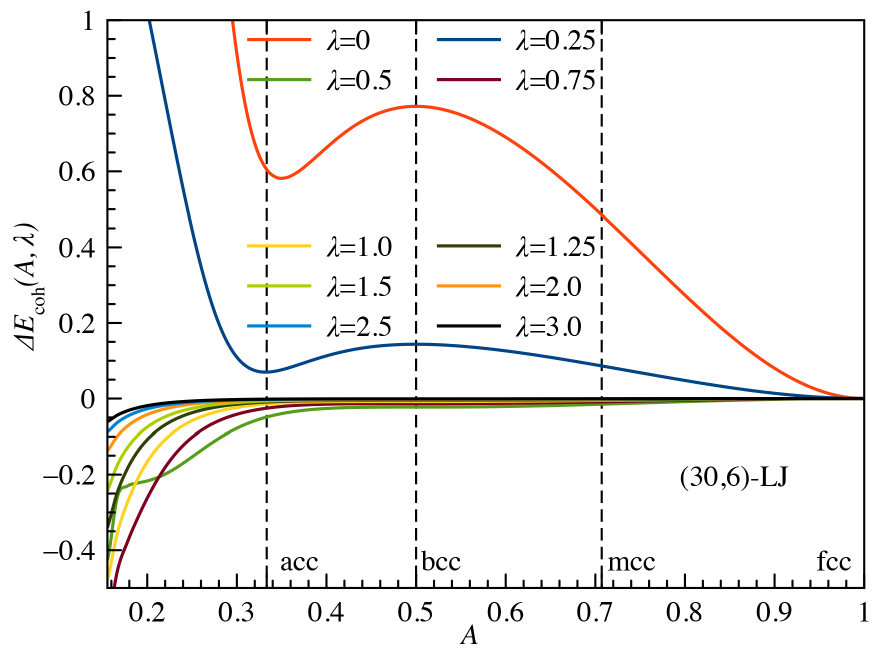}
  \caption{
  Cohesive energy differences, $\Delta E_\text{coh}(A,\lambda)=  E_\text{coh}(A,\lambda)- E_\text{coh}(A=1,\lambda)$ at the corresponding $R_\text{min}(A,\lambda)$ values dependent on the parameters $A$ and $\lambda$ for the $(12,6)$-LJ potential. The four distinct lattices acc ($A=\frac{1}{3}$ on the left, bcc ($A=1$) and mcc ($A=\frac{1}{\sqrt{2}}$) at dashed lines, and fcc ($A=1$) on the right are indicated.
  }
  \label{fig:Ecohlambda1}
  \end{center}
  \end{figure}

Details of the Bain transformation path become more transparent when we plot the difference in cohesive energies with respect to the fcc structure as shown in Figure  \ref{fig:Ecohlambda1}. It was pointed out before that for a certain range of $(n,m)$ values with $m<5.25673, n>m$ and $\lambda=0$ the bcc phase becomes metastable, otherwise it will distort towards lower $A$ values, i.e. the acc structure.\cite{burrows2021b} However, the bcc structure strictly remains an extremum.\cite{burrows2021b} The instability of the bcc phase for certain LJ exponents was already discussed in 1940 by Born and Misra \cite{Born_1940,Misra_1940}, and later by Wallace and Patrick.\cite{Wallace1965} Similar results are obtained for the generalized Morse potential\cite{Milstein1973} indicating that many-body forces have substantial influence on the bcc phase. However, a distortion from the ideal bcc phase was also found by Craievich et al for several elemental metals.\cite{Craievich1997} Adding the ATM potential we see that at a critical coupling strength $\lambda_c$ ($\lambda_c=0.635$ for the (12,6)-LJ potential for example) the bcc phase starts to lie energetically below the fcc phase. The critical $\lambda_c$ values obtained from a polynomial fit are listed in Table \ref{tab:ATM} for the four $(n,m)$-LJ potentials considered.

  \begin{figure}
\centering \includegraphics[width=\linewidth]{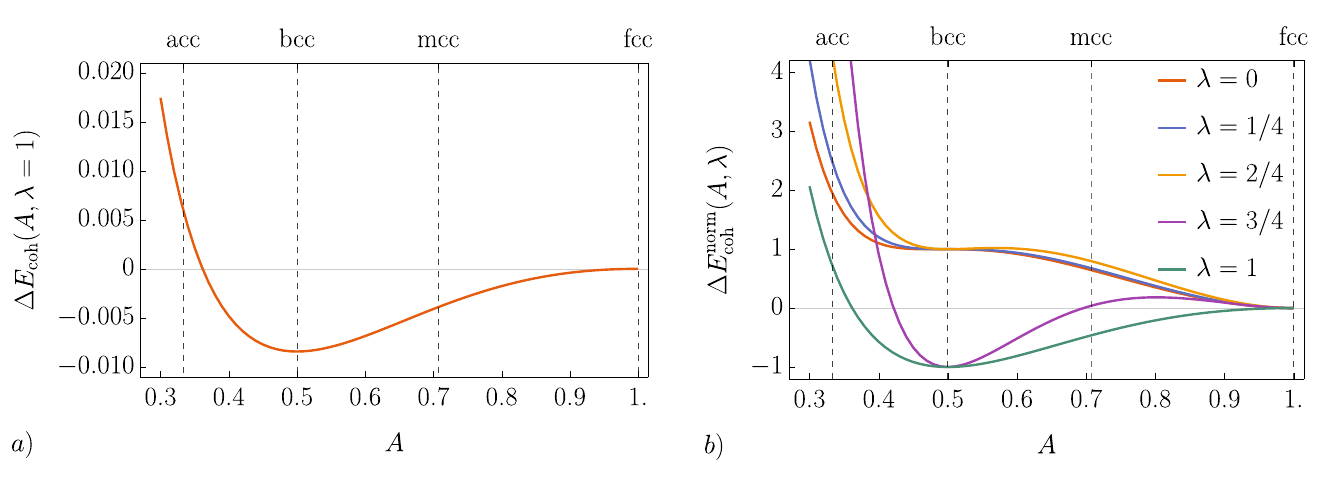} 
  \caption{
  (a) Cohesive energy differences, $\Delta E_\text{coh}(A,\lambda)=  E_\text{coh}(A,\lambda)- E_\text{coh}(A=1,\lambda)$ at the corresponding $R_\text{min}(A,\lambda)$ values for $\lambda=1$ as a function of $A$ for the $(8,4)$-LJ potential. (b)
The normalized cohesive energy differences $\Delta E_\text{coh}^{\rm norm}(A,\lambda)=\Delta E_\text{coh}(A,\lambda)/|\Delta E_\text{coh}(A=1/2,\lambda)|$
for different values of $0\le\lambda\le1$.
Numerically, we observe that bcc becomes energetically favorable compared to fcc for $\lambda>\lambda_c= 0.68740381212384$.
  }
  \label{fig:EcohLJ84}
  \end{figure}

\begin{figure}[htb!]
  \begin{center}
\includegraphics[scale=0.6]{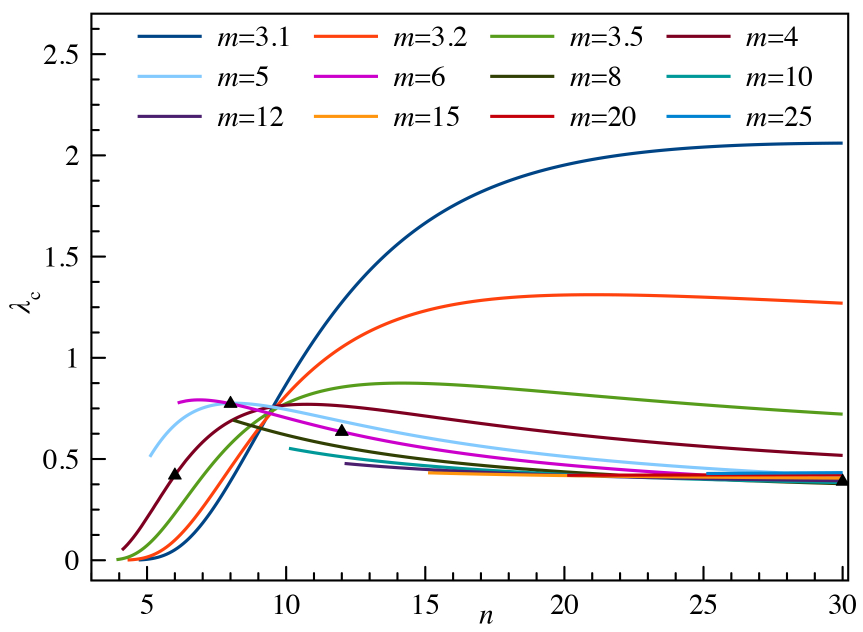} 
  \caption{
  Critical coupling strength $\lambda_c$ for different $(n,m)$-LJ combinations $n>m$. The values for the specific LJ potentials given in Table \ref{tab:ATM} are shown as black triangles. 
  }
  \label{fig:critambda}
  \end{center}
  \end{figure}

Figure \ref{fig:critambda} shows curves of the critical coupling parameters $\lambda_c$ for fixed $m$ and variable $n$ for $(n,m)$-LJ potentials including the ATM potential. The $\lambda_c$ values given in Table \ref{tab:ATM} are indicated as well. For very small values of both exponents $(n,m)$ we see that $\lambda_c$ is zero. At the other end, in the kissing hard-sphere (KHS) limit ($n,m\rightarrow \infty, n>m$) for a LJ potential the cohesive energy is given by $E_{\text{coh}}^{(2)}=N_\text{kiss}/2$, where $N_\text{kiss}$ is the kissing number. Adding the three body term we get the condition for the critical coupling strength considering that $R =1$,
\begin{equation}
-6 + \lambda_c E_{\text{coh}}^{(3)}(A=1)  = -4 +  \lambda_c E_{\text{coh}}^{(3)}(A=\tfrac{1}{2})
\end{equation}
which gives $\lambda_c=0.45430739956758$. This explains the asymptotic behaviour of the curves shown in Figure \ref{fig:critambda} for large $m$-values. However, this coupling parameter results in a purely repulsive force for the KHS limit for all cuboidal structures. The main message of this analysis here is that soft two-body interactions and strong repulsive many-body forces favor the bcc over the fcc phase. A prime example for this is lithium where the two phases are almost energetically degenerate.\cite{Ackland2017} This is also seen in Fig.~\ref{fig:EcohLJ84}, where for a  soft $(8,4)$-LJ potential, bcc becomes energetically favorable compared to fcc for $\lambda>\lambda_c=0.68740381212384$, with bcc forming a local minimum of the cohesive energy along the Bain path.
 
  \begin{figure}[htb!]
  \begin{center}
  a)\includegraphics[scale=0.25]{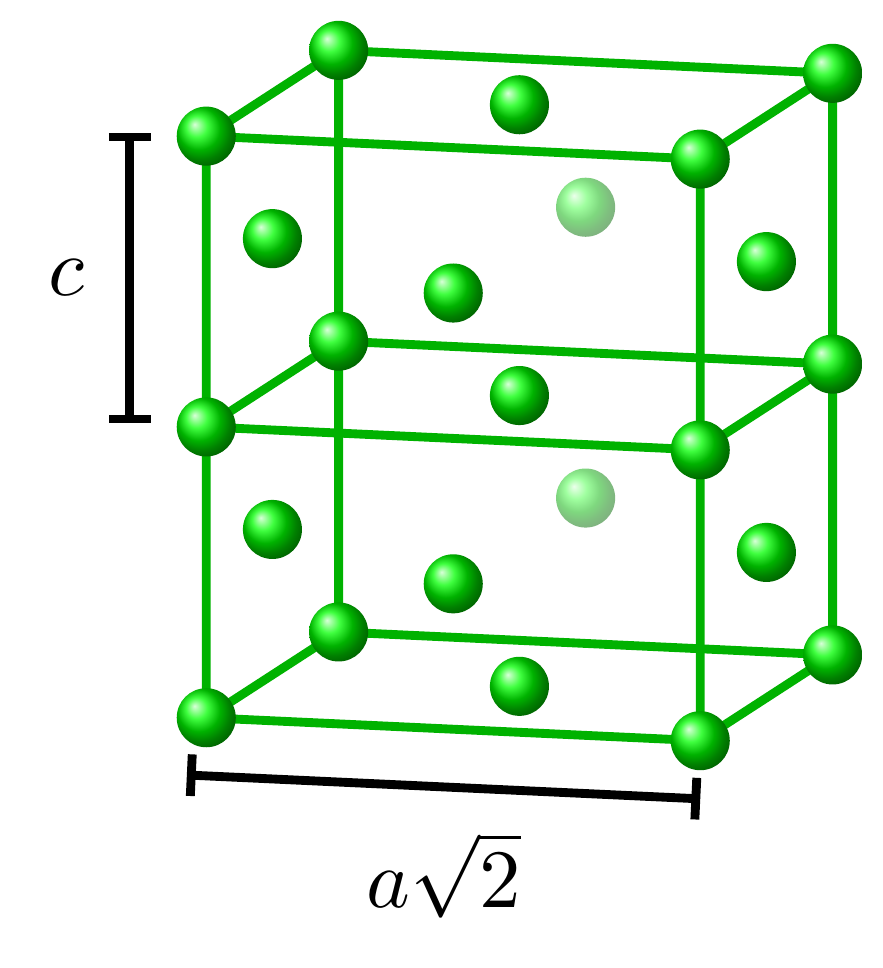}
  b)\includegraphics[scale=0.045]{Cuboidal2} 
  \caption{a) Conventional cell of the acc structure at $A=1/3$ showing the linear chain formation parallel to the $c$-axis, and b) weakly interacting linear chains obtained from a cuboidal structure with $A=0.005$.}
  \label{fig:LinearChainCuboidal}
  \end{center}
  \end{figure}
  
\begin{table}[htb!]
\setlength{\tabcolsep}{0.3cm}
\caption{\label{tab:ATM} Minimum distances and LJ and ATM contributions to the cohesive energy at critical $\lambda_\text{c}$ where $E_\text{coh}(A=\frac{1}{2},\lambda_\text{c})=E_\text{coh}(A=1,\lambda_\text{c})$.}
\begin{center}
\begin{tabular}{|l|r|r|r|r|}
\hline
                                                        & (6,4)     & (8,6)   & (12,6)  & (30,6) \\
\hline 
$\lambda_\text{c}$  									& 0.4193457686 	& 0.7733575166  & 0.6347968492 	& 0.3886346322\\
$R_\text{min}(A\!=\!\frac{1}{2})$ 			            & 0.9852364722	& 1.1958708129  & 1.1721405159	& 1.0793981795\\
$R_\text{min}(A\!=\!1)$ 					            & 1.0137042601 	& 1.2295453143  & 1.2038745402	& 1.1033397959\\
\hline 
$E_\text{coh}^{(2)}(A\!=\!\frac{1}{2})$	                &-22.6422225486	& -4.6655349480 &-4.0473152968	&-4.7402824565\\
$\lambda_\text{c}E_\text{R}^{(3)}(A\!=\!\frac{1}{2})$	& 11.6852182056 &  3.7683462411	& 3.7046445690	& 4.7625029058\\
$\lambda_\text{c}E_\text{A}^{(3)}(A\!=\!\frac{1}{2})$	& -4.5992266773	& -1.4831968267 &-1.4581242585	&-1.8744904913\\
$E_\text{coh}(A\!=\!\frac{1}{2})$                       &-15.5562310202	& -2.3803855336 &-1.8007949863 	&-1.8522700420\\
\hline 
$E_\text{coh}^{(2)}(A\!=\!1)$ 			                &-22.6730462463	& -4.6902916517 &-4.0933436461	&-4.9288827060\\
$\lambda_\text{c}E_\text{R}^{(3)}(A\!=\!1)$			    & 11.6816642186 &  3.7915200539	& 3.7630292199 	& 5.0500055052\\
$\lambda_\text{c}E_\text{A}^{(3)}(A\!=\!1)$			    & -4.5648491451 & -1.4816139852 &-1.4704806092	&-1.9733929072\\
$E_\text{coh}(A\!=\!1)$                                 &-15.5562310202 & -2.3803855336 &-1.8007949863  &-1.8522700420\\
\hline 
$\partial E_\text{coh}(A\!=\!\frac{1}{2})/\partial A~[10^{-5}]$&-3.128265&-1.008829&-0.991775 &-1.275000\\
$\partial^2 E_\text{coh}(A\!=\!\frac{1}{2})/\partial A^2$	   & 1.319068&-0.194625&-0.423196 &-0.811519\\
\hline
\end{tabular}
\end{center}
\end{table}
  
The first derivatives are $\partial E_\text{coh}(A\!=\!\frac{1}{2},\lambda)/\partial A=0$ at any $\lambda$ value and a proof that $\partial f_\mathrm{coh}^{(3)}(A)/\partial A=0$ at $A=\frac{1}{2}$ is given in appendix \ref{ATMLSdata}. This implies that the bcc structure remains an extremum if the ATM term is added. The rather small second derivatives compared to the corresponding values at $\lambda=0$ show the flatness of the cohesive energy curves $E_\text{coh}(A\!=\!\frac{1}{2},\lambda_\text{c})$ clearly seen in Figure \ref{fig:Ecohlambda1}. However, the bcc structure at $\lambda_c$ still remains a minimum for the $(6,4)$-LJ potential, and a maximum for the other three potentials considered. At even higher values, $\lambda\gg \lambda_\text{c})$ the lattice distorts to much lower $A$ values. While for the $(6,4)$-LJ potential we can still locate a very shallow minimum at $\lambda$ values up to the maximum value considered, for the other potentials we change to a monotonically decreasing function to smaller $A$-values, that is the three-body force destabilizes both fcc and bcc. 
  
Kwaadgras et al. discussed in detail the formation of linear chains for finite systems within the induced dipole interaction model,\cite{Kwaadgras2013} and commented on the importance of the ATM potential. Figure \ref{fig:LinearChainCuboidal} depicts the behavior at small $A$-values if the coupling strength $\lambda$ becomes large. We see a formation of linear chains along the $c$-axis. The kissing number is reduced to 2 as already shown in Eq. \eqref{eq:kiss2}. This is easily explained through Eq.~\eqref{eq:relation}: For $A\rightarrow 0$ we have $\gamma\rightarrow 0$ and $R \rightarrow \frac{a}{\sqrt{2}}\rightarrow\infty$, which implies that for increasing coupling strength $\lambda$ keeping $c$ finite we see the formation of largely separated linear chains where the ATM force becomes attractive as explained in section \ref{sec:Linear Chain}. Hence we see exactly the same (unphysical) situation as for the 2D lattice where we allowed for distortion in one direction.

We add some final comments here. First, very large $\lambda$ values are not realistic as shown above. Second, and more importantly, the ATM potential is only valid in the long range. In the very short range, the three-body force becomes even attractive for the rare gas elements.\cite{Ermakova1998,Freiman2077b,SchwerdtfegerHermann2009} Third, the many-body expansion of the total energy of a lattice described by quantum theory does not converge fast at short distances. This is especially the case for metallic systems as already mentioned.\cite{N-body.AH2007} Fourth, if we maintain the use of such a model system we need to make sure that the exponent of the repulsive force in the LJ potential exceeds the one in the ATM potential, otherwise we observe a collapse of the linear chain for $n<9$ as outlined in section \ref{sec:Linear Chain}.

\section{Conclusion}
\label{sec:conclusion}

In this work, we have explored the influence of three-body interactions on the stability of cuboidal lattices. To this end, we have studied the cohesive energy along a Bain phase transformation path connecting the fcc lattice structure to the mcc, bcc, and finally the acc lattice, where we have included both a two-body $(n,m)$-LJ potential and a three-body ATM potential of increasing coupling strength. The two-body lattice sums were computed to full precision using either rapidly converging Bessel function expansions \cite{borwein-2013,burrows-2020} or, alternatively, efficient evaluations based on the Epstein zeta function.\cite{buchheit2024epstein}
The challenging computation of the high-dimensional, slowly converging three-body lattice sums has been successfully achieved using a new representation based on singular integrals involving products of Epstein zeta functions. This approach enables, for the first time, the precise evaluation of three-body lattice sums within minutes on a standard laptop.

Using our advanced numerical framework, we have been able to precisely evaluate the small energy differences between the cuboidal structures along the Bain path. Our results demonstrate that the three-body potential can destabilize the fcc structure for large ATM coupling strengths. We analytically show and numerically confirm that the ATM potential exhibits a minimum along the Bain path at the bcc structure, resulting in the bcc structure becoming a metastable minimum for soft LJ potentials. For hard-wall LJ potentials, the structure distorts towards, and even beyond, the acc phase. Linear chain formation is observed at high ATM coupling strength, which is due to the short-range behavior of the ATM force.

The results indicate that the existence of the bcc lattice structure heavily relies on higher than two-body effects and on the softness of the two-body potential. While this study serves as an initial exploration of the martensitic bcc-to-fcc phase transition mechanism, more realistic systems, such as metals, need to be investigated in future work. This requires a more precise computation of many-body potentials based on density functional theory, as well as the incorporation of temperature and pressure effects.

\subsection*{Acknowledgment}
PS is grateful to Prof. Paul Indelicato (Kastler lab, Paris) for a visiting professorship at Sorbonne.

%\newpage
\appendix
\numberwithin{equation}{section}

\section*{Appendix}
\appendix
\numberwithin{equation}{section}

In this appendix we give a detailed description of the mathematical tools to derive the lattice sums and properties for the cuboidal lattices studied here. We start with defining the required standard functions and theta series used in the theory of lattice sums here. We then introduce the quadratic forms, integral transforms and expansions in terms of Bessel functions. This is followed by a discussion of some important lattice sum properties and their analytic continuation.

%Formulas for special functions
\section{Formulas for special functions}
\label{sec:specialfunctions}
A few special functions have been used in this work. For clarity and ease of use, they are stated here along with references.

The \textit{gamma function} may be defined for $s>0$ by
\begin{equation}
\label{gamma1}
\Gamma(s) = \int_{[0,\infty)}  t^{s-1}\, e^{-t}\,\ud t.
\end{equation}
By the change of variable $t=wx$ this can be rewritten in the form (see~[\onlinecite[(1.1.18)]{andrews1999}]),
\begin{equation}
\label{gamma2}
\frac{1}{w^s} = \frac{1}{\Gamma(s)} \int_{[0,\infty)}  x^{s-1}\,e^{-wx}\,\ud x.
\end{equation}
The following integral may be evaluated in terms of the \textit{modified Bessel function}:
\begin{equation}
\label{bessel1}
\int_{[0,\infty)}  x^{s-1}e^{-ax-b/x} \ud x = 2\left(\frac{b}{a}\right)^{s/2} K_s(2\sqrt{ab}).
\end{equation}
By the change of variable $x=u^{-1}$ it can be shown that
\begin{equation}
\label{bessel1.5}
K_{s}(z) = K_{-s}(z).
\end{equation}
When $s=1/2$ the modified Bessel function reduces to an elementary function
\begin{equation}
\label{bessel2}
K_{1/2}(z) = \sqrt{\frac{\pi}{2z}}\,e^{-z}.
\end{equation}
The asymptotic formula holds
\begin{equation}
\label{bessel3}
K_s(z) \sim \sqrt{\frac{\pi}{2z}} \, e^{-z} \quad \text{as} \quad z\rightarrow \infty, \quad (\,|\text{arg} \,z| < 3\pi/2).
\end{equation}
For all of these properties, see~[\onlinecite[pp. 223, 237]{andrews1999}] or~[\onlinecite[pp. 233--248]{temme1996}].

%and
%\begin{equation}
%\label{char2}
%\chi_{-3}(n) = \frac{\sin(2\pi n/3)}{\sin(2\pi/3)} = \begin{cases}
%1 & \text{if $n \equiv 1 \pmod 3$,} \\
%-1 & \text{if $n \equiv 2 \pmod 3$,} \\
%0 & \text{otherwise.}
%\end{cases}
%\end{equation}

The transformation formula for \textit{theta functions} is~[\onlinecite[p. 119]{andrews1999}], [\onlinecite[(2.2.5)]{borwein1991cubic}]:
\begin{equation}
\label{theta1}
\sum_{n\in\mathbb{Z}} e^{-\pi n^2t+2\pi ina} = \frac{1}{\sqrt{t}} \sum_{n\in\mathbb{Z}} e^{-\pi(n+a)^2/t}, \quad \text{assuming Re$(t)>0$.}
\end{equation}
We will need the special cases $a=0$ and $a=1/2$, which are
\begin{equation}
\label{theta1a}
\sum_{n\in\mathbb{Z}}  e^{-\pi n^2t} = \frac{1}{\sqrt{t}} \sum_{n\in\mathbb{Z}}  e^{-\pi n^2/t}
\end{equation}
and
\begin{equation}
\label{theta1b}
\sum_{n\in\mathbb{Z}}  (-1)^ne^{-\pi n^2t} = \frac{1}{\sqrt{t}} \sum_{n\in\mathbb{Z}}  e^{-\pi(n+\frac12)^2/t},
\end{equation}
respectively.
The sum of two squares formula is~[\onlinecite[(3.111)]{Cooper-2017}]
\begin{equation}
\label{theta2}
\left(\sum_{j\in\mathbb{Z}}  q^{j^2}\right)^2 = \sum_{j,k\in\mathbb{Z}}   q^{j^2+k^2} = \sum_{N\in\mathbb{N}_0}  r_2(N) q^N
\end{equation}
where
\begin{equation}
\label{theta3}
r_2(N) = \#\left\{j^2+k^2=N\right\} = 
\begin{cases} 1 & \text{if $N=0$}, \\ \\
\displaystyle{4\sum_{d|N} \chi_{-4}(d)} &\text{if $N\geq 1$,}
\end{cases}
\end{equation}
the sum being over the positive divisors $d$ of $N$. For example,
\begin{align*}
r_2(18) &= 4\left(\chi_{-4}(1)+\chi_{-4}(2)+\chi_{-4}(3)+\chi_{-4}(6)+\chi_{-4}(9)+\chi_{-4}(18)\right) \\
&= 4 \left(1+0-1+0+1+0\right) = 4.
\end{align*}
By~[\onlinecite[(3.15) and (3.111)]{Cooper-2017}] we also have
\begin{equation}
\label{theta4}
\left(\sum_{j\in\mathbb{Z}}  q^{(j+\frac12)^2}\right)^2 = \sum_{N\in\mathbb{N}_0} r_2(4N+1)q^{(4N+1)/2}.
\end{equation}

%The \textit{cubic theta function} are analogues of the transformation formula ~[\onlinecite[(2.2)]{borwein1991cubic}], [\onlinecite[Cor. 5.19]{cooper2003cubic}]
%\begin{equation}
%\label{theta5}
%\sum_{j,k\in\mathbb{Z}} e^{-2\pi(j^2+jk+k^2)t} = \frac{1}{t\sqrt{3}} \sum_{j,k\in\mathbb{Z}} e^{-2\pi(j^2+jk+k^2)/3t} 
%\end{equation}
%and
%\begin{equation}
%\label{theta5a}
%\sum_{j,k\in\mathbb{Z}} e^{-2\pi((j+\frac13)^2+(j+\frac13)(k+\frac13)+(k+\frac13)^2)t} 
%= \frac{1}{t\sqrt{3}} \sum_{j,k\in\mathbb{Z}} \omega^{j-k}e^{-2\pi(j^2+jk+k^2)/3t} 
%\end{equation}
%where $\omega=\exp(2\pi i/3)$ is a primitive cube root of unity.
%The analogue of the sum of two squares result is~[\onlinecite[(3.124)]{Cooper-2017}]
%\begin{equation}
%\label{theta2a}
%\sum_{j,k\in\mathbb{Z}} q^{j^2+jk+k^2} = \sum_{N\in\mathbb{N}_0} u_2(N) q^N
%\end{equation}
%where
%\begin{equation}
%\label{theta3a}
%u_2(N) = \#\left\{j^2+jk+k^2=N\right\} = 
%\begin{cases} 1 & \text{if $N=0$}, \\ \\
%\displaystyle{6\sum_{d|N} \chi_{-3}(d)} &\text{if $N\geq 1$,}
%\end{cases}
%\end{equation}
%where the sum is again over the positive divisors $d$ of $N$. By~[\onlinecite[(3.18) and (3.124)]{Cooper-2017}] we also have
%\begin{equation}
%\label{theta2b}
%\sum_{j,k\in\mathbb{Z}} q^{(j+\frac13)^2+(j+\frac13)(k+\frac13)+(k+\frac13)^2} = \frac12 \sum_{N\in\mathbb{N}_0}  u_2(3N+1) q^{N+\frac13}
%\end{equation}
%which is the analogue of~\eqref{theta4}.

%The Riemann zeta function and $L$ functions
The \textit{Riemann zeta function} $\zeta(s)$ and \textit{Dirichlet} $L$ \textit{function} are defined by
\begin{align}
\zeta(s) &= \sum_{j\in\mathbb{N}} \frac{1}{j^s} \label{zeta1} \\
L_{-4}(s) &= \sum_{j\in\mathbb{N}} \frac{\chi_{-4}(j)}{j^s} = 1-\frac{1}{3^s}+\frac{1}{5^s}-\frac{1}{7^s}+\cdots.   \label{zeta2}   
%L_{-3}(s) &= \sum_{j=1}^\infty \frac{\chi_{-3}(j)}{j^s}  = 1-\frac{1}{2^s}+\frac{1}{4^s}-\frac{1}{5^s}+\frac1{7^s}-\frac{1}{8^s}+\cdots. \label{zeta3}
\end{align}
For even integers the Riemann zeta function can be expressed as $\zeta(2n)=\pi^{2n}B_n/A_n$ where $A_n$ and $B_n$ are positive integers, e.g. we have $\zeta(2)=\pi^2/6$, $\zeta(4)=\pi^4/90$, $\zeta(6)=\pi^6/945$, $\zeta(8)=\pi^8/9450$,  $\zeta(10)=\pi^{10}/93555$, $\zeta(12)=691\pi^{12}/638512875$, $\zeta(14)=2\pi^{14}/18243225$ and $\zeta(16)=3617\pi^{16}/325641566250$. The coefficients $A_n$ and $B_n$ are listed in the On-Line Encyclopedia of Integer Sequences A002432 and A046988, respectively.\cite{sloane-1995}

For an integer $n$, the \textit{Dirichlet character} $\chi_{-4}(n)$ 
%and $\chi_{-3}(n)$ 
is defined by
\begin{equation}
\label{char1}
\chi_{-4}(n) = \sin(\pi n/2) = \begin{cases}
1 & \text{if $n \equiv 1 \pmod 4$,} \\
-1 & \text{if $n \equiv 3 \pmod 4$,} \\
0 & \text{otherwise}.
\end{cases}
\end{equation}
The Riemann zeta function has a pole of order~$1$ at $s=1$, and in fact
\begin{equation}
\label{zetapole}
\lim_{s\rightarrow 1} (s-1)\zeta(s) = 1.
\end{equation}
This is a consequence of~[\onlinecite[(1.3.2)]{andrews1999}]. See also~[\onlinecite[p. 58]{temme1996}].

We require the following functional equations
\begin{equation}
\label{fe1}
\pi^{-s/2} \Gamma (s/2)\zeta(s) = \pi^{-(1-s)/2} \Gamma((1-s)/2)\zeta(1-s)
\end{equation}
and
\begin{equation}
\label{fe2}
\pi^{-s} \Gamma\left(s\right)\zeta(s)L_{-4}(s) = \pi^{-(1-s)} \Gamma\left(1-s\right)\zeta(1-s)L_{-4}(1-s) 
\end{equation}
and the special values
\begin{equation}
\label{zetavalues}
\zeta(2) = \frac{\pi^2}{6}, \quad \zeta(0)= -\frac12,\quad \zeta(-1) = -\frac{1}{12}, \quad \zeta(-2)=\zeta(-4)=\zeta(-6)=\cdots = 0,
\end{equation}
\begin{equation}
\label{L4values}
L_{-4}(1) = \frac{\pi}{4},  \quad L_{-4}(0) = \frac12, \quad L_{-4}(-1) = L_{-4}(-3) = L_{-4}(-5) = \cdots = 0,
\end{equation}
%and
%\begin{equation}
%\label{L3values}
%L_{-3}(1) = \frac{\pi \sqrt{3}}{9}, \quad L_{-3}(0) =\frac13, \quad L_{-3}(-1) = L_{-3}(-3) = L_{-3}(-5) = \cdots = 0.
%\end{equation}
See~[\onlinecite[Ch. 12]{apostol1998}] or~[\onlinecite{Zucker-1975a}].
Other equalities used are
\begin{equation}
\label{zeta4a}
\sum_{j\in\mathbb{N}_0} \frac{1}{(j+\frac12)^s} = (2^s-1)\zeta(s)
\end{equation}
\begin{equation}
\label{zeta4}
\sum_{j\in\mathbb{N}} \frac{(-1)^j}{j^s} = -(1-2^{1-s})\zeta(s)
\end{equation}
\begin{equation}
\label{zeta5}
{\sum_{j,k\in\mathbb{Z}}}^{\prime} \;\;\; \frac{1}{(j^2+k^2)^s} = 4\zeta(s)L_{-4}(s)
\end{equation}
\begin{equation}
\label{zeta6}
{\sum_{j,k\in\mathbb{Z}}}^{\prime} \;\;\; \frac{(-1)^{j+k}}{(j^2+k^2)^s}= -4(1-2^{1-s})\zeta(s)L_{-4}(s).
\end{equation}
%\begin{equation}
%\label{zeta7}
%{\sum_{i,j}}^{\;\prime} \frac{1}{(i^2+ij+j^2)^{s}}
%= 6\zeta(s) L_{-3}(s)
%\end{equation}
%\begin{equation}
%\label{zeta8}
%{\sum_{i,j}} \frac{1}{((i+\frac13)^2+(i+\frac13)(j+\frac13)+(j+\frac13)^2)^{s}}
%= 3(3^s-1)\zeta(s) L_{-3}(s).
%\end{equation}
The identities~\eqref{zeta4a} and~\eqref{zeta4} follow from the definition of~$\zeta(s)$ by series rearrangements.
For~\eqref{zeta5} and \eqref{zeta6} 
%and \eqref{zeta7}, 
see (1.4.14) and (1.7.5) 
%and (1.4.16), 
respectively, of~[\onlinecite{borwein-2013}].
%The identity~\eqref{zeta8} can be obtained by the method of Mellin transforms, e.g., see~[\onlinecite[Appendix~A]{burrows-2020}], starting with~[\onlinecite[(3.36)]{Cooper-2017}].

Given a positive definite quadratic form $g(i,j,k)$, the corresponding \textit{theta series} is defined for $|q|<1$ by
\begin{equation}
\theta_g(q) = \sum_{i,j,k\in\mathbb{Z}} q^{g(i,j,k)}.
\end{equation}
For the quadratic form in~\eqref{gdef} the theta series is
\begin{equation}
\theta(A;q) = \sum_{i,j,k\in\mathbb{Z}}  q^{(A(i+j)^2+(j+k)^2+(i+k)^2)/(A+1)} \quad \text{ where $1/3\leq A \leq 1$.}
\end{equation}
The first few terms in the theta series for fcc, mcc, bcc and acc as far as~$q^9$ are given respectively by
\begin{align*}
\theta(1;q) &= 1+12q+6q^2+24q^3+12q^4+24q^5+8q^6+48q^7+6q^8+36q^9+\cdots, \\
\theta\left(\tfrac{1}{\sqrt{2}};q\right) &= 1+8q+4q^{4-2\sqrt2} + 2 q^{4 \sqrt2-4} +4 q^{8-4 \sqrt2}+8 q^{2 \sqrt2} + 16q^{-4\sqrt2+9} \\
&\quad +8q^4+8q^{8\sqrt2-7}+4 q^{16-8\sqrt2}+8 q^{-8\sqrt2+17} + 8q^{20-10\sqrt2} + 8q^{-4\sqrt2+12} \\
&\quad+ 2q^{16\sqrt2-16} + 16q^{4\sqrt2+1} + 16q^{-6\sqrt2+16} + 8q^{14\sqrt2-12}+16q^{-12\sqrt2+25} \\
&\quad+8q^{-8+12\sqrt2} + 8q^9 +\cdots, \\
\theta\left(\tfrac12;q\right) &= 1+8q+6q^{4/3}+12q^{8/3}+8q^4+24q^{11/3}+6q^{16/3}+24q^{19/3}+24q^{20/3} \\
&\quad +24q^8+32q^9+\cdots, \\
\theta\left(\tfrac13;q\right) &= 1 + 10q + 4q^{3/2}+8q^{5/2}+12q^3+26q^4+8q^{11/2}+20q^6+32q^7 \\
&\quad +8q^{15/2}+16q^{17/2}+10q^9+\cdots.
\end{align*}
Since the quadratic form $g(A;i,j,k)$ has been normalised to make the minimum distance~$1$, the kissing number occurs in each theta series as the coefficient of~$q$. That is, we have $\textrm{kiss(fcc)}=12$, $\textrm{kiss(mcc)}=8$, $\textrm{kiss(bcc)}=8$ and $\textrm{kiss(acc)}=10$.

Finally, we mention the $d$-dimensional Epstein zeta function\cite{Epstein-1903,crandall1998} in its most general form for a matrix $A$, vectors $\vec{c}$ and $\vec{v}$, and exponent $\rho\in\mathbb{C}$,
\begin{equation}
Z_d(A,\rho)={\sum_{\vec{k}\in\mathbb{Z}^d}}^{'}\frac{e^{2\pi i \vec{c}\cdot A\vec{k}}}{|A\vec{k}-\vec{v}|^\rho}.
\end{equation}
The connection to the generator matrix $B$ in lattices is by setting $A=B^\top$, and the Gram matrix becomes $G=BB^\top=A^\top A$. The vector $\vec{v}$ is often called the shift vector in lattice theory.

\section{Connection to alternative Gram matrices in the literature}

In this subsection, we discuss a few important properties of the generator and Gram matrices used in this work.\cite{conway2013sphere} Two generator matrices $B_1$ and $B_2$ are equivalent if $B_2=cUB_1\mathcal{O}$, $c$ is a non-zero real number, $\mathcal{O}$ a real orthogonal matrix ($\mathcal{O}\mathcal{O}^\top=1$) with $\textrm{det}\mathcal{O}=\pm 1$ describing rotation, reflection or roto-reflection of the lattice, and $U$ a matrix containing integers with $\textrm{det}U=1$ describing for example permutations of the basis vectors. Given two equivalent generator matrices $B_1$ and $B_2$, the corresponding (equivalent) Gram matrices 
$G_1=B_1 B_1^\top$ and $G_2=B_2B_2^\top$
are related by
\begin{equation}
G_2 = B_2 B_2^\top = cUB_1\mathcal{O} \left(cUB_1\mathcal{O}\right)^\top = c^2UB_1\mathcal{O}\mathcal{O}^\top B_1^\top U^\top 
= c^2U G_1 U^\top.
\end{equation}

We now reconcile the Gram matrix $G$ in~\eqref{eq:lattice_and_gram} with two 
matrices given by Conway and Sloane.\cite{Conway1994} Let
\begin{equation}
U_1 = \begin{pmatrix}
1 & 0 & 0 \\
-1 & 0 & 1 \\
0 & -1 & 0 
\end{pmatrix}
\quad\text{and} \quad
U_2 = \begin{pmatrix}
1 & 1 & -1 \\
1 & 0 & 0 \\
0 & 1 & 0 
\end{pmatrix}
\end{equation}
and consider the equivalent matrices $G_1$ and $G_2$ defined by
\begin{equation}
\label{GramG1}
G_1 = U_1\,G\,U_1^\top =  \begin{pmatrix}
u^2+v^2 & -u^2 & -u^2 \\
-u^2 & u^2+v^2 & u^2-v^2 \\
-u^2 & u^2-v^2 & u^2+v^2 
\end{pmatrix}
\end{equation}
and
\begin{equation}
\label{GramG2}
G_2 = U_2\,G\,U_2^\top =  \begin{pmatrix}
4u^2 & 2u^2 & 2u^2 \\
2u^2 & u^2+v^2 & u^2 \\
2u^2 & u^2 & u^2+v^2 
\end{pmatrix}.
\end{equation}
When $u=1/\sqrt{2}$ and $v=1/\sqrt[4]{2}$, the matrix $G_1$ in~\eqref{GramG1} is the Gram matrix
for the mcc lattice given by Conway and Sloane~[\onlinecite[(10)]{Conway1994}].
Moreover, when $u=\sqrt{1/3}$ and $v=\sqrt{2/3}$, the matrix $G_1$ leads to another known
quadratic form for the bcc lattice, e.g., see~[\onlinecite[(8b)]{burrows-2020}].
When $u=1$, $v=\sqrt{3}$, the matrix $G_2$ in~\eqref{GramG2} is the Gram matrix for the acc lattice given
in~[\onlinecite[p. 378]{Conway1994}].
Since $\det U_1^2 = \det U_2^2 = 1$ it follows that
\begin{equation}
\det G_1 = \det G_2 = \det G = (\det B)^2 = 4u^2v^4 = 4v^6A.
\end{equation}
The corresponding quadratic forms $g_1$ and $g_2$ are defined by
\begin{align*}
g_1(i,j,k) &= (i,j,k) \, G_1 (i,j,k)^\top \\
&= (u^2+v^2)i^2 + (u^2+v^2)j^2+(u^2+v^2)k^2 - 2u^2ij - 2u^2ik+2(u^2-v^2)jk 
\intertext{and}
g_2(i,j,k) &= (i,j,k) \, G_2 (i,j,k)^\top \\
&= 4u^2i^2 + (u^2+v^2)j^2+(u^2+v^2)k^2 + 4u^2ij + 4u^2ik+2u^2jk.
\end{align*}
They are related to the quadratic form $g$ in Eq.~\eqref{eq:nearestneighbor} by
\begin{equation}
\label{g1qf}
g_1(i,j,k) = g\left( (i,j,k)U_1 \right) = g(i-j,-k,j) 
\end{equation}
and
\begin{equation}
g_2(i,j,k) = g\left( (i,j,k)U_2 \right) = g(i+j,i+k,-i).
\end{equation}
These quadratic forms are an essential ingredient for the lattice sums $L(G)$ in \eqref{eq:latticesum} used to obtain the cohesive energy of a lattice for the case of inverse power potentials $V(r)=r^{-n}$. This is outlined in the following sections.

\section{Direct summation approach to three-body lattice sums}
\label{appendix:direct_sum}

Unfortunately, the three-body sum is slowly convergent and cannot be analytically expressed in terms of lattice sums containing a single quadratic form as for the Epstein zeta function.\cite{Epstein-1903} It has therefore been treated in the past by  direct summation methods. Along the Bain path, we can express the ATM potential in terms of lattice sums dependent on the parameter $A$ similar to the two-body potential such that
\begin{equation}\label{eq:cohLJATM1}
E_\text{coh}^{(3)}(R ,A,\lambda) =
\lambda \left\{ E_R(R ,A) + E_A(R ,A) \right\} 
\end{equation}
with $E_R=f_r/R^9$ and $E_A=f_a/R^9$.

In our previous work\cite{Schwerdtfeger-2016,Smits2018,Jerabek2019,Smits2020,schwerdtfeger2020,edison2022}, where more complicated forms of three-body forces were used, we produced the cartesian coordinates of the vectors $\vec{R}_{0i}$ for a specific lattice first and stored them for further use in \eqref{eq:nbody}. For this, one could use the fcc lattice vectors $(1,1,0)^\top$, $(1,0,0)^\top$, and $(0,1,1)^\top$, as a starting point and for the different cuboidal lattices scale the cartesian coordinates $(x_n,y_n,z_n)$  such that we have $\frac{R }{\sqrt{A+1}}(x_n\sqrt{A},y_n,z_n)$.\cite{SchwerdtfegerHermann2009,SchwerdtfegerAssadollahzadehHermann2010,Schwerdtfeger-2016} However, for the simple ATM potential this offers no advantage in terms of computer time and memory requirements. Moreover, as we shall see the simple form has some advantage for the determination of the minimum distance for a $\lambda$ dependent energy \eqref{eq:cohLJATM1} at a specific $A$ value. We therefore decided to use \eqref{eq:cohLJATM1} directly by utilizing the permutation symmetry $i_1\leftrightarrow j_1$ for the vectors  $\vec{i}$ and $\vec{j}$.

\begin{figure*}[htbp!]
  \begin{center}
  \includegraphics[scale=0.6]{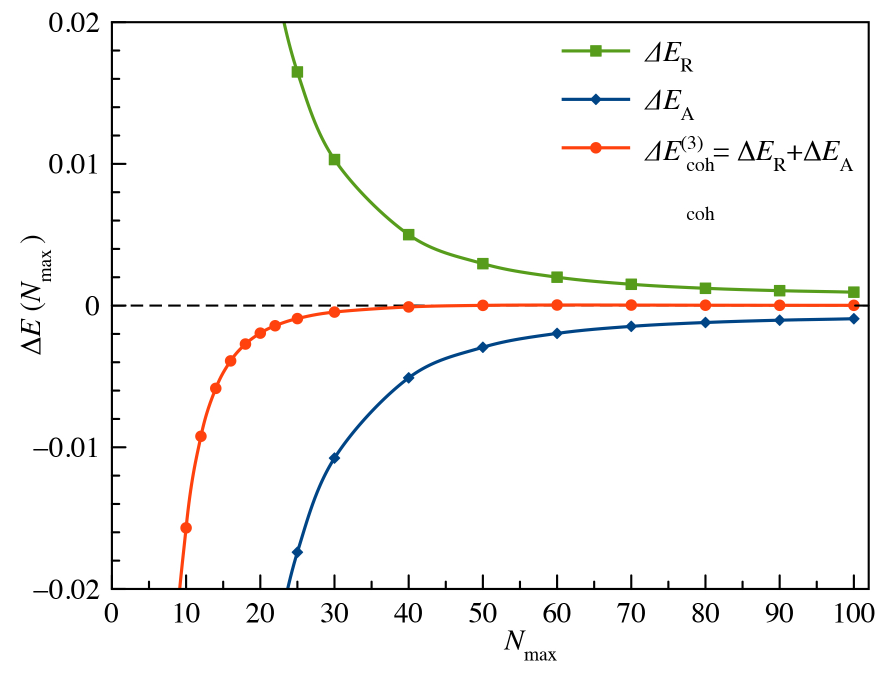}\\
  \caption{
  Convergence of the ATM terms $E_R(N_\text{max},R_\text{min},A)$, $E_A(N_\text{max},R_\text{min},A)$ and $E^{(3)}_\text{coh}(N_\text{max},R_\text{min},A)$ (Eq.\eqref{eq:cohLJATM1}) for the bcc lattice $A=\frac{1}{2}$ using a $(12,6)$-LJ potential. $R_\text{min}$ is set to 0.951864818662439, the minimum distance for the bcc lattice of a $(12,6)$-LJ potential. The values show the difference in energies $\Delta E = E(N_\text{max}\rightarrow\infty) - E(N_\text{max})$ to the extrapolated value $N_\text{max}\rightarrow\infty$. The limit for $N_\text{max}\rightarrow\infty$ was obtained from a linear extrapolation over $N_\text{max}^{-1}$ of the last two values at $N_\text{max}=90$ and 100. This gives $\Delta E_R(N_\text{max}=100)=-9.283159\times 10^{-4}$, $\Delta E_A(N_\text{max}=100)=9.461724\times 10^{-4}$ and $\Delta E_R(N_\text{max}=100)+E_A(N_\text{max}=100)=1.785654\times 10^{-5}$.}
  \label{fig:convergence}
  \end{center}
  \end{figure*}

The convergence for the two individual 3-body terms as well as the sum of both in Eq.~\eqref{eq:cohLJATM1} is shown in Figure \ref{fig:convergence} for the bcc lattice ($A=\frac{1}{2}$) setting $\lambda=1$ and $R_\text{min}$ to the minimum nearest-neighbor distance of a (12,6)-LJ potential. The rather slow convergence of both terms $E_R^{(3)}(R_\text{min},A)$ and $E_A^{(3)}(R_\text{min},A)$ with increasing $N_\text{max}$ is obvious. Nevertheless, the sum of these terms exhibits significantly faster convergence, and $N_\text{max} = 100$ proves in principle to be sufficiently accurate for our analysis. However, due to the computational time scaling of $\mathcal{O}(N_\text{max}^6)$, calculating the ATM term for a specific $A$ value already demands 4 weeks of CPU time on a single processor. We therefore use a far more efficient evaluation of the three-body term through the Epstein zeta function as introduced originally by Crandall and put into a computer efficient form by Buchheit and co-workers.\cite{buchheit2024} This allows to evaluate general three-body lattice sums to machine precision within minutes on a standard laptop.

For our detailed analysis, we tabulate the following properties along the Bain path: $E_\text{coh}(A,R_\text{min})$, $\partial E_\text{coh}(A)/ \partial R |_{R_\text{min}}$, $ \partial^2 E_\text{coh}(A)/ \partial R ^2|_{R_\text{min}}$, $ \partial E_\text{coh}(A)/ \partial A|_{R_\text{min}}$, $ \partial^2 E_\text{coh}(A)/ \partial A^2|_{R_\text{min}}$. The latter two derivatives are obtained analytically for the two-body force (see Ref.\citenum{burrows2021b} for details) and numerically for the three-body force. For the general LJ potential we used the $(n,m)$ combinations $(6,4)$, $(8,6)$, $(12,6)$ and $(30,6)$. The latter represents a hard-wall potential accompanied by a attractive long-range dispersive $r^{-6}$ term. 

\section{Evaluation of the two-body lattice sum for a rectangular 2D lattice}
\label{sec:2bodylatticesums}

For the lattice sum with Gram matrix \eqref{eq:Gramrect} ,
\begin{equation}\label{eq:2Dlatticesum}
	\begin{aligned}
		{\sum_{i\in\mathbb{Z}}}^{'}\left[i^{2} +(\gamma j)^2\right]^{-s} &= 2\zeta(2s)\left\{  1+\gamma^{-2s}   + 2\left( 1+\gamma^2 \right)^{-s} \right\} + 4 {\sum_{i,j\in\mathbb{N},i\ne j}}^{'}\left[i^{2} + (\gamma j)^2\right]^{-s},
	\end{aligned}
\end{equation}
we use Van der Hoff and Benson's original expression derived from a Mellin transformation and the use of theta functions,\cite{Hoff-Benson-1953}
\begin{equation}\label{eq:Hoff-Benson0}
	\begin{aligned}
		{\sum_{i\in\mathbb{Z}}}^{'}\left[x^{2} + (i+a)^{2}\right]^{-s} &= \frac{\sqrt{\pi}\Gamma(s-\frac{1}{2})}{\Gamma(s)\left|x\right|^{2s-1}} + \frac{4\pi^{s}}{\Gamma(s)}\sum_{n\in\mathbb{N}}\left(\frac{n}{\left|x\right|}\right)^{s-\frac{1}{2}}\cos(2\pi n a)K_{s-\frac{1}{2}}(2\pi n|x|)
	\end{aligned}
\end{equation}
with $a \in (0,1)$. Using $a=0$ and $x=\gamma j ~(\gamma>0)$ we get a fast converging series in terms of Bessel functions,
\begin{equation}\label{eq:Hoff-Benson1}
	\begin{aligned}
		{\sum_{i,j\in\mathbb{Z}}}^{'}\left[i^{2}+(\gamma j)^{2} \right]^{-s} &= 2\zeta(2s) + {\sum_{j\in\mathbb{Z}}}^{'} \frac{\sqrt{\pi}\Gamma(s-\frac{1}{2})}{\Gamma(s)\left|\gamma j\right|^{2s-1}} +  \frac{4\pi^{s}}{\Gamma(s)}{\sum_{j\in\mathbb{Z}}}^{'}\sum_{n\in\mathbb{N}}\left(\frac{n}{\left|\gamma j\right|}\right)^{s-\frac{1}{2}} K_{s-\frac{1}{2}}(2\pi n\gamma |j|)
	\end{aligned}
\end{equation}
which simplifies to
\begin{equation}\label{eq:Hoff-Benson2}
	\begin{aligned}
		{\sum_{i,j\in\mathbb{Z}}}^{'}\left[i^{2}+(\gamma j)^{2} \right]^{-s} &= 2\zeta(2s) + \frac{2\sqrt{\pi}\Gamma(s-\frac{1}{2})\zeta(2s-1)}{\Gamma(s)\gamma^{2s-1}} +  \frac{8\pi^{s}}{\Gamma(s)} \sum_{n,j\in\mathbb{N}}\left(\frac{n}{\gamma j}\right)^{s-\frac{1}{2}} K_{s-\frac{1}{2}}(2\pi \gamma n j).
	\end{aligned}
\end{equation}
The additional Riemman zeta function comes from the case when $(j=0,i\neq 0)$. As the Bessel function $K_n(x)$ decays exponentially with the argument $x$, for $\gamma<1$ it is computationally advantageous to rewrite the sum into
\begin{equation}\label{eq:Hoff-Benson3}
	\begin{aligned}
		{\sum_{i,j\in\mathbb{Z}}}^{'}\left[i^{2}+(\gamma j)^{2} \right]^{-s} = \gamma^{-2s} {\sum_{i,j\in\mathbb{Z}}}^{'}\left[\gamma^{-2}i^{2}+j^{2} \right]^{-s} = \gamma^{-2s}  {\sum_{i,j\in\mathbb{Z}}}^{'}\left[i^{2}+(j/\gamma)^{2} \right]^{-s}
	\end{aligned}
    \end{equation}
and we get in a similar fashion
\begin{equation}\label{eq:Hoff-Benson4}
	\begin{aligned}
		\gamma^{-2s}  {\sum_{i,j\in\mathbb{Z}}}^{'}\left[i^{2}+(j/\gamma)^{2} \right]^{-s} &= 2\gamma^{-2s}\zeta(2s) + \frac{2\sqrt{\pi}\Gamma(s-\frac{1}{2})\zeta(2s-1)}{\Gamma(s)\gamma} \\
        &+  \frac{8\pi^{s}}{\Gamma(s)\gamma^{s+\frac{1}{2}}} \sum_{n,j\in\mathbb{N}}\left(\frac{n}{j}\right)^{s-\frac{1}{2}} K_{s-\frac{1}{2}}(2\pi \gamma^{-1} n j)
	\end{aligned}
\end{equation}
This formula is identical to the one given by Bateman and Grosswald.\cite{bateman1964epstein} For computational efficiency, we rewrite Eq.\eqref{eq:Hoff-Benson2},
\begin{equation}\label{eq:Hoff-Benson5}
	\begin{aligned}
		{\sum_{i,j\in\mathbb{Z}}}^{'}\left[i^{2}+(\gamma j)^{2} \right]^{-s} &= \zeta(2s) + \frac{2\sqrt{\pi}\Gamma(s-\frac{1}{2})\zeta(2s-1)}{\Gamma(s)\gamma^{2s-1}} +  \frac{8\pi^{s}}{\Gamma(s)\gamma^{s-\frac{1}{2}}} \sum_{n\in\mathbb{N}} K_{s-\frac{1}{2}}(2\pi \gamma n^2)\\
        &+\frac{8\pi^{s}}{\Gamma(s)\gamma^{s-\frac{1}{2}}} \sum_{n<j\in\mathbb{N}}\left\{ \left(\frac{n}{j}\right)^{s-\frac{1}{2}} + \left(\frac{j}{n}\right)^{s-\frac{1}{2}}\right\} K_{s-\frac{1}{2}}(2\pi \gamma n j)
	\end{aligned}
\end{equation}
and we can do the same for Eq.\eqref{eq:Hoff-Benson4}.

%Evaluation of the sum $L(A;s)$
\section{Evaluation of the lattice sum $L(A;s)$}
\label{sec:latticesumsA}

Consider the quadratic form
\begin{equation}
    g(i,j,k) = (i,j,k) G (i,j,k)^\top,
\end{equation}
with the Gram matrix in Eq.~\eqref{eq:lattice_and_gram}.
To eliminate $v$, we divide the above equation by the squared nearest neighbor distance  $R^2$, yielding
\begin{align}\label{eq:g}
g(A;i,j,k) = \frac{g(i,j,k)}{R^2} 
= \begin{cases}
\displaystyle{\frac{1}{4A}\left(A(i+j)^2+(j+k)^2+(i+k)^2\right)} & \text{if $0<A<1/3$,} \\ \\
\displaystyle{\frac{1}{A+1}\left(A(i+j)^2+(j+k)^2+(i+k)^2\right)} & \text{if $1/3\leq A \leq 1$,} \\ \\
\displaystyle{\frac{1}{2}\left(A(i+j)^2+(j+k)^2+(i+k)^2\right)} & \text{if $A>1$.}
\end{cases} 
\end{align}

The cases which we are mainly interested in are fcc, mcc, bcc, acc, all of which satisfy \mbox{$1/3\leq A \leq 1$}, but we will go slightly beyond this limit mainly to discuss distortions towards the acc structure.
In the important range $1/3\leq A \leq 1$, we have
\begin{equation}
\label{gdef}
g(A;i,j,k) = \frac{1}{A+1}\left(A(i+j)^2+(j+k)^2+(i+k)^2\right),
\end{equation}
corresponding to the rescaled Gram matrix
\begin{equation}\label{eq:Gram}
G(A)= \frac{1}{A+1}\begin{pmatrix}
A+1 & A & 1 \\
A & A+1 & 1 \\
1 & 1 & 2
\end{pmatrix}
\end{equation}
which is used throughout this work.

The lattice sum for inverse power potentials in terms of the quadratic form $g(A;i,j,k)$ defined in \eqref{gdef} is then given by \cite{borwein-2013,burrows-2020}
\begin{equation}
\label{LAsdefinition}
L(A;s) = {\sum_{i,j,k\in\mathbb{Z}}}^{\prime} \left(\frac{1}{g(A;i,j,k)}\right)^s = {\sum_{i,j,k\in\mathbb{Z}}\!\!\!}^{\prime} \left(\frac{A+1}{A(i+j)^2+(j+k)^2+(i+k)^2}\right)^s
\end{equation}
where $1/3\leq A \leq 1$.
Here and throughout this work, a prime on the summation symbol will denote
that the sum ranges over all integer values except for the term when all of the summation indices are simultaneously zero, i.e.,
the sums in~\eqref{LAsdefinition} are over all integer values
 of $i$, $j$ and $k$ except for the term $(i,j,k)=(0,0,0)$, which is omitted.
This lattice sum smoothly connects four different lattices along a cuboidal transition path (the Bain transformation), i.e., when $A=1$, $1/\sqrt{2}$, $1/2$ or $1/3$ we obtain the expressions for the lattice sums of fcc, mcc, bcc and acc respectively (face-centered cubic, mean centred-cuboidal, body-centred cubic, and axial centred cuboidal).
In these cases, we also write $L_3^{\text{fcc}}(s) = L(1;s), L_3^{\text{mcc}}(s) = L(1/\sqrt{2};s), L_3^{\text{bcc}}(s) = L(1/2;s), \text{and}\quad L_3^{\text{acc}}(s) = L(1/3;s)$.

Our objective is to find formulas for $L(A;s)$ that are both simple
and computationally efficient. The formulas we obtain can be used to show that $L(A;s)$ can be analytically continued
to complex values of~$s$, with a simple pole at $s=3/2$ and no other singularities.

One method of evaluating the sum $L(A;s)$ is to use the Terras decomposition.\cite{terras-1973} This was done in our previous work for the fcc and bcc lattices,\cite{burrows-2020} and can in principle also be applied for general $L(A;s)$ with symmetric Gram matrices related to the Epstein zeta function.\cite{Epstein-1903} Here we use an easier method that  works for the entire parameter range $1/3 \leq A \leq 1$ along the Bain transformation path and hence gives the lattice sum for all four lattices fcc, mcc, bcc and acc. 
The advantage is that we obtain two formulas, which not only can be used as checks, but also provide distinct information about their analytic continuation.

We begin by writing the lattice sum in the form
\begin{align}
\label{easier}
L(A;s) &= {\sum_{i,j,k\in\mathbb{Z}}\!\!\!}^{\prime} \left(\frac{A+1}{A(i+j)^2+(j+k)^2+(i+k)^2}\right)^s  
={\sum_{\substack{I,J,K\in\mathbb{Z} \\ I+J+K\, \text{even}}}}^{\!\!\!\!\!\!\!\!\prime} \left(\frac{A+1}{AI^2+J^2+K^2}\right)^s \\
&= \frac{(A+1)^s}{2} {\sum_{i,j,k\in\mathbb{Z}}\!\!\!}^{\prime} \;\;\; \frac{1+(-1)^{i+j+k}}{(Ai^2+j^2+k^2)^s}
=\frac{(A+1)^s}{2} \left(T_1(A;s)  + T_2(A;s)  \right).
\nonumber 
\end{align}
with the two sums 
\begin{equation}
\label{sum1}
T_1(A;s) := {\sum_{i,j,k\in\mathbb{Z}}\!\!\!}^{\prime} \;\;\; \frac{1}{(Ai^2+j^2+k^2)^s}
\end{equation}
and
\begin{equation}
\label{sum2}
T_2(A;s) := {\sum_{i,j,k\in\mathbb{Z}}\!\!\!}^{\prime} \;\;\; \frac{(-1)^{i+j+k}}{(Ai^2+j^2+k^2)^s}
\end{equation}
which we evaluate separately. For $A=1$, $T_1$ is identical to the lattice sum of a simple cubic lattice and $T_2$ to the Madelung constant.\\

\textbf{The lattice sum $T_1(A;s)$}. We shall consider two ways for handling the sum in~\eqref{sum1}. The first is to separate the terms according to whether $i=0$ or $i\neq 0$,
which gives rise to
\begin{equation}
\label{d1}
T_1(A;s) = f(s) + 2F(s)
\end{equation}
where
\begin{align}
f(s) = {\sum_{j,k\in\mathbb{Z}}\!\!}^{\,\prime} \;\;\; \frac{1}{(j^2+k^2)^s}
\quad \text{and} \quad
F(s) = \sum_{i\in\mathbb{N}} \sum_{j,k\in\mathbb{Z}}  \frac{1}{(Ai^2+j^2+k^2)^s}
\end{align}
and $\mathbb{N}$ is the set of positive integers. For simplicity we omit the parameter~$A$ from the notation and just write $f(s)$ and $F(s)$ in place of $f(A;s)$ and $F(A;s)$. This is the starting point of the approach taken by Selberg and Chowla~[\onlinecite[Section~7]{seiberg-1967}].
Using theta series and Mellin transforms, Zucker showed that the double sum can be expressed in terms of standard functions,\cite{Zucker-1974}
\begin{equation}
f(s)= {\sum_{j,k\in\mathbb{Z}}\!\!\!}^{\,\prime} \;\;\; \frac{1}{(j^2+k^2)^s} = 4\zeta(s)L_{-4}(s)
\end{equation}
where $\zeta(s)$ is the Riemann zeta function defined in \eqref{zeta1}, and $L_{-4}(s)$ is the Dirichlet beta series from \eqref{zeta2} described in Appendix \ref{sec:specialfunctions}. It remains to analyze $F(s)$. Using the integral formula for the gamma function~\eqref{gamma2} we get
\begin{align}
\pi^{-s}\Gamma(s)F(s)
&= \int_{[0,\infty)} x^{s-1} \sum_{i\in\mathbb{N}} e^{-\pi A x i^2} 
\sum_{j,k\in\mathbb{Z}} e^{-\pi x(j^2+k^2)} \; \ud x \\
&= \int_{[0,\infty)} x^{s-1} \sum_{i\in\mathbb{N}} e^{-\pi A x i^2} 
\left(\sum_{j\in\mathbb{Z}} e^{-\pi xj^2}\right)^2 \; \ud x.
\nonumber
\end{align}
Now apply the modular transformation for theta functions~\eqref{theta2} to obtain
\begin{align}
\pi^{-s}\Gamma(s)F(s)
&= \int_{[0,\infty)} x^{s-1} \sum_{i\in\mathbb{N}} e^{-\pi A x i^2} 
\left(\frac{1}{\sqrt{x}}\sum_{j\in\mathbb{Z}} e^{-\pi j^2/x}\right)^2 \; \ud x \\
&= \int_{[0,\infty)} x^{s-2} \sum_{i\in\mathbb{N}} e^{-\pi A x i^2} 
\sum_{N\in\mathbb{N}_0} r_2(N) e^{-\pi N/x}\ \; \ud x
\nonumber
\end{align}
where $r_2(N)$ is the number of representations of $N$ as a sum of two squares, e.g., see~\eqref{theta3}, and $\mathbb{N}_0=\mathbb{N} \cup \{0\}$. Separating out the $N=0$ term and evaluating the resulting integrals,  we find that
\begin{align*}
\pi^{-s}\Gamma(s)F(s)
&= \sum_{i\in\mathbb{N}}\int_{[0,\infty)} x^{s-2} e^{-\pi A x i^2}  \; \ud x
+\sum_{i,N\in\mathbb{N}}  r_2(N) 
\int_{[0,\infty)}y x^{s-2}  e^{-\pi A x i^2-\pi N/x}\ \; \ud x \\
&= \frac{\Gamma(s-1)\zeta(2s-2)}{A^{s-1}\pi^{s-1}}
+ 2\sum_{i,N\in\mathbb{N}}  r_2(N) \left(\frac{N}{Ai^2}\right)^{(s-1)/2} K_{s-1}\left(2\pi i \sqrt{AN}\right)
\end{align*}
where we have used the formula~\eqref{bessel1} for the $K$-Bessel function.
On using all of the above back in~\eqref{d1} we obtain
\begin{align}
&{\sum_{i,j,k\in\mathbb{Z}}\!\!\!}^{\prime} \;\;\; \frac{1}{(Ai^2+j^2+k^2)^s}
= 4\zeta(s)L_{-4}(s) +\frac{2\pi}{(s-1)}\frac{ \zeta(2s-2)}{A^{s-1}} \nonumber \\
&\quad +\frac{4\pi^s}{\Gamma(s)}\, A^{(1-s)/2}\, \sum_{i,N\in\mathbb{N}}   r_2(N) \left(\frac{N}{i^2}\right)^{(s-1)/2} K_{s-1}\left(2\pi i \sqrt{AN}\right).
\label{sumpart1}
\end{align}
This is essentially Selberg and Chowla's formula~[\onlinecite[pg.45]{seiberg-1967}], although they write it in terms of a sum over the divisors of $N$ to minimize the number of Bessel function evaluations. We will leave it as it is for simplicity.\\

\textbf{Second formula for the sum $T_1(A;s)$}.
Another way is to separate the terms
according to whether $(j,k)=(0,0)$ or $(j,k) \neq (0,0)$ and write
\begin{equation}
\label{d3}
T_1(A;s) = 2g(s) + G(s)
\end{equation}
where
\begin{align}
g(s)= \sum_{i\in\mathbb{N}} \frac{1}{(Ai^2)^s} 
\quad\text{and}\quad 
G(s) ={\sum_{j,k\in\mathbb{Z}}\!\!\!}^{\,\prime} \; \sum_{i\in\mathbb{Z}} \frac{1}{(Ai^2+j^2+k^2)^s}.
\end{align}
For simplicity we omit the parameter~$A$ from the notation and just write $g(s)$ and $G(s)$ in place of $g(A;s)$ and $G(A;s)$, respectively. Now apply the integral formula for the gamma function~\eqref{gamma2} and then the modular transformation for the theta function~\eqref{theta1} to obtain
\begin{align}
\pi^{-s}\Gamma(s)G(s)
&= \int_{[0,\infty)} x^{s-1} {\sum_{j,k\in\mathbb{Z}}\!\!\!}^{\;\prime} e^{-\pi(j^2+k^2)x}
\sum_{i\in\mathbb{Z}} e^{-\pi i^2 Ax} \, \ud x \\
\nonumber
&= \frac{1}{\sqrt{A}}\int_{[0,\infty)} x^{s-3/2}\, {\sum_{j,k\in\mathbb{Z}}\!\!\!}^{\;\prime} e^{-\pi(j^2+k^2)x}
\sum_{i\in\mathbb{Z}} e^{-\pi i^2/Ax} \, \ud x.
\end{align}
Separate the $i=0$ term, to get
\begin{align}
\pi^{-s}\Gamma(s)G(s)
&=\frac{1}{\sqrt{A}} \int_{[0,\infty)} x^{s-3/2}\, {\sum_{j,k\in\mathbb{Z}}\!\!\!}^{\;\prime} e^{-\pi(j^2+k^2)x}
\, \ud x \\
\nonumber
& \quad + \frac{2}{\sqrt{A}}\int_{[0,\infty)} x^{s-3/2}\, {\sum_{j,k\in\mathbb{Z}}\!\!\!}^{\;\prime} 
e^{-\pi(j^2+k^2)x}
\sum_{i\in\mathbb{N}} e^{-\pi i^2/Ax} \, \ud x.
\end{align}
The first integral can be evaluated in terms of the gamma function by~\eqref{gamma2}, while the second integral
can be expressed in terms of the modified Bessel function by~\eqref{bessel1}. The result is
\begin{align}
\pi^{-s}\Gamma(s)G(s)
&= \frac{\Gamma\left(s-\frac12\right)}{\sqrt{A}\, \pi^{s-\frac12}}
\;{\sum_{j,k\in\mathbb{Z}}\!\!\!}^{\;\prime} \; \frac{1}{(j^2+k^2)^{s-\frac12}} \\
\nonumber
&\quad +\frac{4}{A^{\frac{s}{2}+\frac14}}\, {\sum_{j,k\in\mathbb{Z}}\!\!\!}^{\;\prime} \;\; \sum_{i\in\mathbb{N}}
\left(\frac{i}{\sqrt{j^2+k^2}}\right)^{s-\frac12}
K_{s-\frac12}\left(2\pi i\sqrt{\frac{j^2+k^2}{A}}\right) \\
\nonumber
&= \frac{4}{\sqrt{A}} \, {\pi^{-(s-\frac12)}}\,\Gamma\left(s-\frac12\right)\, \zeta\left(s-\frac12\right)\, L_{-4}\left(s-\frac12\right) \\
\nonumber
&\quad + \frac{4}{A^{\frac{s}{2}+\frac14}} \sum_{N,i\in\mathbb{N}} r_2(N) \left(\frac{i}{\sqrt{N}}\right)^{s-\frac12}
K_{s-\frac12}\left(2\pi i\sqrt{\frac{N}{A}}\right).
\end{align}
On using all of the above back in~\eqref{d3} we  obtain
\begin{align}
\label{sumpart2}
{\sum_{i,j,k\in\mathbb{Z}}\!\!\!}^{\prime} \;\;\; \frac{1}{(Ai^2+j^2+k^2)^s} 
&= 2A^{-s}\zeta(2s)+4 \, \sqrt{\frac{\pi}{A}} \,\frac{\Gamma\left(s-\frac12\right)}{\Gamma(s)}\, \zeta\left(s-\frac12\right)\, L_{-4}\left(s-\frac12\right) \\
\nonumber
& +   \frac{4}{A^{\frac{s}{2}+\frac14}} \, \frac{\pi^s}{\Gamma(s)}\,\sum_{N,i\in\mathbb{N}} r_2(N) \left(\frac{i}{\sqrt{N}}\right)^{s-\frac12}
K_{s-\frac12}\left(2\pi i\sqrt{\frac{N}{A}}\right). 
\end{align}
The terms in~\eqref{sumpart1} involve $K_{s-1}$ Bessel functions whereas $K_{s-\frac{1}{2}}$ Bessel functions occur in~\eqref{sumpart2}.
That is because each application of the theta function transformation formula lowers the subscript in the resulting
Bessel function by $1/2$, due to the creation of a $x^{-1/2}$ factor in the integral. 
The theta function transformation formula is used twice (i.e., the formula is squared) in the derivation of~\eqref{sumpart1} and only once in the derivation of~\eqref{sumpart2}.
Each of~\eqref{sumpart1} and~\eqref{sumpart2} turns out to have its own advantages when it comes to convergence for specific $A$ and $s$ values.\\

\textbf{The alternating lattice sum $T_2(A;s)$}.
The analysis in the previous sections can be modified to handle the alternating series~\eqref{sum2} which has
the term $(-1)^{i+j+k}$ in the numerator, as follows. Separating the terms according to whether $i=0$ or $i\neq 0$ gives
\begin{equation}
T_2(A;s) = h(s) + 2H(S) 
\label{T21}
\end{equation}
where
\begin{equation}
h(s) = {\sum_{j,k\in\mathbb{Z}}\!\!}^{\prime} \;\;\; \frac{(-1)^{j+k}}{(j^2+k^2)^s} 
\quad \text{and} \quad
H(s) = \sum_{i\in\mathbb{N}} \sum_{j,k\in\mathbb{Z}} \frac{(-1)^{i+j+k}}{(Ai^2+j^2+k^2)^s}.
\end{equation}
Using~\eqref{zeta2}, we obtain $h(s) =  -4(1-2^{1-s})\zeta(s)L_{-4}(s)$.
Next, using the integral formula for the gamma function~\eqref{gamma2} we obtain
\begin{align}
\pi^{-s}\Gamma(s)H(s)
&= \int_{[0,\infty)} x^{s-1} \sum_{i\in\mathbb{N}} (-1)^ie^{-\pi A x i^2} 
\sum_{j,k\in\mathbb{Z}} (-1)^{j+k}e^{-\pi x(j^2+k^2)} \; \ud x \\
&= \int_{[0,\infty)} x^{s-1} \sum_{i\in\mathbb{N}} (-1)^ie^{-\pi A x i^2} 
\left(\sum_{j\in\mathbb{Z}} (-1)^je^{-\pi xj^2}\right)^2 \; \ud x.
\end{align}
Applying the modular transformation for theta functions leads to
\begin{align}
\pi^{-s}\Gamma(s)H(s)
&= \int_{[0,\infty)} x^{s-1} \sum_{i\in\mathbb{N}} (-1)^ie^{-\pi A x i^2} 
\left(\frac{1}{\sqrt{x}}\sum_{j\in\mathbb{Z}} e^{-\pi (j+\frac12)^2/x}\right)^2 \; \ud x.
\end{align}
By formula~\eqref{theta4} this can be expressed as
\begin{align}
\pi^{-s}\Gamma(s)H(s)&= \int_{[0,\infty)} x^{s-2} \sum_{i\in\mathbb{N}} (-1)^ie^{-\pi A x i^2} 
\sum_{N\in\mathbb{N}_0} r_2(4N+1) e^{-\pi (4N+1)/2x}\ \; \ud x \\
\nonumber
&= \sum_{i\in\mathbb{N}} \sum_{N\in\mathbb{N}_0}  (-1)^i r_2(4N+1) 
\int_{[0,\infty)} x^{s-2}  e^{-\pi A x i^2-\pi (4N+1)/2x}\ \; \ud x.
\end{align}
The integral can be expressed in terms of Bessel functions using~\eqref{bessel1}
\begin{align}
\pi^{-s}\Gamma(s)H(s)&= 2\sum_{i\in\mathbb{N}} \sum_{N\in\mathbb{N}_0}  (-1)^ir_2(4N+1) \left(\frac{2N+\frac12}{Ai^2}\right)^{(s-1)/2} K_{s-1}\left(2\pi i \sqrt{A(2N+\frac12)}\right).
\end{align}
Incorporating all of the above back in~\eqref{T21} results in
\begin{align}
 \label{sumpart3}
{\sum_{i,j,k\in\mathbb{Z}}\!\!\!}^{\prime} \;\;\; &\frac{(-1)^{i+j+k}}{(Ai^2+j^2+k^2)^s} =  -4(1-2^{1-s})\zeta(s)L_{-4}(s)   \\
&+\frac{4\pi^s}{\Gamma(s)} \, A^{(1-s)/2}\,
\sum_{i\in\mathbb{N}} \sum_{N\in\mathbb{N}_0}  (-1)^ir_2(4N+1) \left(\frac{2N+\frac12}{i^2}\right)^{(s-1)/2} K_{s-1}\left(2\pi i \sqrt{A(2N+\frac12)}\right).
\nonumber
\end{align}

\textbf{Second formula for $T_2(A;s)$}.
This time we separate the terms according to whether $(j,k)=(0,0)$ or $(j,k) \neq (0,0)$ and write
\begin{equation}
\label{dd2}
T_2(A;s) = 2\sum_{i\in\mathbb{N}} \frac{(-1)^i}{(Ai^2)^s} + J(s)
\end{equation}
where
\begin{equation}
J(s) = {\sum_{j,k\in\mathbb{Z}}\!\!\!}^{\,\prime} \; \sum_{i\in\mathbb{Z}} \frac{(-1)^{i+j+k}}{(Ai^2+j^2+k^2)^s}.
\end{equation}
Using \eqref{zeta4} gives
\begin{equation}
2\sum_{i\in\mathbb{N}} \frac{(-1)^i}{(Ai^2)^s}  = -2A^{-s}(1-2^{1-2s})\zeta(2s).
\end{equation}
It remains to analyse the sum for $J(s)$. Using the integral formula for the gamma function \eqref{gamma2} leads to
\begin{align}
\pi^{-s}\Gamma(s)J(s) &= \int_{[0,\infty)} x^{s-1} {\sum_{j,k\in\mathbb{Z}}\!\!\!}^{\;\prime} (-1)^{j+j}e^{-\pi(j^2+k^2)x}
\sum_{i\in\mathbb{Z}} (-1)^ie^{-\pi i^2 Ax} \, \ud x.
\end{align}
Applying the modular transformation~\eqref{theta1b} gives
\begin{align}
\pi^{-s}\Gamma(s)J(s) &= \frac{1}{\sqrt{A}}\int_{[0,\infty)} x^{s-3/2}\, {\sum_{j,k\in\mathbb{Z}}\!\!\!}^{\;\prime} (-1)^{j+k}e^{-\pi(j^2+k^2)x}
\sum_{i\in\mathbb{Z}} e^{-\pi (i+\frac12)^2/Ax} \, \ud x.
\end{align}
Setting $N=j^2+k^2$ and using
\begin{align}
\sum_{i\in\mathbb{Z}} e^{-\pi (i+\frac12)^2/Ax} = 2\sum_{i\in\mathbb{N}_0}^\infty e^{-\pi (i+\frac12)^2/Ax} = 2\sum_{i\in\mathbb{N}} e^{-\pi (i-\frac12)^2/Ax}
\end{align}
gives
\begin{align}
\pi^{-s}\Gamma(s)J(s) &= \frac{2}{\sqrt{A}} \sum_{N,i\in\mathbb{N}} (-1)^Nr_2(N)  \int_{[0,\infty)} x^{s-3/2}\, e^{-\pi Nx - \pi (i-\frac12)^2/Ax} \, \ud x.
\end{align}
The integral can be evaluated in terms of the modified Bessel function, ~\eqref{bessel1},
\begin{align}
\pi^{-s}\Gamma(s)J(s)
&=\frac{4}{A^{\frac{s}{2}+\frac14}} \,
\sum_{N,i\in\mathbb{N}} (-1)^N\,r_2(N) \left(\frac{i-\frac12}{\sqrt{N}}\right)^{s-\frac12}
K_{s-\frac12}\left(2\pi (i-\frac12)\sqrt{\frac{N}{A}}\right). 
\end{align}
It follows that
\begin{align}
\label{sumpart4}
{\sum_{i,j,k\in\mathbb{Z}}\!\!\!}^{\prime} \;\;\; &\frac{(-1)^{i+j+k}}{(Ai^2+j^2+k^2)^s} 
= -2A^{-s}(1-2^{1-{2s}})\zeta(2s)\\
&\quad +   \frac{4}{A^{\frac{s}{2}+\frac14}} \, \frac{\pi^s}{\Gamma(s)}\,
\sum_{N,i\in\mathbb{N}} (-1)^N\,r_2(N) \left(\frac{i-\frac12}{\sqrt{N}}\right)^{s-\frac12} K_{s-\frac12}\left(2\pi (i-\frac12)\sqrt{\frac{N}{A}}\right). 
 \nonumber 
\end{align}

\textbf{Two formulas for $L(A;s)$}.
On substituting the results of~\eqref{sumpart1} and \eqref{sumpart3} back into~\eqref{easier} we obtain a formula for $L(A;s)$
in terms of~$K_{s-1}$ Bessel functions:
\begin{align}
\label{L3formula1}
L&(A;s) = 4\left(\frac{A+1}{2}\right)^s \zeta(s)L_{-4}(s) + \frac{\pi A}{s-1} \left(1+\frac{1}{A}\right)^s \zeta(2s-2) \\
\nonumber 
&\quad + \frac{2\pi^s\sqrt{A}}{\Gamma(s)} \left(\sqrt{A}+\frac{1}{\sqrt{A}}\right)^s \sum_{N,i\in\mathbb{N}}  r_2(N) \left(\frac{N}{i^2}\right)^{(s-1)/2} K_{s-1}\left(2\pi i \sqrt{AN}\right) \\
&\quad + \frac{2\pi^s\sqrt{A}}{\Gamma(s)} \left(\sqrt{A}+\frac{1}{\sqrt{A}}\right)^s
\sum_{i\in\mathbb{N}}\sum_{N\in\mathbb{N}_0}  (-1)^ir_2(4N+1)  \left(\frac{2N+\frac12}{i^2}\right)^{(s-1)/2}  K_{s-1}\left(2\pi i \sqrt{A(2N+\frac12)}\right).
\nonumber 
\end{align}
\noindent
On the other hand, if the results of~\eqref{sumpart2} and~\eqref{sumpart4} are used in~\eqref{easier}, the resulting formula for $L(A;s)$ involves
$K_{s-1/2}$ Bessel functions:
\begin{align}
\label{L3formula2}
L(A;s&)= 2\left(\frac{A+1}{4A}\right)^s\zeta(2s)
+2\,\sqrt{\frac{\pi}{A}}\,(A+1)^s\,\frac{\Gamma\left(s-\frac12\right)}{\Gamma(s)}\, \zeta\left(s-\frac12\right)\, L_{-4}\left(s-\frac12\right) \nonumber \\
&\quad + \frac{2}{A^{1/4}}\left(\sqrt{A}+\frac{1}{\sqrt{A}}\right)^s\, \frac{\pi^s}{\Gamma(s)}\, \,\sum_{N,i\in\mathbb{N}} N^{(1-2s)/4}\,r_2(N)  \\
&\qquad \times \left\{ i^{s-\frac12}
K_{s-\frac12}\left(2\pi i\sqrt{\frac{N}{A}}\right) + (-1)^N \left(i-\frac12\right)^{s-\frac12}
K_{s-\frac12}\left(2\pi (i-\frac12)\sqrt{\frac{N}{A}}\right) \right\}.
\nonumber
\end{align}
The formulas \eqref{L3formula1} and \eqref{L3formula2} can be used as checks against one another.
Moreover, the formulas offer different information about special values of the lattice sum, as will be seen in Section~\ref{pole}.

\section{A minimum property of the lattice sum $L(A;s)$}
\label{sec:MinimumProperty}

It was noted that on the interval $1/3 \leq A \leq 1$, the packing density function~$\Delta_\Lambda$
has a minimum value when~$A=1/2$. Provided that $s>3/2$, the corresponding lattice sum $L(A;s)$ also attains a minimum at the same value $A=1/2$.
 The proof for this condition is provided in ref.~\onlinecite{cooper2025}.
\begin{Theorem}
\label{theorem3point1}
Let $L(A;s)$ be the lattice defined by~\eqref{LAsdefinition}, that is,
\begin{equation}
\label{LAsdefinition1}
L(A;s) = {\sum_{i,j,k\in\mathbb{Z}}\!\!\!}^{\prime} \left(\frac{1}{g(A;i,j,k)}\right)^s = {\sum_{i,j,k\in\mathbb{Z}}\!\!\!}^{\prime} \left(\frac{A+1}{A(i+j)^2+(j+k)^2+(i+k)^2}\right)^s
\end{equation}
where $s>3/2$ and $1/3\leq A \leq 1$. 
Then 
\begin{equation}
\left.\frac{\partial}{\partial A} L(A;s) \right|_{A=1/2} =0
\quad \text{and}\quad \left.\frac{\partial^2}{\partial A^2} L(A;s) \right|_{A=1/2} >0.
\end{equation}
\end{Theorem}

\begin{figure}[htbp]
  \begin{center}
  \includegraphics[scale=0.8]{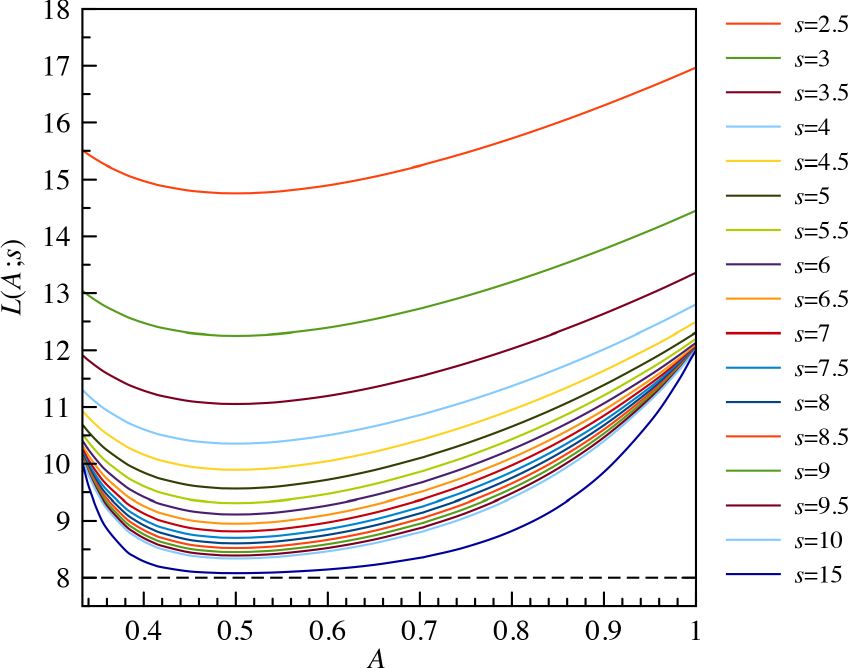}\\
  \caption{
  Graph of $L(A;s)$ versus $A$ for various values of $s$ For $s\rightarrow \infty$ we have at both ends of the interval $L(A=\frac{1}{3};\infty)=10$ and $L(A=1;\infty)=12$.
  }
  \label{fig:LatticeSums}
  \end{center}
  \end{figure}

As consequence, for any fixed value $s>3/2$, the lattice sum $L(A;s)$ attains a minimum when~\mbox{$A=1/2$.} 
Graphs of $L(A;s)$ to illustrate this minimum property are shown in Fig.\ref{fig:LatticeSums}.
In the limiting case $s\rightarrow\infty$ we have
\begin{align}
L(A;\infty) = \lim_{s\rightarrow\infty}L(A;s) = \textrm{kiss}(\Lambda) 
=  \begin{cases}
10 & \text{if $A=1/3$,} \\
8 & \text{if $1/3 < A < 1$,} \\
12 & \text{if $A=1$.}
\end{cases} 
\end{align}
  
We find an interesting relation between the lattice sum and its first derivative,
\begin{align}
\label{dLdAequalsL}
\left. \frac{\partial}{\partial A} L(A;s)\right|_{A=1} = \left.\frac{s}{6} L(A;s)\right|_{A=1}
\end{align}
which can be proved as follows. On calculating the derivative using~\eqref{LAsdefinition1} we obtain
\begin{align*}
\frac{\partial}{\partial A}L(A;s) = 
\frac{2s}{(A+1)^2} {\sum_{i,j,k\in\mathbb{Z}}\!\!\!}^{\prime} (k^2+ik+jk-ij)\left(\frac{A+1}{A(i+j)^2+(j+k)^2+(i+k)^2}\right)^{s+1}
\end{align*}
and on setting $A=1$ it follows that
\begin{align*}
\left.\frac{\partial}{\partial A}L(A;s)\right|_{A=1} = 
\frac{s}{2} {\sum_{i,j,k\in\mathbb{Z}}\!\!\!}^{\prime} \frac{k^2+ik+jk-ij}{\left( i^2+j^2+k^2+ij+jk+ki \right)^{s+1}}.
\end{align*}
Now replace the summation indices $(i,j,k)$ with the cyclic permutations $(j,k,i)$ and $(k,i,j)$ and add the three equations to obtain
\begin{align*}
3\left.\frac{\partial}{\partial A}L(A;s)\right|_{A=1} &= 
\frac{s}{2} {\sum_{i,j,k\in\mathbb{Z}}\!\!\!}^{\prime} \frac{i^2+j^2+k^2+ij+jk+ki}{\left( i^2+j^2+k^2+ij+jk+ki \right)^{s+1}} \\
&=\frac{s}{2} {\sum_{i,j,k\in\mathbb{Z}}\!\!\!}^{\prime} \frac{1}{\left( i^2+j^2+k^2+ij+jk+ki \right)^{s}} 
=  \left.\frac{s}{2} L(A;s)\right|_{A=1}.
\end{align*}
This proves~\eqref{dLdAequalsL}.

\section{Analytic continuations of the lattice sums $L(A;s)$}
\label{pole}
We will now show that the lattice sums $L(A;s)$ can be continued analytically to the whole $s$-plane, and that the resulting functions have a single simple pole
at $s=3/2$ and no other singularities. We do this in steps. First, we show that
the lattice sums each have a simple pole at $s=3/2$ and determine the residue. Then, 
we show that the analytic continuations obtained are valid for the whole $s$-plane and there are no other singularities.
Finally, values of the analytic continuations at the points $s=1/2$ and $s=1,\,0,\,-1,\,-2,\ldots$
are computed. In particular, the evaluation of $T_2(A;s)$ at $s=1/2$ in the case $A=1$ gives the Madelung constant,
e.g., see~[\onlinecite{borwein1998convergence}],
~[\onlinecite[pp. xiii, 39--51]{borwein-2013}],~[\onlinecite{madelung1918}].

We start by showing that $L(A;s)$ has a simple pole at $s=3/2$ and determine the residue.
In the formula~\eqref{L3formula1}, all of the terms are analytic at $s=3/2$ except for the term involving $\zeta(2s-2)$.
It follows that
\begin{align}
\lim_{s\rightarrow 3/2}& (s-3/2)L(A;s) 
= \lim_{s\rightarrow 3/2} (s-3/2)\frac{\pi A}{s-1} \left(1+\frac{1}{A}\right)^s \zeta(2s-2)  \\
\nonumber
&=2\pi A \left(1+\frac{1}{A}\right)^{3/2} \lim_{s\rightarrow 3/2} (s-3/2) \zeta(2s-2) \\
\nonumber
&=  \frac{2\pi}{\sqrt{A}} \left(A+1\right)^{3/2} \times \frac12 \; \lim_{u\rightarrow 1} (u-1) \zeta(u) 
= \frac{\pi}{\sqrt{A}}  \left(A+1\right)^{3/2}
\end{align}
where~\eqref{zetapole} was used in the last step of the calculation.
It follows further that $L(A;s)$ has a simple pole at $s=3/2$ and the residue is given by
\begin{align}
\text{Res}(L(A;s),3/2) = \frac{\pi}{\sqrt{A}}  \left(A+1\right)^{3/2}.
\end{align}
This corresponds to 12 times the packing density, i.e.,
\begin{align}
\text{Res}(L(A;s),3/2) = 12\Delta_\Lambda.
\end{align}
For example, taking $A=1$ gives
\begin{equation}
\label{residue1}
\text{Res}(L_3^{\text{FCC}}(s),3/2)  = 2\sqrt{2}\,\pi
\end{equation}
while taking $A=1/2$ gives
\begin{align}
\text{Res}(L_3^{\text{BCC}}(s),3/2)  = 3\sqrt{3}\,\pi/2.
\end{align}
Laurent's theorem implies there is an expansion of the form
\begin{equation}
\label{laurent1}
L(A;s) = \frac{c_{-1}}{s-3/2}+c_0+\sum_{n=1}^\infty c_n (s-3/2)^n
\end{equation}
where
\begin{equation}
c_{-1} = \text{Res}(L(A;s),3/2) = \frac{\pi}{\sqrt{A}}  \left(A+1\right)^{3/2}
\end{equation}
and the coefficients $c_0$, $c_1$, $c_2,\ldots$ depend on $A$ but not on $s$.
To calculate~$c_0$, start with the fact that
\begin{equation}
\lim_{s\rightarrow 3/2} \left(\frac{\pi A}{s-1} \left(1+\frac{1}{A}\right)^s \zeta(2s-2) - \frac{c_{-1}}{s-3/2}\right) 
= \frac{\pi}{\sqrt{A}} \left(A+1\right)^{3/2} \left(2\gamma - 2 + \log\left(1+\frac{1}{A}\right)\right)
\end{equation}
where $\gamma = 0.57721\,56649\,01532\,86060\,\cdots$ is the Euler–Mascheroni constant. Then use~\eqref{L3formula1} and~\eqref{bessel2} to deduce
\begin{align}
\nonumber
c_0 = &\lim_{s\rightarrow 3/2} \left( L(A;s) - \frac{c_{-1}}{s-3/2}\right) 
=\sqrt{2}\left(A+1\right)^{3/2} \zeta\left(\frac32\right)L_{-4}\left(\frac32\right) \\
& + \frac{\pi}{\sqrt{A}} \left(A+1\right)^{3/2} \left(2\gamma - 2 + \log\left(1+\frac{1}{A}\right) \right) \\
\nonumber
& + \frac{2\pi}{\sqrt{A}} \left(A+1\right)^{3/2} \sum_{k,N\in\mathbb{N}}  \frac{1}{k} \, r_2(N) \exp\left(-2\pi k \sqrt{AN}\right) \\
\nonumber
& + \frac{2\pi}{\sqrt{A}} \left(A+1\right)^{3/2} \sum_{k,N\in\mathbb{N}}  \frac{(-1)^k}{k}\,r_2(4N+1)\, \exp\left(-2\pi k \sqrt{A\left(2N+\frac12\right)}\right).
\end{align}
Interchanging the order of summation and evaluating the sum over~$k$ gives
\begin{align}
c_0 =&\sqrt{2}\left(A+1\right)^{3/2} \zeta\left(\frac32\right)L_{-4}\left(\frac32\right) \\
\nonumber
&+ \frac{\pi}{\sqrt{A}} \left(A+1\right)^{3/2} \left(2\gamma - 2 + \log\left(1+\frac{1}{A}\right) \right) \\
\nonumber
& - \frac{2\pi}{\sqrt{A}} \left(A+1\right)^{3/2} \sum_{N\in\mathbb{N}}   r_2(N) \log\left(1- e^{-2\pi \sqrt{AN}}\right) \\
\nonumber
& - \frac{2\pi}{\sqrt{A}} \left(A+1\right)^{3/2} \sum_{N\in\mathbb{N}_0}  r_2(4N+1)\, \log\left(1+ e^{-\pi  \sqrt{2A\left(4N+1\right)}}\right).
\end{align}
Numerical evaluation in the case $A=1$ gives
\begin{equation}
\label{c0}
\left. c_0\right|_{A=1} = 6.98405\,25503\,22247\,93406\dots.
\end{equation}

We now evaluate the analyticity of the lattice sums at other values of $s$.
By~\eqref{bessel3}, the double series of Bessel functions in~\eqref{L3formula1} converges absolutely and uniformly on compact subsets of the $s$-plane
and therefore represents an entire function of~$s$. It follows that $L(A;s)$ has an analytic continuation to a meromorphic function
which is analytic except possibly at the singularities of the terms
\begin{equation}
\label{sing1}
4\left(\frac{A+1}{2}\right)^s \zeta(s)L_{-4}(s)
\end{equation}
and
\begin{equation}
\label{sing2}
\frac{\pi A}{s-1} \left(1+\frac{1}{A}\right)^s \zeta(2s-2).
\end{equation}
The function in \eqref{sing1} is analytic except at $s=1$ due to the pole of~$\zeta(s)$. The function $L_{-4}(s)$ and the exponential function are both entire.
The function in~\eqref{sing2} is analytic except at $s=1$ and $s=3/2$. 

The singularity at $s=3/2$ was already studied before.\cite{Schwerdtfeger-2006,burrows-2020} Using~\eqref{zetapole} and the values of $\zeta(0)$ and $L_{-4}(1)$ in~\eqref{zetavalues} and~\eqref{L4values} we find that
\begin{equation}
4\left(\frac{A+1}{2}\right)^s \zeta(s)L_{-4}(s)  = \frac{(A+1)\pi}{2(s-1)}+O(1) \quad\text{as $s\rightarrow 1$}
\end{equation}
and 
\begin{equation}
\frac{\pi A}{s-1} \left(1+\frac{1}{A}\right)^s \zeta(2s-2)= -\frac{(A+1)\pi}{2(s-1)}+O(1) \quad\text{as $s\rightarrow 1$}.
\end{equation}
It follows that the sum of the functions in~\eqref{sing1} and~\eqref{sing2} has a removable singularity at~$s=1$ and thus $L(A;s)$ is also analytic at~$s=1$. The analyticity at $s=1$ can also be seen directly from the alternative formula for~$L(A;s)$ in~\eqref{L3formula2}.
We thus showed that $L(A;s)$ has an analytic continuation to a meromorphic function of~$s$ which has a simple pole at~$s=3/2$
and no other singularities. Because $L(A;s)$ has only one singularity, namely $s=3/2$, the Laurent expansion~\eqref{laurent1} is valid
in the annulus $0<|s-3/2|<\infty$, i.e., for all $s \neq 3/2$.

By the theory of complex variables, the analytic continuation, if one exists, is unique, e.g., see~[\onlinecite[p. 147, Th. 1]{levinson1970}].
Therefore analytic continuation formulas can be used to assign values to divergent series.
For example, the Madelung constant is defined by
\begin{equation}
\label{madelungdefinition}
M={\sum_{i,j,k\in\mathbb{Z}}\!\!\!}^{\prime} \;\;\;\left. \frac{(-1)^{i+j+k}}{(i^2+j^2+k^2)^{s}}\right|_{s=1/2}. 
\end{equation}
This is interpreted as being the value of the analytic continuation of the series at $s=1/2$, because the sum diverges if $s=1/2$. From now on, we shall use the expression ``the value of a series at a point~$s$'' to mean ``the value of the
analytic continuation of the series at the point~$s$''. 

For the $A$-dependent case, on putting $s=1/2$ in~\eqref{sumpart3} we obtain an analytic expression for the value of 
\begin{align}
M(A)={\sum_{i,j,k\in\mathbb{Z}}\!\!\!}^{\prime} \;\;\;\left. \frac{(-1)^{i+j+k}}{(Ai^2+j^2+k^2)^{s}}\right|_{s=1/2}
\end{align}
which specialises to the Madelung constant in the case $A=1$. We have
\begin{align}
M(A)&=  -4(1-2^{1-s})\zeta(s)L_{-4}(s) \Bigg|_{s=1/2} \\
\nonumber 
&+\frac{4\pi^s}{\Gamma(s)} \,A^{(1-s)/2}
\sum_{i\in\mathbb{N}}\sum_{N\in\mathbb{N}_0}  (-1)^ir_2(4N+1) \left(\frac{2N+\frac12}{i^2}\right)^{(s-1)/2} 
 K_{s-1}\left(2\pi i \sqrt{A(2N+\frac12)}\,\right) \Bigg|_{s=1/2}.
\end{align}
Now use~\eqref{bessel1.5} and \eqref{bessel2} to express the Bessel functions in terms of exponential functions. The result simplifies to 
\begin{align}
M(A)&=4(\sqrt2-1)\zeta\left(\frac12\right)L_{-4}\left(\frac12\right)
+2\sum_{i\in\mathbb{N}} \sum_{N\in\mathbb{N}_0}  (-1)^i\, \frac{r_2(4N+1)}{\sqrt{2N+\frac12}}\,e^{-2\pi i \sqrt{A(2N+1/2)}}.
\intertext{On interchanging the order of summation and summing the geometric series, we obtain}
M(A)&=4(\sqrt2-1)\zeta\left(\frac12\right)L_{-4}\left(\frac12\right)
-2\sqrt{2}\sum_{N\in\mathbb{N}_0}  \frac{r_2(4N+1)}{\sqrt{4N+1}}\left(\frac{1}{e^{\pi  \sqrt{2A(4N+1)}}+1}\right).
\end{align}
When $A=1$ this gives the Madelung constant defined by~\eqref{madelungdefinition}.
Numerical evaluation gives
\begin{equation}
\label{Mminus}
M=M(1)=-1.74756\;45946\;33182\;19063\dots
\end{equation}
which is in agreement with~[\onlinecite[p. xiii]{borwein-2013}] (apart from the minus sign which we have corrected here) and matches the value of $d(1)$
in~[\onlinecite[pp 39--51]{borwein-2013}].

In a similar way, starting from~\eqref{sumpart1} and using~\eqref{bessel2} and~\eqref{zetavalues} we obtain
\begin{align}
{\sum_{i,j,k\in\mathbb{Z}}}^{\prime}& \;\;\;\left. \frac{1}{(Ai^2+j^2+k^2)^{s}}\right|_{s=1/2}  =4 \zeta\left(\frac12\right)L_{-4}\left(\frac12\right)+\frac{\pi\sqrt{A}}{3} +2\sum_{i,N\in\mathbb{N}}  \frac{r_2(N)}{\sqrt{N}}\,e^{-2\pi i \sqrt{AN}} \nonumber \\
&=4 \zeta\left(\frac12\right)L_{-4}\left(\frac12\right)+\frac{\pi\sqrt{A}}{3} +2\sum_{N\in\mathbb{N}}  \frac{r_2(N)}{\sqrt{N}}\left(\frac{1}{e^{2\pi \sqrt{AN}}-1}\right).
\label{remarkable1}
\end{align}
Numerical evaluation in the case $A=1$ gives
\begin{equation}
{\sum_{i,j,k\in\mathbb{Z}}\!\!\!}^{\prime} \;\;\;\left. \frac{1}{(i^2+j^2+k^2)^{s}}\right|_{s=1/2} = -2.83729\;74794\;80619\;47666\dots. \label{remarkable3}
\end{equation}
Now, from~\eqref{easier} we have for the fcc lattice
\begin{align}
L\left(A=1;\tfrac{1}{2}\right)&= \frac{1}{\sqrt{2}}{\sum_{i,j,k\in\mathbb{Z}}\!\!\!}^{\prime} \left. \frac{1}{(i^2+j^2+k^2)^{s}}\right|_{s=1/2} 
+  \frac{1}{\sqrt{2}}{\sum_{i,j,k\in\mathbb{Z}}\!\!\!}^{\prime} \left. \frac{(-1)^{i+j+k}}{(i^2+j^2+k^2)^{s}}\right|_{s=1/2}.
\end{align}
Hence, using the values from~\eqref{Mminus} and~\eqref{remarkable3} we obtain
\begin{align}
L\left(A=1;\tfrac{1}{2}\right)&= -3.24198\;70634\;10888\;39428\dots.
\end{align}

We now turn to the value of the lattice sum at $s=1$. It was noted above that~\eqref{L3formula1}, which involves $K_{s-1}$ Bessel functions, contains terms with
singularities at $s=1$ and therefore is not suitable for calculations at that value of~$s$.
Instead we can use~\eqref{L3formula2}, which involves $K_{s-1/2}$ Bessel functions.
As in the previous section, two steps are involved. First, the the $K_{1/2}$ Bessel functions can be expressed in terms of the exponential function by~\eqref{bessel2}. Then,
the double sum can be reduced to a single sum by geometric series. We omit the details and just record the final results and corresponding numerical
values. From~\eqref{sumpart2} we have
\begin{align}
{\sum_{i,j,k\in\mathbb{Z}}\!\!\!}^{\prime} \;\;\;\left. \frac{1}{(Ai^2+j^2+k^2)^{s}}\right|_{s=1}
=\frac{\pi^2}{3A} + \frac{4\pi}{\sqrt{A}} \zeta\left(\frac12\right) L_{-4}\left(\frac12\right)
+\frac{2\pi}{\sqrt{A}} \sum_{N\in\mathbb{N}} \frac{r_2(N)}{\sqrt{N}} \left(\frac{1}{e^{2\pi \sqrt{N/A}}-1}\right) \label{remarkable2}
\end{align}
while~\eqref{sumpart4}  gives
\begin{align}
{\sum_{i,j,k\in\mathbb{Z}}\!\!\!}^{\prime} \;\;\;\left. \frac{(-1)^{i+j+k}}{(Ai^2+j^2+k^2)^{s}}\right|_{s=1}
=\frac{-\pi^2}{6A}  +\frac{2\pi}{\sqrt{A}} \sum_{N\in\mathbb{N}} (-1)^N\, \frac{r_2(N)}{\sqrt{N}} \left(\frac{1}{e^{\pi \sqrt{N/A}}-e^{-\pi \sqrt{N/A}}}\right).
\end{align}
Then~\eqref{easier} can be used to write down the value of $L(A;s)$.
For example, when $A=1$ the above formulas give
\begin{align}
{\sum_{i,j,k\in\mathbb{Z}}\!\!\!}^{\prime}  \;\;\;\left. \frac{1}{(i^2+j^2+k^2)^{s}}\right|_{s=1}
&=-8.91363\;29175\;85151\;27268\dots \label{remarkable4}
\end{align}
and
\begin{align}
{\sum_{i,j,k\in\mathbb{Z}}\!\!\!}^{\prime} \;\;\;\left. \frac{(-1)^{i+j+k}}{(i^2+j^2+k^2)^{s}}\right|_{s=1}
&=-2.51935\;61520\;89445\;31334\dots. 
\label{unremarkable2}
\end{align}
Then taking $s=1$ in~\eqref{easier} gives for the fcc lattice
\begin{align}
L(A=1,1) &={\sum_{i,j,k}}^{\prime} \;\;\;\left. \frac{1}{(i^2+j^2+k^2)^{s}}\right|_{s=1} + {\sum_{i,j,k}}^{\prime} \;\;\;\left. \frac{(-1)^{i+j+k}}{(i^2+j^2+k^2)^{s}}\right|_{s=1} \\
&=  -11.43298\;90696\;74596\;58602\dots.
\end{align}
We note a connection between two of the values in the above analysis.
By setting~$A=1$ in each of~\eqref{remarkable1} and~\eqref{remarkable2} we obtain the remarkable result
\begin{equation}
{\sum_{i,j,k\in\mathbb{Z}}\!\!\!}^{\prime}  \;\;\;\left. \frac{1}{(i^2+j^2+k^2)^{s}}\right|_{s=1}
=\pi \; {\sum_{i,j,k\in\mathbb{Z}}\!\!\!}^{\prime}  \;\;\;\left. \frac{1}{(i^2+j^2+k^2)^{s}}\right|_{s=1/2}.
\end{equation}
This is consistent with~[\onlinecite[p. 46 (1.3.44)]{borwein-2013}] and is the special case~$s=1$ of the functional equation
\begin{equation}
\label{fe0}
\pi^{-s} \Gamma(s) T_1(1;s) = \pi^{-(\frac32-s)} \Gamma\left(\frac32-s\right) T_1\left(1;\frac32-s\right).
\end{equation}
This functional equation can be deduced from the two formulas for $T_1(A;s)$ in~\eqref{sumpart1} and~\eqref{sumpart2}, as follows.
Replace $s$ with $\frac32-s$ in~\eqref{sumpart1}, then multiply by $\pi^{s-\frac32}\Gamma(\frac32-s)$ and set $A=1$ to get
\begin{align}
\lefteqn{\pi^{s-\frac32}\Gamma\left(\frac32-s\right)T_1\left(1;\frac32-s\right)} \\
\nonumber
&= 4\pi^{s-\frac32}\Gamma\left(\frac32-s\right)\zeta\left(\frac32-s\right)L_{-4}\left(\frac32-s\right)  +2\pi^{s-\frac12}\Gamma\left(\frac12-s\right) \zeta(1-2s) \nonumber \\
&\quad +4 \sum_{i,N\in\mathbb{N}}  r_2(N) \left(\frac{N}{i^2}\right)^{(\frac12-s)/2} K_{\frac12-s}\left(2\pi i \sqrt{N}\right),
\nonumber
\end{align}
where we have used the functional equation for the gamma function in the form $\Gamma(3/2-s) = (1/2-s)\Gamma(1/2-s)$
to obtain the second term on the right hand side.
Now apply the functional equations~\eqref{bessel1.5},~\eqref{fe1} and \eqref{fe2} to deduce
\begin{align}
\lefteqn{\pi^{-(\frac32-s)}\Gamma\left(\frac32-s\right)T_1\left(1;\frac32-s\right)} \\
\nonumber
&= 4\pi^{\frac12-s}\,\Gamma\left(s-\frac12\right)\zeta\left(s-\frac12\right)L_{-4}\left(s-\frac12\right)  +2\pi^{-s}\,\Gamma(s) \zeta(2s) \\
\nonumber
&\quad +4 \sum_{i,N\in\mathbb{N}}  r_2(N) \left(\frac{i}{\sqrt{N}}\right)^{s-\frac12} K_{s-\frac12}\left(2\pi i \sqrt{N}\right).
\end{align}
The functional equation~\eqref{fe0} follows from this by using~\eqref{sumpart2}.
In addition to providing another proof of the functional equation, the calculation above also demonstrates the interconnection between the
formulas~\eqref{sumpart1} and~\eqref{sumpart2}. Further functional equations of this type are considered in~[\onlinecite[p. 46]{borwein-2013}].
%\begin{align}
%L_{3}^{\text{HCP}}(1)  
%&=\frac{\pi^2}{8}
%+ \pi\;\sqrt\frac{27}{8} \;(\sqrt{3}+1)\;
%\zeta\left(\frac12\right)\,L_{-3}\left(\frac12\right)
%\nonumber \\
%&\quad
%+ \pi\, \sqrt{\frac{3}{2}}\,
%\sum_{k=1}^\infty \sum_{N=1}^\infty \frac{u_2(N)}{\sqrt{N}}\,
%\exp\left(-\pi k \sqrt{\frac{3N}{2}} \right)
%\nonumber \\
%& \quad + \pi\, \sqrt{\frac{3}{8}}\,
%\sum_{k=1}^\infty  \sum_{N=0}^\infty(-1)^k\, \frac{u_2(3N+1)}{\sqrt{N+\frac13}} 
%\exp\left(-\pi k \sqrt{\frac{3N+1}{2}}\right). \label{HCPs1}
%\end{align}

%\noindent
%From~\eqref{sumpart2} and~\eqref{sumpart4} we also have
%{\color{blue}{
%$$
%{\sum_{i,j,k}}^{\prime} \;\;\;\left. \frac{1}{(i^2+j^2+k^2)^{s}}\right|_{s=1}=...
%$$
%and
%$$
%{\sum_{i,j,k}}^{\prime} \;\;\;\left. \frac{(-1)^{i+j+k}}{(i^2+j^2+k^2)^{s}}\right|_{s=1}=...
%$$
%As a check, taking $A=1$ and $s=1$ in~\eqref{sum3} gives
%\begin{align*}
%L_3^{\text{FCC}}(1) &={\sum_{i,j,k}}^{\prime} \;\;\;\left. \frac{1}{(i^2+j^2+k^2)^{s}}\right|_{s=1} + {\sum_{i,j,k}}^{\prime} \;\;\;\left. \frac{(-1)^{i+j+k}}{(i^2+j^2+k^2)^{s}}\right|_{s=1} \\
%&=  \text{compute the values and put here}
%\end{align*}
%in agreement with~\eqref{LFCC1}.
%}}

We now evaluate the values at $s=0,\,-1,\,-2,\,-3,\ldots$ for the lattice sum. Recall from~\eqref{L3formula1} that
\begin{align}
L(A;s) &= 4\left(\frac{A+1}{2}\right)^s \zeta(s)L_{-4}(s) + \frac{\pi A}{s-1} \left(1+\frac{1}{A}\right)^s \zeta(2s-2) \\
\nonumber
&\quad + \frac{2\pi^s\sqrt{A}}{\Gamma(s)} \left(\sqrt{A}+\frac{1}{\sqrt{A}}\right)^s \sum_{i,N\in\mathbb{N}}  r_2(N) \left(\frac{N}{i^2}\right)^{(s-1)/2} K_{s-1}\left(2\pi i \sqrt{AN}\right) \\
\nonumber
&\quad + \frac{2\pi^s\sqrt{A}}{\Gamma(s)} \left(\sqrt{A}+\frac{1}{\sqrt{A}}\right)^s
\sum_{i\in\mathbb{N}} \sum_{N\in\mathbb{N}_0} (-1)^ir_2(4N+1)
 \left(\frac{2N+\frac12}{i^2}\right)^{(s-1)/2} K_{s-1}\left(2\pi i \sqrt{A(2N+\frac12)}\right).
\end{align}
On using the values $\zeta(0) = -\frac12$, $\zeta(-2) = 0$, $L_{-4}(0) = \frac12$ and the limiting value
$\lim_{s\rightarrow 0} 1/ \Gamma(s) = 0$
we readily obtain the result $L(A;0) = -1$.
Moreover, since
\begin{align}
\zeta(-2)&=\zeta(-4)=\zeta(-6) = \dots = 0, \\
L_{-4}(-1)&=L_{-4}(-3)=\zeta(-5) = \dots = 0, \\
\text{and} \quad
\lim_{s\rightarrow N} \frac{1}{\Gamma(s)} &= 0\quad \text{if $N=0,\, -1,\, -2,\, \dots$}
\end{align}
it follows that
\begin{align}
L(A;-1) = L(A;-2) = L(A;-3) = \dots = 0.
\end{align}

\begin{figure}[htbp!]
\begin{center}
\includegraphics[scale=0.4, trim=4cm 0 0 0]{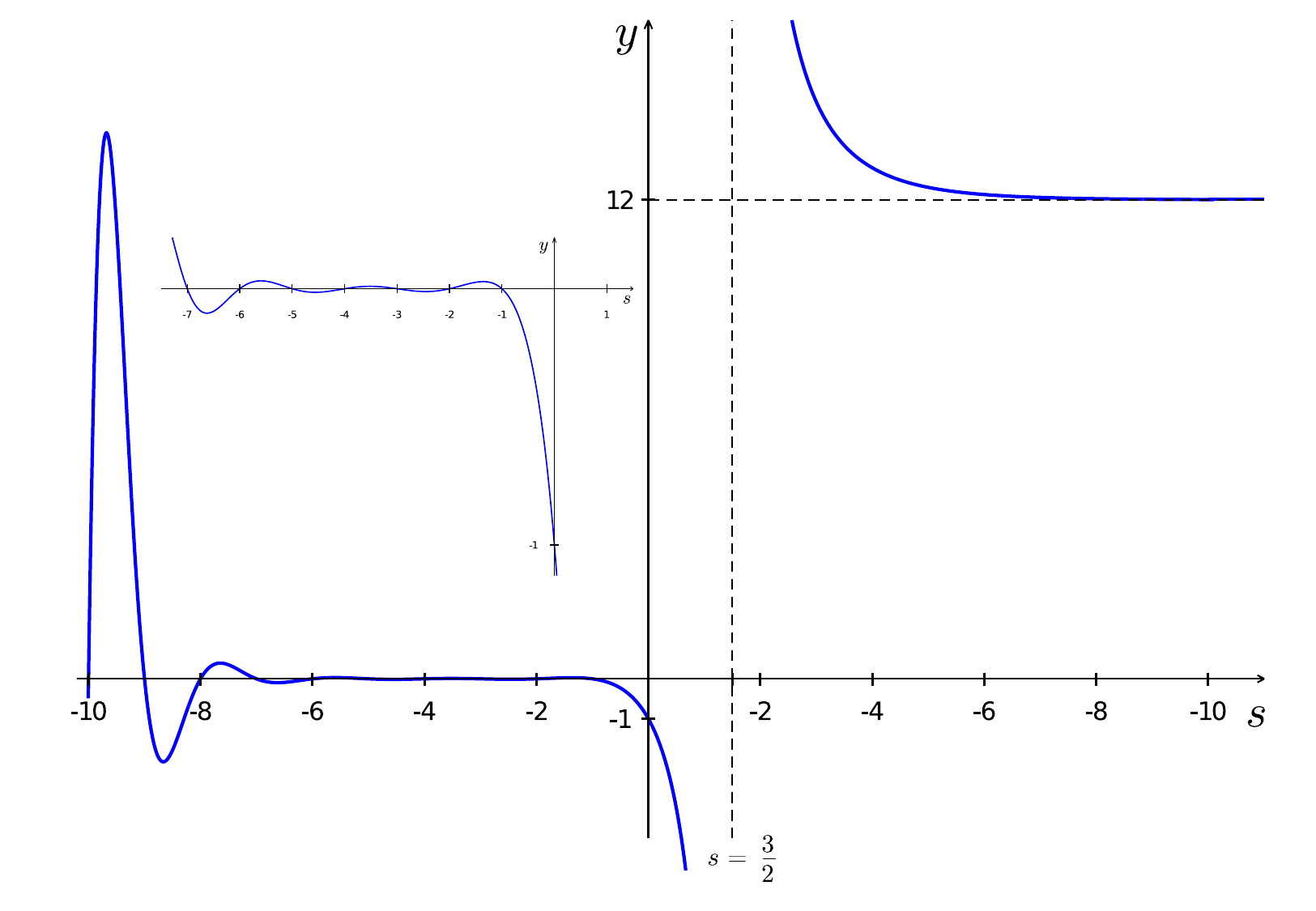}
\caption{Graph of $y=L(A=1;s)$ for $-10<s<10$ for the fcc structure. Inlet shows $y=L(A=1;s)$ for $-7<s<0$}
\label{fig3}
\end{center}
\end{figure}

The graph of $L(A=1,s)$ obtained from the formulas ~\eqref{L3formula1} and \eqref{L3formula2} on the intervals $-10<s<10$ and $-7<s<0$ is shown in Figure~\ref{fig3}, which illustrates the properties discussed in this section.
 
We briefly consider the behaviour of the lattices in the limiting cases $A\rightarrow 0^+$ and $A\rightarrow+\infty$. For example, from Eq.\eqref{Bain1} we can easily see that one of the basis vectors become zero in the limit $A\rightarrow 0^+$, leaving a sub-lattice of lower dimension. We therefore discuss each case
$A\rightarrow 0^+$ and $A\rightarrow+\infty$ both in terms of theta functions
and then in terms of the basis vectors.

First, consider the limit $A\rightarrow 0^+$. In the interval $0<A<1/3$ the theta function is
\begin{align}
\theta(A;q) = {\sum_{i,j,k\in\mathbb{Z}}}~q^{g(A;i,j,k)} ={\sum_{i,j,k\in\mathbb{Z}}}~ q^{(A(i+j)^2+(j+k)^2+(i+k)^2)/4A}.
\end{align}
As $A\rightarrow 0^+$ we have $q^{(j+k)^2/4A} \rightarrow 0$ and $q^{(i+k)^2/4A} \rightarrow 0$
unless $j=-k$ and $i=-k$, respectively. Hence,
\begin{align}
\lim_{A\rightarrow 0^+}\theta(A;q)
&= \lim_{A\rightarrow 0^+} \sum_{k\in\mathbb{Z}} \left(\sum_{i=-k} \sum_{j=-k}q^{(A(i+j)^2+(j+k)^2+(i+k)^2)/4A} \right) \\
\nonumber
&= \lim_{A\rightarrow 0^+} \sum_{k\in\mathbb{Z}} q^{A(-k-k)^2/4A} =\sum_{k=-\infty}^\infty q^{k^2}.
 \end{align}
This corresponds to the one-dimensional lattice with minimal distance~$1$. The kissing number is~2, which is in agreement with the other lattices
in the range \mbox{$0<A<1/3$}, as indicated in Table~1.
In terms of the basis vectors for the bct lattice we have
\begin{align}
\vec{b}_1 = \left(\frac12,\frac{1}{2\sqrt{A}},0\right)^\top, \quad
\vec{b}_2 = \left(\frac12,0,\frac{1}{2\sqrt{A}}\right)^\top, \quad
\vec{b}_3 = \left(0,\frac{1}{2\sqrt{A}},\frac{1}{2\sqrt{A}}\right)^\top.
\end{align}
The only linear combinations $\vec{v} = i \vec{b}_1 + j \vec{b}_2 + k \vec{b}_3$ (for $i,j,k\in \mathbb{Z}$) that remain finite in the limit~\mbox{$A\rightarrow 0^+$} occur when $i=-k$, $j=-k$ in which case we obtain $\vec{v} = -k \vec{b}_1 -k \vec{b}_2 + k \vec{b}_3 = -k(1,0,0)^\top.$ That is, the limiting lattice is just the one-dimensional lattice consisting of integer multiples of $(1,0,0)^\top$.

Now consider the limit $A\rightarrow +\infty$. For $A>1$ the theta function is 
\begin{align}
\theta(A;q) &={\sum_{i,j,k\in\mathbb{Z}}} q^{g(A;i,j,k)} ={\sum_{i,j,k\in\mathbb{Z}}}q^{(A(i+j)^2+(j+k)^2+(i+k)^2)/2}.
\end{align}
Since $q^{A(i+j)^2/2} \rightarrow 0$ as $A\rightarrow +\infty$ unless $i=-j$, it follows that
\begin{align}
\lim_{A\rightarrow +\infty}\theta(A;q)
&={\sum_{j,k\in\mathbb{Z}}} \left( \sum_{i=-j}  q^{(A(i+j)^2+(j+k)^2+(i+k)^2)/2}\right) \\
\nonumber
&={\sum_{j,k\in\mathbb{Z}}}  q^{((j+k)^2+(-j+k)^2)/2} ={\sum_{j,k\in\mathbb{Z}}} q^{j^2+k^2}.
\end{align}
This is the theta series for the two-dimensional square close packing lattice with minimal distance~$1$. The kissing number is~$4$, in agreement
with other values in the range~$A>1$ given by Table 1. In terms of the basis vectors we have
\begin{align}
\vec{b}_1 = \frac{1}{\sqrt{2}}(\sqrt{A},1,0)^\top, \quad
\vec{b}_2 = \frac{1}{\sqrt{2}}(\sqrt{A},0,1)^\top, \quad
\vec{b}_3 = \frac{1}{\sqrt{2}}(0,1,1)^\top.
\end{align}
The only linear combinations $\vec{v} = i \vec{b}_1 + j \vec{b}_2 + k \vec{b}_3$ (for $i,j,k\in \mathbb{Z}$) that remain finite in the limit~\mbox{$A\rightarrow +\infty$}
occur when $i=-j$, in which case we obtain
\begin{align}
\vec{v} = -j \vec{b}_1 +j \vec{b}_2 + k \vec{b}_3 = \frac{1}{\sqrt{2}} \left[ j(0,-1,1)^\top + k(0,1,1)^\top\right].
\end{align}
This is isomorphic to the two-dimensional square close packing lattice with minimal distance~$1$, rotated from the coordinate axes by 45 degrees.

\section{Three-body Lattice sums}
\label{ATMLSdata}

The three-body lattice sums for the ATM potential for different $A$-values according to Eq.~\eqref{eq:f_coh} are listed in Table \ref{tab:ATMdata}. The data was obtained from the treatment of the Epstein zeta function as described in section \ref{sec:Epstein zeta}.

\begin{table}[htb!]
\setlength{\tabcolsep}{0.4cm}
\renewcommand{\arraystretch}{.55}
\caption{\label{tab:ATMdata} Values for ATM three-body lattice sums. 
%\ABComment{(TODO: Insert new, even more precise values.)}
}
\begin{center}
\scriptsize{}
\begin{tabular}{|l|r|r|r|}
\hline
$A$ & $f_r^{(3)}(A)$ & $f_a^{(3)}(A)$ & $f_\mathrm{coh}^{(3)}=f_r^{(3)}(A)+f_r^{(3)}(A)$\\
\hline
0.10000000000000000  &	94.323511425615860  &	-69.582608619579300  &	24.740902806036560  \\
0.11111111111111111  &	78.996234807153090  &	-51.772409380650180  &	27.223825426502913  \\
0.12222222222222222  &	68.141987707984340  &	-40.523230176857204  &	27.618757531127130  \\
0.13333333333333333  &	60.127931145011650  &	-33.031920833977730  &	27.096010311033922  \\
0.14444444444444444  &	54.014423818716900  &	-27.821816896120083  &	26.192606922596810  \\
0.15555555555555555  &	49.227632547044850  &	-24.065452623566330  &	25.162179923478533  \\
0.16666666666666666  &	45.399641069773320  &	-21.274409710807973  &	24.125231358965350  \\
0.17777777777777777  &	42.284826583051180  &	-19.147330114086900  &	23.137496468964287  \\
0.18888888888888888  &	39.713541043931286  &	-17.491030258502235  &	22.222510785429050  \\
0.20000000000000000  &	37.565176004582040  &	-16.177483733401676  &	21.387692271180384  \\
0.21111111111111111  &	35.751832524862300  &	-15.119334299395149  &	20.632498225467145  \\
0.22222222222222222  &	34.208057391103196  &	-14.255421758422894  &	19.952635632680300  \\
0.23333333333333333  &	32.884186353006996  &	-13.541936344866237  &	19.342250008140766  \\
0.24444444444444444  &	31.741906110675053  &	-12.946849917296078  &	18.795056193378983  \\
0.25555555555555555  &	30.751222581342716  &	-12.446315486424176  &	18.304907094918548  \\
0.26666666666666666  &	29.888344501346580  &	-12.022282828652422  &	17.866061672694160  \\
0.27777777777777777  &	29.134177122870604  &	-11.660884764889580  &	17.473292357981023  \\
0.28888888888888888  &	28.473231311321676  &	-11.351323228504182  &	17.121908082817498  \\
0.30000000000000000  &	27.892820967578835  &	-11.085086344259825  &	16.807734623319007  \\
0.31111111111111111  &	27.382464087618917  &	-10.855388996987124  &	16.527075090631797  \\
0.32222222222222222  &	26.933429942612158  &	-10.656766985644794  &	16.276662956967370  \\
0.33333333333333333  &	26.538392635746654  &	-10.484778456088293  &	16.053614179658368  \\
0.34444444444444444  &	26.191163135730186  &	-10.335781403352645  &	15.855381732377538  \\
0.35555555555555555  &	25.886479914810245  &	-10.206765870266250  &	15.679714044544000  \\
0.36666666666666666  &	25.619843846530660  &	-10.095225986419386  &	15.524617860111277  \\
0.37777777777777777  &	25.387386879598203  &	-9.9990613779312730  &	15.388325501666927  \\
0.38888888888888888  &	25.185766737800538  &	-9.9165004740401060  &	15.269266263760429  \\
0.40000000000000000  &	25.012081855389425  &	-9.8460403104795870  &	15.166041544909831  \\
0.41111111111111111  &	24.863802178265250  &	-9.7863988838762060  &	15.077403294389043  \\
0.42222222222222222  &	24.738712502866125  &	-9.7364771433200620  &	15.002235359546063  \\
0.43333333333333333  &	24.634865795888576  &	-9.6953284459936560  &	14.939537349894920  \\
0.44444444444444444  &	24.550544514439906  &	-9.6621338408291200  &	14.888410673610792  \\
0.45555555555555555  &	24.484228380961902  &	-9.6361819377060480  &	14.848046443255853  \\
0.46666666666666666  &	24.434567397857776  &	-9.6168524107358490  &	14.817714987121931  \\
0.47777777777777777  &	24.400359140141877  &	-9.6036024011992400  &	14.796756738942634  \\
0.48888888888888888  &	24.380529560080320  &	-9.5959552491335140  &	14.784574310946809  \\
0.50000000000000000  &	24.374116689926883  &	-9.5934911064150800  &	14.780625583511807  \\
0.51111111111111111  &	24.380256747943890  &	-9.5958390788033300  &	14.784417669140566  \\
0.52222222222222222  &	24.398172246700100  &	-9.6026706173911090  &	14.795501629308987  \\
0.53333333333333333  &	24.427161776971960  &	-9.6136939362561870  &	14.813467840715774  \\
0.54444444444444444  &	24.466591199811035  &	-9.6286492771770470  &	14.837941922633988  \\
0.55555555555555555  &	24.515886026805056  &	-9.6473048769029430  &	14.868581149902113  \\
0.56666666666666666  &	24.574524806789046  &	-9.6694535196819120  &	14.905071287107134  \\
0.57777777777777777  &	24.642033368202980  &	-9.6949095794417950  &	14.947123788761190  \\
0.58888888888888888  &	24.717979791452187  &	-9.7235064733021200  &	14.994473318150071  \\
0.60000000000000000  &	24.801970006178507  &	-9.7550944620320370  &	15.046875544146474  \\
0.61111111111111111  &	24.893643925210420  &	-9.7895387441317570  &	15.104105181078666  \\
0.62222222222222222  &	24.992672040848475  &	-9.8267177994030350  &	15.165954241445434  \\
0.63333333333333333  &	25.098752420626415  &	-9.8665219451811340  &	15.232230475445277  \\
0.64444444444444444  &	25.211608049226257  &	-9.9088520744194820  &	15.302755974806772  \\
0.65555555555555555  &	25.330984471164940  &	-9.9536185498382640  &	15.377365921326671  \\
0.66666666666666666  &	25.456647695517326  &	-10.000740232325285  &	15.455907463192048  \\
0.67777777777777777  &	25.588382329509372  &	-10.050143625200935  &	15.538238704308440  \\
0.68888888888888888  &	25.725989912504650  &	-10.101762118796010  &	15.624227793708641  \\
0.70000000000000000  &	25.869287425869180  &	-10.155535322025365  &	15.713752103843817  \\
0.71111111111111111  &	26.018105957551953  &	-10.211408469676325  &	15.806697487875624  \\
0.72222222222222222  &	26.172289503063816  &	-10.269331895722843  &	15.902957607340980  \\
0.73333333333333333  &	26.331693886967138  &	-10.329260564360926  &	16.002433322606210  \\
0.74444444444444444  &	26.496185791055442  &	-10.391153651638422  &	16.105032139417020  \\
0.75555555555555555  &	26.665641877180228  &	-10.454974171488043  &	16.210667705692188  \\
0.76666666666666666  &	26.839947994200763  &	-10.520688640873030  &	16.319259353327737  \\
0.77777777777777777  &	27.018998459845170  &	-10.588266779406624  &	16.430731680438555  \\
0.78888888888888888  &	27.202695409400450  &	-10.657681239450945  &	16.545014169949510  \\
0.80000000000000000  &	27.390948204127140  &	-10.728907363179943  &	16.662040840947200  \\
0.81111111111111111  &	27.583672893138015  &	-10.801922963620449  &	16.781749929517560  \\
0.82222222222222222  &	27.780791723221473  &	-10.876708126908810  &	16.904083596312660  \\
0.83333333333333333  &	27.982232691721270  &	-10.953245033538540  &	17.028987658182730  \\
0.84444444444444444  &	28.187929138152157  &	-11.031517796468115  &	17.156411341684050  \\
0.85555555555555555  &	28.397819370709860  &	-11.111512314338800  &	17.286307056371058  \\
0.86666666666666666  &	28.611846324266285  &	-11.193216138171746  &	17.418630186094546  \\
0.87777777777777777  &	28.829957246812050  &	-11.276618350194180  &	17.553338896617873  \\
0.88888888888888888  &	29.052103411639130  &	-11.361709453496548  &	17.690393958142580  \\
0.90000000000000000  &	29.278239852844173  &	-11.448481271471891  &	17.829758581372290  \\
0.91111111111111111  &	29.508325121990370  &	-11.536926856036661  &	17.971398265953710  \\
0.92222222222222222  &	29.742321063989510  &	-11.627040403757988  &	18.115280660231520  \\
0.93333333333333333  &	29.980192610464800  &	-11.718817179160165  &	18.261375431304643  \\
0.94444444444444444  &	30.221907589035680  &	-11.812253444459664  &	18.409654144576024  \\
0.95555555555555555  &	30.467436547115813  &	-11.907346395178287  &	18.560090151937530  \\
0.96666666666666666  &	30.716752588962620  &	-12.004094101029036  &	18.712658487933580  \\
0.97777777777777777  &	30.969831224834856  &	-12.102495451618665  &	18.867335773216197  \\
0.98888888888888888  &	31.226650231227172  &	-12.202550106530040  &	19.024100124697128  \\
1.00000000000000000  &	31.487189521251523  &	-12.304258449363747  &	19.182931071887780  \\
\hline
\end{tabular}
\end{center}
\end{table}

The total lattice sum shows a minimum at the bcc structure ($A=\frac{1}{2}$), see Figure \ref{fig:f3(A)}. In order to prove this, we show that
for any $\nu_i\in \mathbb C$
\begin{equation}
    \frac{\partial}{\partial A} \zeta_{\Lambda(A)}^{(3)}(\vec{\nu}) \bigg\vert_{A=1/2} = 0
\end{equation}
holds. As the ATM potential is a finite sum of three-body zeta functions, its derivative therefore also vanishes. 
Let $\vec{x}(A) = B^\top (A) \vec{n}$ with $\vec{n}\in \mathbb Z^d$. Then
\[
\frac{\partial}{\partial A} \frac{1}{\vert B^\top (A) \vec{n} \vert^{\nu}}\bigg\vert_{A=1/2} = -\nu \frac{\vec{x}(1/2)^T D \vec{x}(1/2)}{\vert \vec{x}(1/2) \vert^{\nu+1}},
\]
with the diagonal traceless matrix
\[
D= {B^\top}'(1/2) \big(B^\top(1/2)\big)^{-1}=\left(\begin{matrix}
 -2/3 & 0 & 0\\
 0 & 1/3 & 0 \\
 0 & 0 & 1/3
\end{matrix}\right).
\]
which is convenient for our proof as we shall see. Thus 
\begin{equation}
\label{eq:zeta_derivative}
\frac{\partial}{\partial A} \zeta_{\Lambda(A)}^{(3)}(\vec{\nu}) \bigg\vert_{A=1/2} = -\,\sideset{}{'}\sum_{\vec{x},\vec{y}\in {\Lambda(1/2)}} \bigg(\nu_1 \frac{\vec{x}^T D \vec{x}}{\vert \vec{x} \vert^{\nu_1+1}}\frac{1}{\vert \vec{y}\vert^{\nu_2}}\frac{1}{\vert \vec{z}\vert^{\nu_3}}+ \nu_2 \frac{1}{\vert \vec{x} \vert^{\nu_1}}\frac{\vec{y}^T D \vec{y}}{\vert \vec{y}\vert^{\nu_2+1}}\frac{1}{\vert \vec{z}\vert^{\nu_3}}+\nu_3 \frac{1}{\vert \vec{x} \vert^{\nu_1}}\frac{1}{\vert \vec{y} \vert^{\nu_2}}\frac{\vec{z}^T D \vec{z}}{\vert \vec{z}\vert^{\nu_3+1}}\bigg)
\end{equation}
with the convention $\vec{z}= \vec{y}-\vec{x}$. As $\Lambda(1/2)$ is the bcc lattice, we can choose a rotated lattice $\Lambda_0$ such that $\Lambda_0 = c \Big(\mathbb Z^d\cup (\mathbb Z^d+1/2)\Big)$ 
%\textcolor{blue}{(AB: Maybe it would be useful to define the lattice matrix $B(A)$ directly in such a way, that $B(1/2)\mathbb Z^d=c(\mathbb Z^d\cup (\mathbb Z^d+1/2))$ without the need of an additional rotation.)} 
and the resulting lattice sums do of course not depend on this particular choice. 

The bcc lattice $\Lambda_0$ in this representation now exhibits the property that for $\vec{z} \in \Lambda_0$ also its cyclic permutation
\[
\sigma \vec{z} = (z_2, \dots, z_d, z_1)^T,
\]
is an element of $\Lambda_0$. Thus we have $\sigma^n \Lambda_0=\Lambda_0$ for any $n\in \mathbb N$. 

We now show that sum over the first term in Eq. \eqref{eq:zeta_derivative} vanishes and thus, in complete analogy, the two remaining sums as well. 
Averaging over cyclic permutations and using that permutations of the elements of $\vec{z}$ do not change the norm, we find
\[
-\nu_1\sum_{\vec{x},\vec{y}\in {\Lambda_0}} \frac{\vec{x}^T D \vec{x}}{\vert \vec{x} \vert^{\nu_1+1}}\frac{1}{\vert \vec{y}\vert^{\nu_2}}\frac{1}{\vert \vec{z}\vert^{\nu_3}}=-\nu_1\frac{1}{3}~ \sideset{}{'}\sum_{\vec{x},\vec{y}\in {\Lambda_0}} \sum_{n=0}^2 \Big((\sigma^n \vec{x})^T D (\sigma ^n\vec{x})\Big)\frac{1}{\vert \vec{x}\vert^{\nu_1+1}}\frac{1}{\vert \vec{y}\vert^{\nu_2}}\frac{1}{\vert \vec{z}\vert^{\nu_3}}.
\]
But as $D$ is diagonal, we have
\[
\sum_{n=0}^2 \Big((\sigma^n \vec{x})^T D (\sigma ^n\vec{x})\Big) = \vert \vec{x}\vert^2 \,\mathrm{Tr}(D) = 0,
\]
as $D$ is traceless. With the same argument for the remaining two sums, we have thus shown that all three terms in Eq.~\eqref{eq:zeta_derivative} vanish. Thus also
\[
\frac{\partial}{\partial A} \zeta_{\Lambda(A)}^{(3)}(\vec{\nu}) \bigg\vert_{A=1/2} = 0.
\]
Finally, recall that the three-body cohesive energy is a recombination of three-body zeta functions
\[
E_{\mathrm{coh}}^{(3)}/\lambda=\frac{1}{24}\zeta_{\Lambda(A)}^{(3)}(3,3,3) - \frac{3}{16} \zeta_{\Lambda(A)}^{(3)}(-1,5,5)+\frac{3}{8} \zeta_{\Lambda(A)}^{(3)}(1,3,5),
\]
and therefore,
\[
\frac{\partial}{\partial A} E_{\mathrm{coh}}^{(3)}\bigg\vert_{A=1/2}= 0.
\]
\qed 

The defining integral for $\zeta_\Lambda^{(3)}$ can be meromorphically continued to $\nu_i \in \mathbb C$ by means of the Hadamard integral. This, however, requires the computation of derivatives of the Epstein zeta function, which can be avoided for the special case of the ATM potential. Here, only the $\vec{\nu} = (-1,3,5)^T$ term leads to a hypersingular Hadamard integral, which can be reduced to a standard integral as follows. We readily find that
\[
\int_\mathrm{BZ} Z_{\Lambda,\nu}(\vec{k})\,\mathrm d \vec{k} = 0,\quad \nu>d,
\]
and thus also the meromorphic continuation to   $\nu \in \mathbb C$ equals zero. Hence, we have
\[
\zeta_\Lambda^{(3)}(-1,1,3)= \frac{1}{V_\Lambda}\int_{\mathrm BZ} Z_{\Lambda,-1}(\vec{k})\Big( Z_{\Lambda,1}(\vec{k})Z_{\Lambda,3}(\vec{k}) - Z_{\Lambda,1}(\vec{0})Z_{\Lambda,3}(\vec{0}) \Big)\,\mathrm d \vec{k},
\]
where the right-hand side is defined as a regular integral as the term in brackets scales as $\vec{k}^2$ around $\vec{k}=0$, due to reflection symmetry as $\vec{k}\to -\vec{k}$. In conclusion, the ATM potential for any lattice and any dimension can be written in terms of three generalized zeta functions that can, in turn, be efficiently computed to machine precision from singular integrals that involve products of Epstein zeta functions.

\begin{figure} \centering \includegraphics[width=\linewidth]{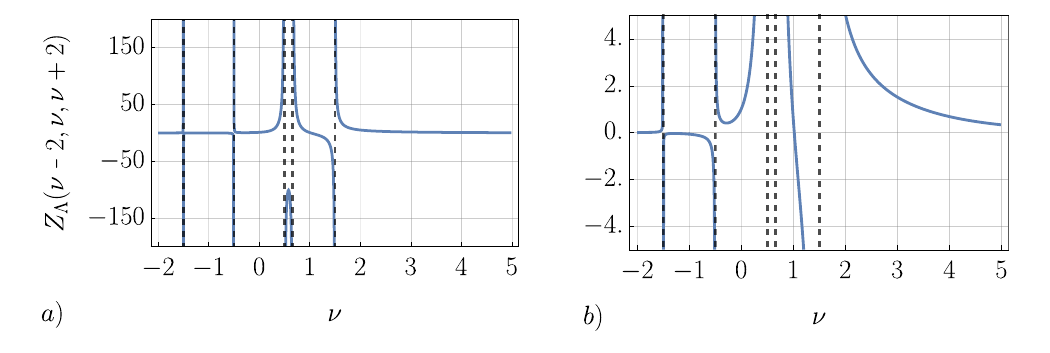} \caption{Three-body zeta function for $\Lambda=\mathbb Z$ computed via the Epstein integral representation for $\vec \nu=(\nu-2,\nu,\nu+2)^T$ including its meromorphic continuation (a).
The dashed gray lines indicate the simple poles at $\nu \in 3/2 -\mathbb N$, corresponding to the condition $\nu_i+\nu_j\in d+2\mathbb N$ for $i\neq j$, and $\nu = 2/3$, corresponding to $\nu_1+\nu_2+\nu_3 = 2d$.
Panel (b) offers a magnified view of the region close to the origin.
} 
\label{fig:threeBodyZeta1D} \end{figure}

\begin{figure} \centering 
\includegraphics[width=\linewidth]{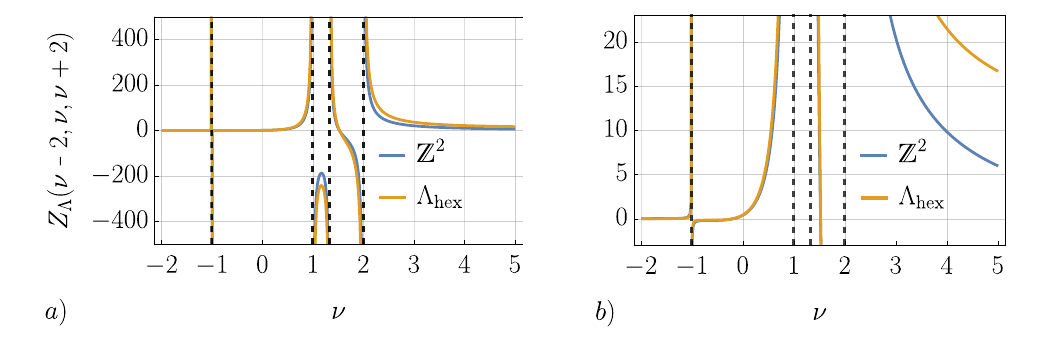} % version 
\caption{Two dimensional three-body zeta function a) for the square lattice $\Lambda=\rm{SL}=\mathbb{Z}^2$ (blue) and the hexagonal lattice $\Lambda=\rm{HL}$ (orange) as shown in Figure \ref{fig:squarelattice} for $R=1$ via the Epstein integral representation for $\vec \nu=(\nu-2,\nu,\nu+2)^T$ including its meromorphic continuation.
For $\nu\le 1$, the three-body zeta function for the square lattice and the three-body zeta function for the hexagonal lattice are visually indistinguishable.
The dashed gray lines indicate the simple poles at $\nu \in 1 -2\mathbb N$, corresponding to the condition $\nu_i+\nu_j\in d+2\mathbb N$ for $i\neq j$, and $\nu = 4/3$, corresponding to $\nu_1+\nu_2+\nu_3 = 2d$.
Panel (b) offers a magnified view of the region close to the origin.
} 
\label{fig:threeBodyZeta2D} \end{figure}

We display the behavior of the one-dimensional three-body zeta function $Z_{\mathbb Z}^{(3)}$ and two-dimensional three-body zeta functions
$Z_{\rm{SL}}^{(3)}$ for $\rm{SL}=\mathbb Z^2$ and $Z_{\rm{HL}}^{(3)}$ for 
$$
\rm{HL}=
\begin{bmatrix}
1 & 1/2 \\
0 & \sqrt{3}/2
\end{bmatrix}
\mathbb{Z}^2
$$
as a function of $\vec \nu$, including its meromorphic continuation, in Fig.~\ref{fig:threeBodyZeta1D}
and Fig.~\ref{fig:threeBodyZeta2D}.
We observe simple poles at $\nu_1+\nu_2+\nu_3=2d$ and $\nu_i+\nu_j = d-2n$, $n\in \mathbb N$, where $i\neq j$.

\section{Convergence of the lattice sum of the three-body zeta function}
In this section, we show that the the defining lattice sum for the three-body zeta function
\[
\zeta_\Lambda^{(3)}(\vec{\nu}) = \,\sideset{}{'}\sum_{\vec{x},\vec{y} \in \Lambda} \vert \vec{x} \vert^{-\nu_1}\vert \vec{y} \vert^{-\nu_2} \vert \vec{y}-\vec{x} \vert^{-\nu_3}
\]
converges if all of the following conditions hold
\begin{align*}
&\nu_i+\nu_j>d,\quad i\neq j,\quad  \mathrm{and}\quad \nu_1+\nu_2+\nu_3>2d,
\end{align*}
for $i\in \{1,2,3\}$.
Note that all summands are non-negative, so convergence of the sum does not depend on the order of summation. We first investigate the sum over $\vec{y}$, which converges if and only if
\begin{align}
\label{eq:cond1}
\nu_2+\nu_3>d.
\end{align}
We then cast the sum over $\vec{y}$ in terms of Epstein zeta functions
\[
\sideset{}{'}\sum_{\vec{y} \in \Lambda} \vert \vec{y} \vert^{-\nu_2} \vert \vec{y}-\vec{x} \vert^{-\nu_3} = \sideset{}{'}\sum_{\vec{y},\vec z \in \Lambda} \vert \vec{y} \vert^{-\nu_2} \vert \vec z \vert^{-\nu_3} \delta_{\vec y-\vec x,\vec z}  =  V_\Lambda\int_{\mathrm{BZ}} Z_{\Lambda,\nu_2}(\vec k)Z_{\Lambda,\nu_3}(\vec k) e^{2\pi i \vec y \cdot \vec k}\,\mathrm d \vec k
\]
using that
\[
V_\Lambda \int_{\mathrm BZ} e^{-2\pi i \vec z\cdot \vec k} = \delta_{\vec z,0},
\]
for a lattice vector $\vec z \in \Lambda$, as well as the lattice symmetry $\Lambda=-\Lambda$. We will now use knowledge of the singularity of the Epstein zeta function at $\vec k=0$ as well as standard results from Fourier analysis to derive the asymptotic decay of the above sum in $\vec x$.

Let $\chi(\vec k)$ be a smooth cutoff function with
\[
\chi(\vec k) = \left\{ \begin{matrix}
    1,\quad \vert \vec k\vert<r/2,\\
    0,\quad \vert \vec k \vert>r
\end{matrix}
\right.
\]
and $r>0$ chosen small enough that the support lies within an open subset of the first Brillouin zone. Adding and subtracting the cutoff in the integrand, we find
\[
V_\Lambda\int_{\mathrm{BZ}} Z_{\Lambda,\nu_2}(\vec k)Z_{\Lambda,\nu_3}(\vec k) e^{2\pi i \vec y \cdot \vec k}\,\mathrm d \vec k = f(\vec x) + V_\Lambda\int_{\mathbb R^d} \chi(\vec k)Z_{\Lambda,\nu_2}(\vec k)Z_{\Lambda,\nu_3}(\vec k) e^{2\pi i \vec y \cdot \vec k}\,\mathrm d \vec k,
\]
with $f$ decaying superalgebraically as the Fourier integral of a smooth function and where we could extend the integral on the right to $\mathbb R^d$ due to the compact support of the cutoff.  The term on the right hand side is then a standard inverse Fourier transform. 

Now separate the Epstein zeta function into an analytic function and the singularity $\hat s_\nu$, see Eq.~\eqref{eq:epstein_decomposition},
yielding 
\[
V_\Lambda \mathcal F^{-1} \Big(\chi(\vec k) (Z_{\Lambda,\nu_2}^{\mathrm{reg}}(\vec k)+c_{\nu_2} \vert  \vec k\vert^{\nu_2-d})( Z_{\Lambda,\nu_3}^{\mathrm{reg}}(\vec k)+c_{\nu_3} \vert  \vec k\vert^{\nu_3-d})\Big),
\]
with constants $c_\nu\in \mathbb R$ and $\nu_2,\nu_3\not \in {d+2\mathbb N}$. If $\nu_2$ or $\nu_3\in {d+2\mathbb N}$, then powers of $\log(\vec k)$ need to be included that do not alter convergence behavior and the proof proceeds in complete analogy. We then find that the above Fourier integral can be rewritten as
\[
\mathcal F^{-1}(h_0)(\vec x) + \mathcal F^{-1}(h_1  \vert\cdot \vert^{\nu_2-d} )(\vec x)+\mathcal F^{-1} (h_2  \vert\cdot \vert^{\nu_3-d} )(\vec x)+\mathcal F^{-1} (h_3 \vert\cdot \vert^{(\nu_2+\nu_3-d)-d} )(\vec x)
\]
with $h_0,\dots,h_3$ smooth compactly supported functions, whose Fourier transforms decay superalgebraically. Thus we only need to analyze the asymptotic decay in $\vec x $ of 
\[
\mathcal F^{-1}(h\, \vert\cdot \vert^{\nu -d}),
\]
for $h$ a smooth compactly supported function. This is however a standard result, 
\[
\Big\vert \mathcal F^{-1}(h\, \vert\cdot \vert^{\nu -d})\Big \vert\le C \vert \vec x\vert^{-\nu}, \quad |\vec x|>R
\]
for some $C,R>0$. Inserting these bounds into the sum over $\vec x$, we obtain the addition constraints 
\begin{align}    \nu_1+\nu_2>d,\quad \nu_1+\nu_3>d,\quad \nu_1+\nu_2+\nu_3 -d>d .
\label{eq:condition2}
\end{align}
The conditions in Eqs.~\eqref{eq:cond1} and \eqref{eq:condition2} then yield the convergence criteria for the three-body zeta lattice sum.

%\newpage
\bibliography{references}

\end{document}